\documentclass[usenatbib,usegraphicx]{mn2e}
\usepackage{times}

\newcommand{\kms}{\ensuremath{\mathrm{km\,s^{-1}}}}
\newcommand{\fe}{\ensuremath{\langle\mathrm{Fe}\rangle}}
\newcommand{\mgb}{\ensuremath{\mathrm{Mg}\,b}}
\newcommand{\mgfe}{\ensuremath{[\mathrm{MgFe}]}}
\newcommand{\mgone}{\ensuremath{\mathrm{Mg}_1}}
\newcommand{\mgtwo}{\ensuremath{\mathrm{Mg}_2}}
\newcommand{\cnone}{\ensuremath{\mathrm{CN}_1}}
\newcommand{\cntwo}{\ensuremath{\mathrm{CN}_2}}
\newcommand{\hbeta}{\ensuremath{\mathrm{H}\beta}}
\newcommand{\hbetag}{\ensuremath{\mathrm{H}\beta_G}}
\newcommand{\ctwo}{\ensuremath{\mathrm{C}_2 4668}}
\newcommand{\hga}{\ensuremath{\mathrm{H}\gamma_A}}
\newcommand{\hda}{\ensuremath{\mathrm{H}\delta_A}}
\newcommand{\hgf}{\ensuremath{\mathrm{H}\gamma_F}}
\newcommand{\hdf}{\ensuremath{\mathrm{H}\delta_F}}
\newcommand{\logt}{\ensuremath{\log t}}
\newcommand{\z}{\ensuremath{\mathrm{[Z/H]}}}
\newcommand{\feh}{\ensuremath{\mathrm{[Fe/H]}}}
\newcommand{\enh}{\ensuremath{\mathrm{[E/Fe]}}}
\newcommand{\afe}{\ensuremath{\mathrm{[\alpha/Fe]}}}

\newcommand{\reo}[1]{\ensuremath{r_{e}/#1}}

\newcommand\phn{\phantom{0}}

\newcommand\phm[1]{\phantom{#1}}

\begin{document}

\pubyear{2008}
\pagerange{1--37}
 
\title[Stellar population histories of ETGs. III. Coma]{The stellar
  population histories of early-type galaxies. III. The Coma
  Cluster\thanks{The data presented herein were obtained at the
  W.M.~Keck Observatory, which is operated as a scientific partnership
  among the California Institute of Technology, the University of
  California and the National Aeronautics and Space
  Administration. The Observatory was made possible by the generous
  financial support of the W.M.~Keck Foundation.}}
 
\author[S.~C. Trager, S.~M. Faber \&
  A. Dressler]{S. C. Trager$^1$\thanks{email: sctrager@astro.rug.nl},
  S. M. Faber$^2$ and Alan Dressler$^3$\\
$^1$Kapteyn Astronomical Institute, University of Groningen, Postbus
  800, NL-9700 AV Groningen, The Netherlands\\
$^2$UCO/Lick Observatory and Department of Astronomy and Astrophysics,
  University of California, Santa Cruz, Santa Cruz, CA 95064, USA\\
$^3$The Observatories of the Carnegie Institution of Washington, 813
  Santa Barbara Street, Pasadena, CA 91101, USA}

\maketitle

\begin{abstract}
We present stellar population parameters of twelve elliptical and S0
galaxies in the Coma Cluster around and including the cD galaxy NGC
4874, based on spectra obtained using the Low Resolution Imaging
Spectrograph on the Keck II Telescope.  Our data are among the most
precise and accurate absorption-line strengths yet obtained for
cluster galaxies, allowing us to examine in detail the zero-point and
scatter in the stellar population properties of Coma Cluster
early-type galaxies (ETGs).  Recent observations of red-sequence
galaxies in the high-redshift Universe and generic hierarchical
galaxy-formation models lead to the following expectations for the
stellar populations of local ETGs.  (1) In all environments, bigger
ETGs should have older stellar populations than smaller ETGs
(`downsizing'); (2) ETGs at fixed stellar mass form stars earlier and
thus should have older stellar population ages in the highest-density
environments than those in lower-density environments; and (3) the
most-massive ETGs in the densest environments should have a small
spread in stellar population ages.  We find the following surprising
results using our sample.  (1) Our ETGs have
single-stellar-population-equivalent (SSP-equivalent) ages of on
average 5--8 Gyr with the models used here, with the oldest galaxies
having SSP-equivalent ages of $\la10$ Gyr old.  This average age is
identical to the mean age of field ETGs.  (2) The ETGs in our sample
span a large range in velocity dispersion (mass) but are consistent
with being drawn from a population with a single age.  Specifically,
ten of the twelve ETGs in our sample are consistent within their
formal errors of having the same SSP-equivalent age, $5.2\pm0.2$ Gyr,
over a factor of more than 750 in mass.  We therefore find no evidence
for downsizing of the stellar populations of ETGs in the core of the
Coma Cluster.  We confirm the lack of a trend of SSP-equivalent age
with mass in the core of the Coma Cluster from \emph{all} other
samples of Coma Cluster ETG absorption-line strengths available in the
literature, but we do find from the largest samples that the
dispersion in age increases with decreasing mass.  These conclusions
stand in stark contrast to the expectations from observations of
high-redshift red-sequence galaxies and model predictions.  We suggest
that Coma Cluster ETGs may have formed the majority of their mass at
high redshifts but suffered small but detectable star formation events
at $z\approx0.1$--0.3.  In this case, previous detections of
`downsizing' from stellar populations of local ETGs may not reflect
the same downsizing seen in lookback studies of RSGs, as the young
ages of the local ETGs represent only a small fraction of their total
masses.
\end{abstract}


\section{Introduction}
\label{sec:introduction}

Our understanding of the stellar populations of early-type galaxies --
elliptical and S0 galaxies, hereafter called ETGs -- once thought to
be simple, static, and old \citep{Baade}, has undergone a revolution
in the past decades \citep[cf.][for two different views of this
revolution]{Renzini06,Faber05}.  We now understand that ETGs are a
complex, mutable population of objects with a variety of stellar
population histories.  This revolution has arisen from many different,
convergent lines of evidence: detailed studies of local ETGs,
large-area surveys of the local and distant Universe, and
semi-analytic and numerical simulations of galaxy formation.

The recent explosion of data from large-area, high-quality galaxy
surveys, from the local Universe (e.g., the Sloan Digital Sky Survey,
\citealt{SDSS}, and the 2dF Survey, \citealt{2dF}) to $z\ga1$ (e.g.,
COMBO-17: \citealt{COMBO17}; DEEP-2: \citealt{DEEP2}; VVDS:
\citealt{VVDS}; and COSMOS: \citealt{COSMOS}), has allowed us to study
the cosmic evolution of the star-formation and mass-accretion
histories of galaxies.

A fundamental discovery of these surveys has been that there exists a
strong bi-modality in the colour distribution of galaxies
\citep{Strateva01}, which take the form in the colour-magnitude or
colour-mass diagrams as a `red sequence' -- a sequence because the
distribution is narrow in colour -- and a `blue cloud' -- a cloud
because the colour dispersion is large.  This bi-modality persists out
to $z\sim1.3$ or beyond (e.g.,
\citealt{Bell04,Weiner05,Willmer06,Bundy06,Cooper06}).  Hereafter, we
call galaxies on the red-sequence RSGs (for `red-sequence galaxies').
We want to impress upon the reader that these objects are defined only
by their \emph{red colours}, not by their morphologies.

\citet{Bell04}, using data from COMBO-17, and \citet{Faber05}, using
data from DEEP2, have analysed the colour-magnitude diagrams and
luminosity functions of RSGs from $z\sim1$ to the present and found
that the stellar mass of galaxies in this sequence has increased by a
factor of 2--5 since $z\sim1$.  This result has been confirmed by many
other authors using these and other surveys (e.g.,
\citealt{Ferreras05,Bundy06,CDR06,Cooper06,Ilbert06,Zucca06,Brown07}).
This growth in mass can arise from either the growth of mass of
objects already on the red sequence and/or by the addition of
once-blue galaxies which have become red by ceasing to form stars
\citep[see e.g.,][for many references to past work relating to these
ideas]{Bell04,Faber05}.

This growth of stellar mass on the red-sequence is not distributed
uniformly along this sequence.  \citet{Bundy06}, \citet*{CDR06}, and
\citet{Brown07} have claimed that the luminosity density of the
most-massive RSGs (galaxies with $L\ga4L_*$ or
$M\ga10^{11}\,M_{\odot}$) has not evolved significantly since
$z\sim0.8$ \citep[although see][]{Faber05}.  At lower and lower
masses, the red sequence is fully populated (compared with today's red
sequence) as the Universe ages.  This appears to be confirmed by the
evolution of the Fundamental Plane (FP) of early-type galaxies (ETGs),
galaxies selected to have elliptical or S0 morphologies \citet{Treu05}
find that massive ETGs appear on the FP first, followed by the later
arrival of smaller ETGs.  In the current work we call this process --
the increasing appearance of lower-mass galaxies on the red-sequence
(and its FP) with decreasing redshift -- `downsizing' by analogy to
the decrease in specific star formation rate with decreasing redshift
\citep[e.g.,][]{Cowie96,Drory04,Juneau05,Noeske07,Zheng07}.

The growth of stellar mass on the red sequence also depends on
environment.  The earliest observational evidence for the growth of
the red sequence was the Butcher-Oemler effect \citep{BO78,BO84}, the
systematic increase of the fraction of blue galaxies relative to red
galaxies in rich clusters with increasing redshift out to $z\sim0.5$.
\citet{Cooper06} and \citet{Bundy06} have shown that the speed of
galaxies joining the red sequence is faster in higher density
environments, so that at a given mass, red-sequence galaxies in denser
environments have ceased their star formation earlier -- and thus have
older stars -- than those in less-dense environments.  It is worth
pointing out here that the DEEP2 survey, as noted by both
\citet{Bundy06} and \citet{Cooper06}, contains no rich clusters, so
the evolution of red-sequence galaxies in the densest environments
have not been probed in the same way as `field' galaxies, but FP
studies have shown that ETGs in the densest environments have reached
the FP more quickly than those in lower-density environments (e.g.,
\citealt{Gebhardt03,Treu05}, although selection effects and neglect of
rotation may play a significant r\^ole: \citealt{vdW04,vdM06b}).

In order for a blue galaxy to become red, it must stop forming stars.
This can be accomplished by consumption or removal of (cold) gas in
the galaxy.  Slow consumption of gas can turn a blue, star-forming
disc galaxy into a red S0 \citep*{LTC80}.  Early-type disc galaxies
like M31 are becoming RSGs (in fact, M31's bulge already is an RSG,
and its stellar population parameters are tabulated in
\citealt{G93,T00a}) and perhaps even ETGs by moving to earlier type in
the Hubble sequence.  \citet{RT05} and \citet{Ball06} have shown that
more than half of ETGs at the knee of the galaxy luminosity function
(i.e., at $L_{\ast}$) in SDSS are S0s.  Perhaps many S0s were once
star-formating disc galaxies -- an idea that dates back at least as
far as \citet{BO78}, if not earlier.

If the removal of gas is rapid, we call the removal process
`quenching' and say that the galaxy has `quenched' its star formation
(this is the sense used by \citealt{Faber05}; \citealt{Bell04} refer
to this as `truncation').  Quenching can be the result of different
processes: for example, the truncation of cold flows onto galaxies in
high-mass halos \citep[e.g.,][]{Cattaneo06}, explicit AGN feedback
\citep[e.g.,][]{Croton06}, or mergers that consume all the
projenitors' gas resevoirs in a large starburst
\citep[e.g.,][]{MH94a,MH94b}; a more complete list can be found in
\citet{Faber05}.  Unfortunately, quenching may be any and all of these
things.  We use `quenching' in this paper to mean \emph{rapid}
cessation of star formation to distinguish it from more gradual gas
consumption.

A heuristic model in which blue galaxies `quench', and thus end up on
the red sequence, at an epoch that depends on mass and environment
serves to provide a framework to explain these results
\citep{Bell04,Faber05}.  To be precise, if the quenching redshift of a
galaxy increases with increasing mass and increasing environmental
density, the observations described above are naturally explained.  It
is important to note however that the quenching time is \emph{not} the
stellar age of the galaxy (as will be shown in Sec.~\ref{sec:ages}
below).  In fact, the quenching time, the dominant star formation
epoch, and the mass assembly epoch are likely to be different for
red-sequence galaxies \citep{Faber05,deLucia06}.  A clear
demonstration of this effect is given by \citet{deLucia06}, who follow
the evolution of ETGs in the (to date) largest simulations of
structure formation in the Universe.  In these simulations, the most
massive galaxies in the densest environments form their stars
\emph{earlier}, but assemble \emph{later}, than lower-mass galaxies in
lower-density environments.  This is simply because the
highest-density environments are at the sites of the highest-$\sigma$
fluctuations in the primordial density field, and therefore collapse
first and merge more rapidly \citep[see, e.g.,][]{BFPR84,deLucia06};
the stars in these dark matter haloes form early and quickly but
assemble over a longer period as the haloes themselves accrete more
(sub)haloes.  Therefore, the most massive galaxies should have the
oldest stellar ages even though they assembled more recently
\citep{Kaviraj06,Faber05}; this is a generic prediction of all
hierarchical galaxy formation models.  Note however that without some
form of suppression of star formation (by whatever method), the
galaxies in the most massive haloes will continue to form stars for
far too long and will be too blue \citep{Croton06,deLucia06}.

These results suggest three predictions for red-sequence galaxies in
the local Universe:
\begin{enumerate}
  \item In all environments, lower-mass galaxies form their stars
  later -- or at least have a much larger dispersion in quenching
  redshift -- than more-massive galaxies in the same environment
  (downsizing).  In other words, the typical quenching redshift is
  higher in high-mass galaxies than in low-mass galaxies.
  \item Red-sequence galaxies of a given mass in the highest-density
  environments form their stars earlier and thus have older stars than
  galaxies of the same mass in lower-density environments.  In other
  words, the quenching redshift is higher in high-density environments
  than in low-density environments.
  \item The most-massive red-sequence galaxies in high-density
  environments should have a small spread in stellar population ages.
\end{enumerate}

In order to test these predictions for the stellar population of local
RSGs using currently available data, we need to make a crucial
assumption: that local RSGs are represented by ETGs, that is, galaxies
that have been morphologically classified as elliptical and S0
galaxies. This is only required because local samples of galaxy
absorption-line strengths are largely restricted to galaxies selected
(primarily) by morphology, with the notable exception of the NOAO
Fundamental Plane Survey \citep{Smith04,Smith06,Nelan05}\footnote{We
note that although it is possible to select true RSG samples from
SDSS, to date only the line strengths of ETG samples drawn from SDSS
have been studied, as far as we are aware
\citep[e.g.,][]{Bernardi06}.}.  That this is a reasonable assumption
in the local Universe can be seen in \citet{Blanton05}, who show that
the local red sequence is dominated by galaxies with large S\'ersic
index $n$.  This is even true at $z\approx0.8$, where
\citet{BellGEMS04} find that 85 per cent of red-sequence galaxies at
this redshift are ETGs \citep[but see][who find a lower estimate of 62
per cent]{vdW07}.  On the other hand, \citet{Renzini06} states that 70
per cent of an ETG sample selected by M.~Bernardi from the SDSS are
RSGs, while only 58 per cent of RSGs are ETGs \citep[see also the
Appendix of][]{Mitchell05}.  One should therefore keep in mind that
red-sequence galaxies are \emph{not necessarily} ETGs and that
\emph{not all} ETGs are red-sequence galaxies -- as can be seen in
\S\ref{sec:downsizing} -- but the differences ought to be small.  With
this assumption and its caveat in mind, we make three predictions for
the stellar populations of local ETGs.

Prediction (i): \emph{Lower-mass ETGs in all environments have younger
stellar population ages than high-mass ETGs.}  Downsizing of stellar
population ages of ETGs appears to have been first suggested by
\citet[although hints of the effect were also briefly discussed by
\citealt{G93}]{TFGW93}, who examined the highest-quality spectra of
ETGs available in the Lick/IDS galaxy database \citep{TWFBG98} and
found that low-$\sigma$ (low-mass) elliptical galaxies, in both the
Virgo Cluster and the general field, have younger ages on average than
high-$\sigma$ (high-mass) ellipticals.  Given the large uncertainties
in the Lick/IDS galaxy line strengths \citep[cf.][]{T00b}, further
probing of this result was difficult from that sample.  Suggestions of
downsizing can be seen from the high-quality data in the small dataset
of Paper II (Fig.~7a).  The clearest indications of downsizing in the
stellar populations of ETGs come from \citet{TMBO05} and from the NOAO
Fundamental Plane Survey, as shown in \citet{Nelan05}.  \citet{TMBO05}
find $t\propto\sigma^{0.24}$ for high-density environments and
$t\propto\sigma^{0.32}$ for low-density environments, while
\citet{Nelan05} find the very strong relation $t\propto\sigma^{0.59}$
for their (high-density) sample.  \citet{Clemens06} find a somewhat
more complicated pattern, with age increasing with $\sigma$ until the
relation saturates at moderate $\sigma$.  These studies all point
towards downsizing (as defined above) occurring in the stellar
populations of ETGs in all environments.  On the contrary,
\citet{SB06b} in the Coma Cluster and \citet{KIFvD06} in a cluster at
$z=0.33$ find \emph{no} evidence for an age--$\sigma$ relation in ETGs
using similar techniques, casting some doubt on the universality of
downsizing, at least in ETGs with $\sigma>125\,\kms$.  We return this
point in \S\ref{sec:downsizing}.

Prediction (ii): \emph{ETGs in high-density environments are older than
those in low-density environments.}  \cite{TMBO05} were the first to
claim that galaxies in high-density environments are $\sim2$ Gyr older
than those of the same mass in low-density environments.
\citet{Bernardi06}, \citet{Clemens06}, and \citet{SB06b} have all
claimed that field ETGs are on average 1--2 Gyr younger than cluster
galaxies, as expected from recent galaxy formation models
\citep[e.g.,][]{deLucia06}.

Prediction (iii): \emph{Massive ETGs in high-density environments have
a small stellar population age spread compared with lower-mass ETGs
and those in lower-density environments.}  Models
\citep[e.g.,][]{deLucia06} imply that these galaxies should have
formed their stars most quickly of all ETGs.  It is unclear with the
current samples if this is the case.  \citet[hereafter Paper II]{T00b}
seem to see a hint of a smaller spread in the ages of the most-massive
`cluster' ellipticals (in that sample, `cluster' refers to galaxies in
the Fornax and Virgo Clusters).  On the other hand, \citet{TMBO05},
who combined the Coma Cluster data of \citet{Mehlert00,Mehlert03} with
cluster galaxies from \citet{G93} and \citet{Beuing02} to create a
high-density sample and used field galaxies from \citet{G93} and
\citet{Beuing02} to create a low-density sample, find a smaller
scatter in the ages of their high-density sample galaxies than in
their low-density sample galaxies, although they do not report a
narrowing of the age--$\sigma$ relation with increasing velocity
dispersion $\sigma$ in their data.  \citet{Nelan05} show a convincing
narrowing of the age--$\sigma$ relation with increasing $\sigma$ in
the NOAO Fundamental Plane Survey cluster galaxies, but they do not
present a comparison field sample.  Further, it is not clear if the
enhancement ratios of cluster galaxies are convincingly higher than
those of field galaxies \citep{TMBO05,Bernardi06}.

In the current paper, we test the predictions presented above for ETGs
in the Coma Cluster.  We present and analyse very high signal-to-noise
spectra of twelve elliptical and S0 galaxies in the centre of the Coma
Cluster (\S\ref{sec:data}).  We combine these data with high-quality
data from the literature to explore the stellar populations of ETGs in
the high-density environment of the Coma Cluster using newly-modified
stellar population models and a new grid-inversion method described in
\S\ref{sec:parammethod}.  In \S\ref{sec:results} we find (1) the mean
single-stellar-population-equivalent (SSP-equivalent) age of Coma
Cluster galaxies in the LRIS sample is 5--7 Gyr, depending on
calibration and emission-line fill-in correction, and (2) ten of the
twelve ETGs in the LRIS sample are consistent with having the same age
of $5.2\pm0.2$ Gyr within their formal errors (ignoring systematic
calibration, emission-line correction, and stellar-population
modelling uncertainties, which amount to roughly 25 per cent).  This
age is identical within the formal errors of that of field ETGs
($5.0\pm0.1$ Gyr).  Futhermore, we see no evidence of downsizing in
the LRIS sample, but the sample is admittedly very small.  But we find
no evidence of downsizing in any sample of ETGs in the Coma Cluster
\emph{except} the red-sequence selected sample of \citet{Nelan05},
which is likely to be due a difference in emission-line corrections of
the Balmer lines.  These results imply that predictions (i) and (ii)
are violated in the stellar populations of ETGs in the Coma Cluster.
Finally, we also find that the stellar population hyperplane -- the
$Z$-plane, a correlation between age, metallicity, and velocity
dispersion; and the \enh--$\sigma$ relation, a correlation between
$\log\sigma$ and \enh\ \citep{T00b} -- exists in the Coma Cluster.  We
discuss these results, their implications, and their connection to the
formation of ETGs in general in \S\ref{sec:discussion}.  In
particular, we find that models in which stars have formed
continuously in the galaxies from high redshift and then recently
quenched to be a poor explanation of our results, as such models
violate the known fraction of red galaxies in intermediate-redshift
clusters and the present-day mass-to-light ratios of our sample
galaxies.  Instead, models with small, recent bursts (or `frostings')
of star formation on top of massive, old populations are more tenable.
We summarise our findings in \S\ref{sec:conclusions}.  Finally, two
appendixes discuss the calibration of the data and comparison of the
LRIS data with literature data.

\section{Data}
\label{sec:data}

Our intent is to determine the stellar population parameters -- ages,
metallicities, and abundance ratios -- of ETGs in the Coma cluster.
For this purpose, we have observed twelve ETGs in the core of the Coma
cluster and have also collected high-quality line-strength data from
the literature.  In this section, we discuss the acquisition and
reduction of Keck/LRIS spectroscopy, the derivation of systemic
velocities and velocity dispersion, and the extraction of Lick/IDS
line strengths.  A full description of the calibration of the line
strengths is deferred to Appendix~\ref{sec:appcal}.  At the end of
this section we briefly discuss data taken from the literature; a full
comparison with the LRIS data and presentation of the data is deferred
to Appendix~\ref{sec:others}.

\subsection{LRIS spectroscopy}
\label{sec:selection}

\begin{figure}
\caption{Field around NGC 4874, with the positions of the slitlets
overlaid.  This is a 15 second `white light' exposure taken with
LRIS directly before the spectral data described in the text.  North
is at the top, and east is to the left.  Seeing in this image is
approximately $0\farcs8$.\label{fig:coma}}
\end{figure}

\begin{table*}
  \begin{minipage}{178mm}
    \caption{Observed Coma galaxies}
    \label{tbl:sample}
    \begin{tabular}{rlccrrccccl}
      \hline \multicolumn{1}{c}{GMP}&\multicolumn{1}{c}{Other name}&
      \multicolumn{1}{c}{RA}&\multicolumn{1}{c}{DEC}&
      \multicolumn{1}{c}{$cz_{hel}$}& \multicolumn{1}{c}{$\sigma_0$}&
      \multicolumn{1}{c}{$\log(r_e/\arcsec)$}&
      \multicolumn{1}{c}{$\langle\mu_e\rangle$}&
      \multicolumn{1}{c}{$B$}& \multicolumn{1}{c}{$B-R_c$}&
      \multicolumn{1}{c}{Morph.}\\
      &&\multicolumn{2}{c}{(J2000.0)}&\multicolumn{1}{c}{(\kms)}&
      \multicolumn{1}{c}{(\kms)}&& \multicolumn{1}{c}{($r$
        mag/$\sq\arcsec$)}&\multicolumn{1}{c}{(mag)}&
      \multicolumn{1}{c}{(mag)}&\multicolumn{1}{c}{type}\\
      \hline 
      3254&D127, RB042&12:59:40.3&$+$27:58:06&$7531\pm2$&$117\pm\phn3$&0.54&20.20&17.01&1.36&S0\\
      3269&D128, RB040&12:59:39.7&$+$27:57:14&$8029\pm2$&$111\pm\phn4$&0.40&19.30&16.71&1.31&S0\\
      3291&D154, RB038&12:59:38.3&$+$27:59:15&$6776\pm2$&$\phn67^\mathrm{a}\pm\phn6$&1.08&22.25&16.77&1.28&S0\\
      3329&NGC 4874&12:59:35.9&$+$27:57:33&$7176\pm3$&$270\pm\phn4$&1.85&22.13&13.48&1.42&D\\
      3352&NGC 4872&12:59:34.2&$+$27:56:48&$7193\pm2$&$209\pm\phn3$&0.48&18.53&15.32&1.38&SB0\\
      3367&NGC 4873&12:59:32.7&$+$27:59:01&$5789\pm2$&$179\pm\phn3$&0.87&20.09&15.12&1.33&S0\\
      3414&NGC 4871&12:59:30.0&$+$27:57:22&$6729\pm2$&$164\pm\phn3$&0.92&20.24&15.02&1.38&S0\\
      3484&D157, RB014&12:59:25.5&$+$27:58:23&$6112\pm2$&$115\pm\phn3$&0.49&19.48&16.43&1.32&S0\\
      3534&D158, RB007&12:59:21.5&$+$27:58:25&$6020\pm2$&$\phn58^\mathrm{a}\pm\phn6$&0.64$^\mathrm{b}$&20.48$^\mathrm{b}$&17.25&1.22&SA0\\
      3565&&12:59:19.8&$+$27:58:26&$7206\pm3$&$\phn41^\mathrm{a}\pm10$&0.60$^\mathrm{b}$&21.67$^\mathrm{b}$&18.32&1.26&E/S0$^\mathrm{c}$\\
      3639&NGC 4867&12:59:15.2&$+$27:58:16&$4786\pm3$&$224\pm\phn3$&0.49&18.53&15.10&1.28&E\\
      3664&NGC 4864&12:59:13.1&$+$27:58:38&$6755\pm2$&$221\pm\phn3$&0.89&19.78&14.91&1.42&E\\
      \hline
    \end{tabular}

    Col.\ 1: \citet{GMP83} ID.  Col.\ 2: other names (NGC,
    \citealt{D80}, and/or \citealt{RB67} ID). Cols.\ 3 and 4:
    Coordinates.  Cols.\ 5 and 6: Heliocentric velocity and velocity
    dispersion measured through synthesised $2\farcs7$-diameter
    circular aperture (see text).  Cols.\ 7 and 8: Effective radius
    $r_e$ and mean surface brightness within $r_e$ in Gunn $r$ from
    \citet{JF94}, except as noted.  Cols.\ 9 and 10: $B$ magnitude and
    $B-R_c$ colour from \citet{Eisenhardt07}.  Col.\ 11: Morphology
    from \citet{D80}, except as noted.

    $^\mathrm{a}$Significantly below instrumental resolution
    limit; see \S\ref{sec:sigma}.

    $^\mathrm{b}$Derived from images described in
    \citet{Beijersbergen} using GALFIT \citep{GALFIT}

    $^\mathrm{c}$Morphology from \citet{Beijersbergen}
  \end{minipage}
\end{table*}

The spectra were collected with the Low-Resolution Imaging
Spectrograph \citep[LRIS:][]{LRIS} on the Keck II Telescope, which has
a $7\farcm7$ long slit.  We selected galaxies from Palomar Observatory
Sky Survey\footnote{The National Geographic Society--Palomar
Observatory Sky Atlas (POSS-I) was made by the California Institute of
Technology with grants from the National Geographic Society.} prints
of the centre of the Coma Cluster.  Galaxies were determined to be
morphologically ETGs by SMF directly from the plate material.  Several
multislit mask designs were generated using software kindly provided
by Dr.~A. Phillips at Lick Observatory.  The design that preserved the
preferred east-west orientation of the slit (to minimise atmospheric
refraction effects) and also maximised the number of ETGs along the
slit length covered a region around the cD galaxy GMP 3329 (=NGC
4874).

\begin{figure}
  \includegraphics[width=89mm]{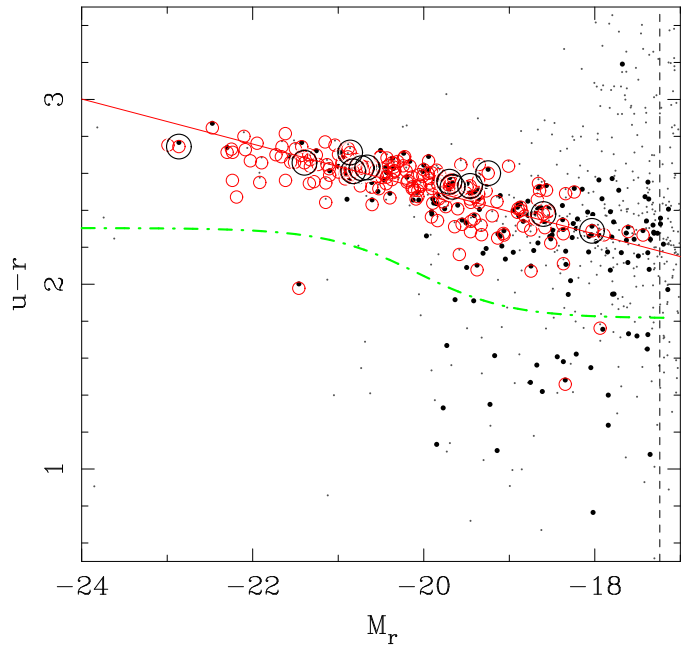}
  \caption{Sloan Digital Sky Survey (SDSS) DR6 \citep{DR6}
  colour--magnitude diagram of the central 1\fdg5-diameter region of
  the Coma Cluster.  Large black circles are the LRIS sample of
  galaxies presented here; red circles are early-type galaxies with
  line strengths analysed in the current work from several different
  samples (\S\ref{sec:literature}), some of which are outside the
  central region; black dots are galaxies with redshifts from DR6
  placing them at the distance of the Coma Cluster; and small grey
  points are other galaxies in the same field without DR6 redshifts,
  assuming that they are also in the cluster.  The dashed vertical
  line is the magnitude limit of SDSS spectroscopy; the dot-dashed
  green line demarcates the blue-red galaxy division of
  \citet{Baldry04}; and the red solid line is a fit to the brighter
  ($-23<M_r<-18$ mag), red, redshift-selected cluster members to guide
  the eye as to the location of the red sequence.}
  \label{fig:comacmd}
\end{figure}

Twelve ETGs with $-22\la M_{b_J}\la-16$ (assuming a cluster velocity
of $cz_{hel}=7007\;\kms$, \citealt{Hudson02}, and
$H_0=72\;\kms\mathrm{Mpc^{-1}}$, \citealt{F01}) were observed
(Fig.~\ref{fig:coma}; Table~\ref{tbl:sample}).  Eight objects are
typed as `S0', two are typed as `E', and one (GMP 3329=NGC 4874) is
typed as `D' by \citet{D80}; the twelfth object, GMP 3565, is typed as
`E/S0' by P.~van Dokkum in \citet{Beijersbergen}.  We therefore have
observed an ETG sample, albeit one dominated by S0
galaxies\footnote{Note that the Coma Cluster is particularly rich in
S0's \citep{D80a}.}.  All of these galaxies lie on the cluster red
sequence (Fig.~\ref{fig:comacmd}).

Spectra were obtained in three consecutive 30-minute exposures on 1997
April 7 UT with the red side of LRIS (LRIS-B was not yet available),
with seeing $\mathrm{FWHM}\approx0\farcs8$ (as determined from the
image in Fig.~\ref{fig:coma}, taken directly before the spectrographic
exposures), through clouds.  A slit width of 1\arcsec\ was used in
conjunction with the 600 line $\mathrm{mm^{-1}}$ grating blazed at
5000 \AA, giving a resolution of 4.4 \AA\ FWHM ($\sigma=1.9$ \AA,
corresponding to a velocity dispersion resolution of
$\sigma\sim125\;\kms$) and a wavelength coverage of typically
3500--6000 \AA, depending on slit placement.  Stellar spectra of five
Lick/IDS standard G and K giant stars (HR 6018, HR 6770, HR 6872, HR
7429, and HR 7576) and four F9--G0 dwarfs (HD 157089, HD 160693, HR
5968, and HR 6458) also from the Lick/IDS stellar sample
\citep{WFGB94} were observed on the same and subsequent nights through
the LRIS 1\arcsec\ long slit using the same grating.  However, the
wavelength coverage of the stellar spectra was restricted to the
region 3500--5530 \AA, preventing the calibration of indexes redder
than Fe5406 present in the galaxy spectra (such as NaD).

\subsubsection{Data reduction}
\label{sec:reduction}

The spectral data were reduced using a method that combined the
geometric rectification procedures described by \citet{KIvDF00} and
the sky-subtraction methodology of \citet{Kelson03}.  Namely, after
basic calibrations (overscan correction, bias removal, dark
correction, and flat field correction), a mapping of the geometric
distortions and wavelength calibrations were made using a suite of
Python scripts written by Dr.~Kelson, following the precepts of
\citet{KIvDF00}.  Arc lamps were used for wavelength calibration,
which was adequate (but not perfect; see Appendix~\ref{sec:appcal})
for wavelengths longer than 3900 \AA.  No slits were tilted, so
geometric rectification was generally simple.  However, these mappings
were not applied until \emph{after} the sky subtraction, for reasons
detailed by \citet{Kelson03}.  For nine of the galaxies, sky spectra
were interpolated from the slit edges, as the galaxies did not fill
the slitlets, using Python scripts written by Dr.~Kelson implementing
his sky-subtraction method.  However, for NGC 4874, which did fill its
slitlet, and for D128 and NGC 4872, whose spectra were contaminated by
that of NGC 4874, sky subtraction was performed first using the `sky'
information at the edge of their slitlets and then corrected by
comparing this sky spectrum to the average sky from all other
slitlets.

Extraction of one-dimensional spectra from the two-dimensional,
sky-subtracted long-slit images involved a simultaneous
variance-weighted extraction of the objects in a central aperture from
all three images while preserving the best possible spectral
resolution \citep{Kelson06}.  This involved using the geometric and
wavelength mappings and interpolating the spectra to preserve the
spectral resolution in the summed spectra.  This extraction also
serves as an excellent cosmic ray rejection scheme.  Variance spectra
were computed from the extracted signal and noise spectra.  To
understand the level of random errors, one-dimensional spectra were
also extracted from each image separately (after a separate cosmic-ray
cleaning step).

Various apertures were used to extract the spectra: apertures with
equivalent circular diameters of $2\farcs0$, $2\farcs7$, $3\farcs2$,
$3\farcs4$, $3\farcs6$, $3\farcs8$, $4\farcs5$, and a `physical'
aperture of \reo{2}\ diameter for all galaxies except GMP 3329 (=NGC
4874). For this galaxy, an aperture of \reo{8}\ diameter was used due
its large projected size (note that an aperture of \reo{8}\ is too
small to be extracted reliably for many of the galaxies in our sample:
for example, $\reo{8}=0\farcs33$ for GMP 3534;
Table~\ref{tbl:sample}).  These circular-aperture-equivalent
extraction apertures were chosen to match closely existing
line-strength measurements of these galaxies in the literature (see
Appendix~\ref{sec:others}).  We use only indexes from the
$2\farcs7$-diameter aperture in the analysis in this study, however.
At the distance of the Coma Cluster and assuming again
$H_0=72\;\kms\mathrm{Mpc^{-1}}$, this corresponds to a physical
diameter of 637 pc.  We note that the results given here do not change
significantly when using the `physical' \reo{2}\ aperture instead of
the 2\farcs7 aperture: for example, the mean ages of the LRIS galaxies
(ignoring GMP 3329) are $\langle\logt\rangle=0.69\pm0.02$ for the
2\farcs7 aperture and $\langle\logt\rangle=0.67\pm0.02$ for the
\reo{2}\ aperture.  When comparing other studies to ours in the
analysis, we use available gradient information to transform their
indexes to an equivalent circular aperture of the same diameter
($2\farcs7$), when possible.  We postpone discussion of the gradients
in our data to future work.

To account for the transformation of the rectangular extraction
aperture to an equivalent circular aperture, the extracted spectra
were weighted by $|r\,\Delta r|$, where $r$ is the distance from the
object centre to the object row being extracted and $\Delta r$ is the
pixel width \citep[cf.][]{G93}.  Note that the spectra were
sub-sampled along the spatial direction during the extraction process
to account for the geometric and wavelength distortions, so $\Delta r$
was typically less than one in pixel space.  The variance spectra were
weighted by $|r^2\Delta r|$ in order to preserve the noise properties.

\begin{figure}
\includegraphics[width=89mm]{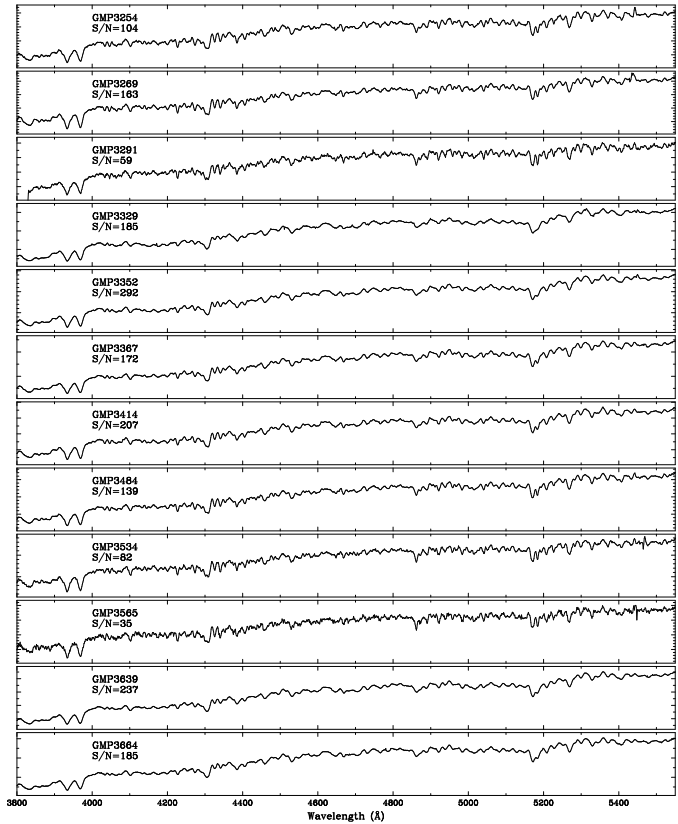}
\caption{Spectra of Coma ETGs through a central aperture of diameter
  $2\farcs7$.  Each spectrum has been flux-calibrated and shifted to
  zero systemic velocity; no smoothing has been applied.  The `noise'
  around 5450 \AA\ is the result of imperfect subtraction of the
  extremely bright [\mbox{O\,\sc{i}}]$\lambda5577$ night-sky line.
  The median S/N per $75\,\kms$ pixel in the observed wavelength range
  4285--5200 \AA\ is given.  See Fig.~\ref{fig:residuals} for an
  expanded view in the restframe region 4800--5300 \AA.
\label{fig:spectra}}
\end{figure}

In order to remove the instrumental response function from the
galaxies, flux standard stars can be used to calibrate the object
spectra onto a relative flux scale.  As the spectra were taken through
clouds, it is not possible to calibrate them to an \emph{absolute}
flux scale.  However, an absolute flux measurement is unnecessary when
our purpose is to measure line-strength indexes, as these are
\emph{relative} measurements of the absorption line fluxes with
respect to the level of the nearby continuum.  The flux standard star
BD$+33\degr2642$ \citep{Oke90} was taken through both the longslit
setup and slitless at different detector locations through the
multislit setup to cover the full wavelength range.  The extracted
spectrum was first normalised by dividing by the median count level of
the spectrum and then smoothed with a wavelet filter to derive a
sensitivity curve for each observation of the flux standard.  The
flux-calibrated spectrum of BD$+33\degr2642$ from \citet{Oke90} was
also normalised and smoothed and then divided by the normalised,
smoothed LRIS flux standard spectra to create a sensitivity spectrum
$\mathcal{F}(\lambda)$.  The sensitivity curves derived from slitless
spectra taken at the most extreme positions perpendicular to the slit
direction (i.e., at the largest wavelength spread) were combined into
a single sensitivity spectrum after flat fielding by joining them at a
convenient matching point in order to derive a sensitivity curve for
the multislit spectra (which covered a larger wavelength span than the
longslit spectra).  The final fluxed spectra through the central
$2\farcs7$-diameter equivalent circular aperture of the twelve
galaxies observed are shown in Figure~\ref{fig:spectra}.

\subsection{Velocities, velocity dispersions, and line strengths}
\label{sec:sigmaindex}

To test the predictions made in \S1, we require both the stellar
population parameters of ETGs in the Coma cluster and their velocity
dispersions.  We measure line strengths of the galaxies and compare
them with stellar population models such as \citet{W94}.  Therefore we
must know the systemic velocity of the object to place the bandpasses
on the spectra properly and we must know the velocity dispersion of
the object \citep[e.g.,][]{G93,TWFBG98} to place line strengths onto
the Lick/IDS \emph{stellar} system on which the models are defined
(see \S\ref{sec:calibration} and Appendix~\ref{sec:appcal}).

\subsubsection{Systemic velocities and velocity dispersions}
\label{sec:sigma}

We begin with a discussion of the determination of systemic velocities
$v$ and velocity dispersions $\sigma$.  Following \citet[and earlier
  work by \citealt{RW92}]{KIvDF00}, we first build a pixel-space model
$M$ of the galaxy spectrum $G$ from a stellar or stellar population
model template $T$ convolved with a broadening function $B(v,\sigma)$:
$M=B(v,\sigma)\circ T$.  In its simplest form, we want to minimise the
residuals between the galaxy and model $\chi^2=|G-M|^2=|G-B\circ
T|^2$.  However, as both noise and continuum mismatches (both
multiplicative and additive) between the galaxy and model will be
present in any practical situation, we instead write
\begin{equation}
\chi^2=\left|{\left\{G-\left[P_M(B\circ T) +
\sum_{j=0}^{K}a_j H_j\right]\right\}\times W}\right| ^2.
\end{equation}
Here $P_M$ is a multiplicative polynomial used to remove large-scale
fluxing differences between the galaxy and template spectra (which
here is \emph{not} continuum-subtracted before fitting).  In this
study, we use a fourth-order Legendre polynomial for $P_M$ to remove
the multiplicative continuum mismatch between the galaxy and template.
The zeroth order term of $P_M$ is equivalent to $\gamma$, the `line
strength parameter', found in the literature \citep{KIvDF00}.  The
additive continuum mismatch is controlled by the collection of sines
and cosines $H_j$ up to order $K$.  This is effectively a low-pass
filter used to minimise continuum mismatch.  We use
$K=1.5\Delta\lambda/100\,$\AA\ in the current study, where
$\Delta\lambda$ is the wavelength coverage (in the restframe) of the
fitting region.  $W$ is the pixel-space weight vector, which can be a
combination of the variance spectrum and any masking of `bad' regions
(e.g., poorly-subtracted strong night sky lines) desired.  (Note that
we ignore the additive polynomial functions described by
\citealt{KIvDF00}.)  The coefficients of $P_M$ and $H_j$ as well as
the desired quantities $v$ and $\sigma$ are solved for in the fitting
process, which is described in detail by \citet{KIvDF00}.  Dr.~Kelson
has kindly provided us with LOSVD, a Python script that implements
this algorithm.

For ten galaxies, the K1 giant star HR 6018 proved to be the best
velocity dispersion template, as judged by the reduced-$\chi^2$ of the
fit.  For the galaxies GMP 3534 and GMP 3565, the G0 dwarf HR 6458
provided a somewhat better fit.  Tests using the \citet{V99} spectral
models as templates suggest that the use of an well-matched template
never changes the derived velocity dispersion by more than 2 per cent.
This is negligible for our purposes of correcting the Lick/IDS line
strengths onto the stellar system (below) or for determining
correlations of velocity dispersion with stellar population
parameters.  We fit the galaxy spectra in the observed wavelength
region 4285--5200 \AA\ (roughly 4180--5080 \AA\ in the rest frame, and
therefore $K=13$), which covers the strong G band feature, H$\gamma$,
\hbeta, and many other weaker lines.  We do not fit the MgH and
\mbox{Mg\,\sc{i}} triplet region at 5100--5300 \AA\ in the rest frame
due to the strong but variable continuum depression from the dense
forest of MgH lines.  In all cases, the template stars were set to
zero recessional velocity and derived velocities were corrected to
heliocentric velocities.

We note that for three galaxies, GMP 3291, GMP 3534, and GMP 3565, the
measured velocity dispersions are significantly below the resolution
limit of $\sim125\,\kms$ and thus may be significantly in error, even
given the high signal-to-noise of the present spectra.  Two of these
galaxies, GMP 3534 and GMP 3565, were recently observed at
$\sim35\,\kms$ resolution by \citet{MG05}, and our measured velocity
dispersions match theirs within the $1\sigma$ joint errors for each of
these two objects.  While this does not guarantee that our velocity
dispersion measurement of GMP 3291 is correct, it does suggest that
our measurement is not far from the true value (see
Appendix~\ref{sec:others} for more detailed comparisons).

\subsection{Line strengths on the Lick/IDS system}
\label{sec:calibration}

Once the systemic velocity of the object is known, the bandpasses can
be placed on the spectrum and line strengths can be measured.  We give
a brief description here and leave a detailed description for
Appendix~\ref{sec:appcal}.

First, any emission lines in the spectra are corrected using GANDALF
\citep{Sarzi06}.  These corrections only affect the Fe5015 indexes, as
significant \hbeta\ emission is not detected using this procedure in
any galaxy, even though significant [O\textsc{iii}] emission is
detected in ten of the twelve.  We therefore use the uncorrected
\hbeta\ strengths throughout this paper; we discuss this further in
\S\ref{sec:agecaveats} below.  Then the spectra are smoothed to the
Lick/IDS resolution, which varies with wavelength \citep{WO97}.  Next,
the wavelengths of the Lick/IDS index bandpasses are defined using a
template star.  These bandpasses are then shifted to match the
velocity of each object.  Corrections for non-zero velocity dispersion
are made for each index of each galaxy.  Stellar indexes are then
compared to those of the same stars in the Lick/IDS stellar library
\citep{WFGB94} to determine the offsets required to bring each index
onto the Lick/IDS system.

\setcounter{table}{1}
\begin{table*}
  \vbox to220mm{\vfil Landscape table to go here
  \vfil}
  \caption{}
  \label{tbl:linestrengths}
\end{table*}

The fully-corrected (emission-, Lick/IDS system-, and velocity
dispersion-corrected) line strengths for the $2\farcs7$-diameter
equivalent circular aperture are given in
Table~\ref{tbl:linestrengths}.  We summarise this subsection (and by
extension Appendixes~\ref{sec:appcal} and \ref{sec:others}) by stating
that the LRIS data are fully corrected and well-calibrated onto the
Lick/IDS system for all indexes of interest to the current study.

\subsection{Literature data: Coma and `field' galaxies}
\label{sec:literature}

We briefly describe other high-quality line strength data of Coma
Cluster galaxies available in the literature.  A full comparison of
these data with our LRIS data is given in Appendix~\ref{sec:others}.

\begin{table}
  \caption{Literature data sources of absorption-line strengths in the
    Coma Cluster on the Lick/IDS system}
  \label{tbl:literature}
  \begin{tabular}{llc}
    \hline
    &&\multicolumn{1}{c}{Effective circular}\\
    Reference&Abbreviation&\multicolumn{1}{c}{aperture diameter}\\
    \hline
    \citet{D84}&D84&4\farcs5\\
    \citet*{FFI95}&FFI&3\farcs2\\
    \citet{GLCT92}&G92&3\farcs8\\
    \citet{Hudson02}&H01&2\farcs7\\
    \citet{J99}&J99&3\farcs4\\
    \citet{K01}&K01&3\farcs6\\
    \citet{MG05}&MG05&3\farcs0\\
    \citet{Mehlert00}&M00&2\farcs7\\
    \citet{M02}&M02&2\farcs7\\
    \citet{Nelan05}&NFPS&2\farcs0\\
    \citet{P01}&P01&2\farcs7\\
    \citet{SB06a}&SB06&2\farcs7\\
    \citet{TKBCS99}&T99&2\farcs0\\
    \citet{TWFBG98};&IDS&2\farcs7\\
    \quad \citet{LW05}&&\\
    \hline
  \end{tabular}
\end{table}

In Table~\ref{tbl:literature} we list all of the sources of
absorption-line strength data calibrated onto the Lick/IDS system (as
well as heliocentric velocities and velocity dispersions) for the Coma
Cluster that we have found in the literature.  For the Lick/IDS (IDS)
sample, the higher-order Balmer line strengths of NGC 4864 and NGC
4874 were taken from \citet{LW05}.  Most of these line strengths were
measured through fibres of various apertures, or in the case of the
Lick/IDS sample, a rectangular slit; in those cases where long slit
data were obtained, an equivalent circular aperture was synthesised
from published gradient data.  For the \citet{M02} sample, \hbeta\
line strengths were corrected for emission using the equivalent width
of the [\mbox{O\,\sc{iii}}]$\lambda5007$ \AA\ line following the
procedure detailed in \citet{T00a}; that is, we correct \hbeta\ by
adding $-0.6\times\mathrm{EW([O\,\textsc{iii}])}$ when [O\textsc{iii}]
is in emission (i.e., $\mathrm{EW([O\,\textsc{iii}])}<0$).  We note
that this correction was not made for the \citet{J99},
\citet{Mehlert00}, \citet{Nelan05}, or \citet{SB06a} samples, nor even
our own LRIS sample; we return this point in \S\ref{sec:downsizing}
below.  In Table~\ref{tbl:allcoma} in Appendix~\ref{sec:others} we
present the line strengths and stellar population parameters for all
Coma Cluster galaxies for which line strengths were available in the
literature that were taken through or could be synthesised to form a
$2\farcs7$-diameter aperture.

We also use the samples of \citet[field and Virgo cluster
ellipticals]{G93}, \citet*[field and Virgo cluster S0's]{FFI96}, and
\citet[Fornax Cluster ETGs]{K00} in our analysis.  In each case we
computed line strengths through synthesised apertures of diameter
$2\farcs7$ \emph{projected to the distance of Coma} using the
published gradient data.  That is, we measured line strengths through
a fixed physical aperture of radius $637\;\mathrm{pc}$ (assuming
$H_0=72\;\kms\,\mathrm{Mpc^{-1}}$, as above).We have combined these
three samples, excluding ETGs in the Virgo Cluster, to create a
low-density environment sample that we refer to as our `field' sample.
We can then directly compare the stellar populations of these ETGs in
less-dense environments to those of Coma ETGs.

\subsection{Galaxy masses and mass-to-light ratios}
\label{sec:masses}

\begin{table}
  \caption{Virial $M_{\mathrm{vir}}$ mass-to-light ratios
    and masses of the LRIS sample galaxies}
  \label{tbl:masses}
  \begin{tabular}{lrr}
    \hline
    GMP&
    \multicolumn{1}{c}{$\log M_{\mathrm{vir}}/L_B$}&
    \multicolumn{1}{c}{$\log M_{\mathrm{vir}}$}\\
    \hline
    3254&$0.66\pm0.08$&$10.04\pm0.08$\\
    3269&$0.36\pm0.09$&$ 9.86\pm0.09$\\
    3291&$0.55\pm0.11$&$10.02\pm0.11$\\
    3329&$1.07\pm0.08$&$11.90\pm0.08$\\
    3352&$0.44\pm0.08$&$10.49\pm0.08$\\
    3367&$0.56\pm0.08$&$10.69\pm0.08$\\
    3414&$0.49\pm0.08$&$10.66\pm0.08$\\
    3484&$0.37\pm0.08$&$ 9.98\pm0.08$\\
    3534&$0.24\pm0.12$&$ 9.52\pm0.12$\\
    3565&$0.33\pm0.23$&$ 9.18\pm0.23$\\
    3639&$0.41\pm0.08$&$10.56\pm0.08$\\
    3664&$0.67\pm0.08$&$10.89\pm0.08$\\
    \hline
  \end{tabular}
\end{table}

We are further interested in the masses and mass-to-light ratios of
ETGs in the Coma Cluster to examine the variation of line strengths
and stellar population parameters as a function of mass and to probe
for complex star-formation histories.  We have determined a `virial
mass' $M_{\mathrm{vir}}=500\,\sigma_e^2r_e\,\mathrm{M_{\odot}}$
\citep{Cappellari06}, where $\sigma_e$ is the light-weighted velocity
dispersion within the effective radius in \kms, computed as
$\sigma_e=\sigma_{2\farcs7}[r_e(\arcsec)/2\farcs7]^{-0.066}$, and
$r_e$ is the effective radius in parsecs derived from \citet[or
Table~\ref{tbl:sample} when necessary]{JF94}.  The virial
mass-to-light $M_{\mathrm{vir}}/L_B$ ratio in the $B$-band is computed
from $M_{\mathrm{vir}}$ using $L_B$.  We have corrected $B$-band
magnitudes for our galaxies from \citet{Eisenhardt07} using
$k$-corrections appropriate for a 13 Gyr-old, solar-metallicity SSP
\citep{BC03} model\footnote{Using a 4 Gyr-old, solar-metallicity SSP
model changes the $k$-corrections by less than $0.02$ mag.} and using
extinctions computed using \citet*{SFD98} for the KPNO $B$ filter and
finally assumed a distance modulus of 34.94 to the Coma Cluster to
determine $L_B$.  We do not use the direct scaling of $M/L$ with
$\sigma_e$ from \citet[their Eq.~(7)]{Cappellari06}, as the galaxies
with the lowest velocity dispersions in our sample have resulting
mass-to-light ratios much lower than their stellar populations would
suggest, which is unphysical.

\section{Derivation of stellar population parameters}
\label{sec:parammethod}

Our methodology for inferring stellar population parameters from
absorption-line strengths has changed since \citet{T00a,T00b,TWFD05}
due to improvements in the models and to increasing computer
speeds\footnote{The SSP-equivalent parameters given in \citet{TWFD05}
were based on an earlier version of our models that used the
\citet{TB95} response functions as detailed in Paper I.  However, the
method used for grid inversion is that described in
\S\ref{sec:method}.  The SSP-equivalent parameters given here
supersede those in \citet{TWFD05}.}.  We first describe our new models
and then the method used to infer SSP-equivalent
(single-stellar-population-equivalent) parameters from the observed
line strengths.  By `SSP-equivalent', we mean that the stellar
population parameters we determine are those the object would have
\emph{if} formed at a single age with a single chemical composition.
As discussed at length in Paper II and in \S\ref{sec:ages}, we do not
believe that early-type galaxies are composed of single stellar
populations.  For convenience and because of the degeneracies
discussed in \citet{W94}, Paper II and \citet[among many
others]{ST06}, our analysis is however conducted using SSP-equivalent
parameters.

\subsection{Models}
\label{sec:models}

In the current paper we follow our past practise and analyse stellar
populations with the aid of the \citet{W94} models.  We have however
improved our previous models and methods in two ways: (1) the
treatment of non-solar abundance ratios has been improved, and (2) the
`grid inversion' scheme used to infer stellar population parameters in
this paper is a significant improvement on our previous scheme.

The first major improvement in the method is the improved treatment of
the effect of non-solar abundance ratios on the line strengths.  In
the past, we \citep[hereafter Paper I]{T00a} and others
\citep*[e.g.,][hereafter TMB03]{TMB03} used the response functions of
\citet{TB95} to account for these effects in the original 21 Lick/IDS
indexes \citep{WFGB94}.  The \citet{TB95} response functions were
computed for only three stars along a 5 Gyr old, solar-metallicity
isochrone, leaving some doubt about their applicability to
significantly different populations.  These were superseded by the
response functions of \citet{KMT05}, who used three stars on 5 Gyr
isochrones at many different metallicities and also computed the
response functions for the higher-order Balmer-line indexes of
\citet{WO97}.

Recently, Worthey (priv.~comm.) has produced new response functions
for non-solar abundance ratios.  These are based on newly-computed
synthetic spectra of model stellar atmospheres for \emph{all} of the
stars in \emph{all} of the isochrones in the `vanilla' W94 (i.e.,
using original W94 isochrones) and the `Padova' W94 models (i.e.,
using the \citealt{Padova} isochrones).  One element at a time is
altered in each spectrum, extending the work of \citet{SWB05}.  Each
new spectrum is subtracted from the synthetic scaled-solar spectrum to
compute a response function for each star along the isochrone; these
are then summed to alter the model line strengths for each
single-stellar-population model.  Dr.~Worthey kindly sent us model
indexes for an elemental mixture of fixed $\enh=+0.3$ (mixture 4 of
Paper I) for the full grids of both the W94 and Padova models.
Because of the close similarity of the Padova1994 plus Salpeter IMF
version of the \citet[hereafter `BC03']{BC03} models and the Padova
W94 models, we can use the deviation of the indexes of the $\enh=+0.3$
Padova W94 models from the scaled-solar mixture Padova W94 models to
correct the BC03 models for non-solar abundance ratios.  We do not
give detailed results for stellar populations inferred from the
`Padova' W94 and BC03 models in this paper.  However, we will point
out the ranges in stellar population parameters that result from using
different models when necessary, as the entire analysis has been
carried out with the Padova W94 and BC03 models in parallel with the
vanilla W94 models.

\begin{figure*}
\includegraphics[width=178mm]{hbmgfe_coma_ap2p7gi_w94newalpha.eps}
\caption{Stellar populations of Coma ETGs observed with LRIS in
  (\hbeta, \mgfe) and (\mgb, \fe) space, where
  $\mgfe=\sqrt{\mgb\times\fe}$ and
  $\fe=(\mathrm{Fe5270}+\mathrm{Fe5335})/2$.  Line strengths in this
  figure are measured through the synthesised $2\farcs7$-diameter
  aperture.  Triangles are S0's, squares are ellipticals.  Model grids
  come from the vanilla \citet{W94} models, modified for \enh\ as
  described in the \S\ref{sec:models}.  In both panels, solid lines
  are isochrones (constant age) and dashed lines are isofers (constant
  metallicity \z).  In the left panels, the models are for solar \enh;
  models with higher \enh\ have slightly lower \hbeta\ but similar
  \mgfe.  Therefore this an appropriate grid from which to visually
  assess age and metallicity, although accurate determinations are
  made in (\hbeta, \mgb, Fe5270, Fe5335) space (see text).  In the
  right panel, grids have $\enh=0$, $+0.3$ (upper and lower,
  respectively).  This is an appropriate diagram from which to
  visually asses \enh.
  \label{fig:hbmgbfe_lris_w94}}
\end{figure*}

\begin{figure*}
\includegraphics[width=178mm]{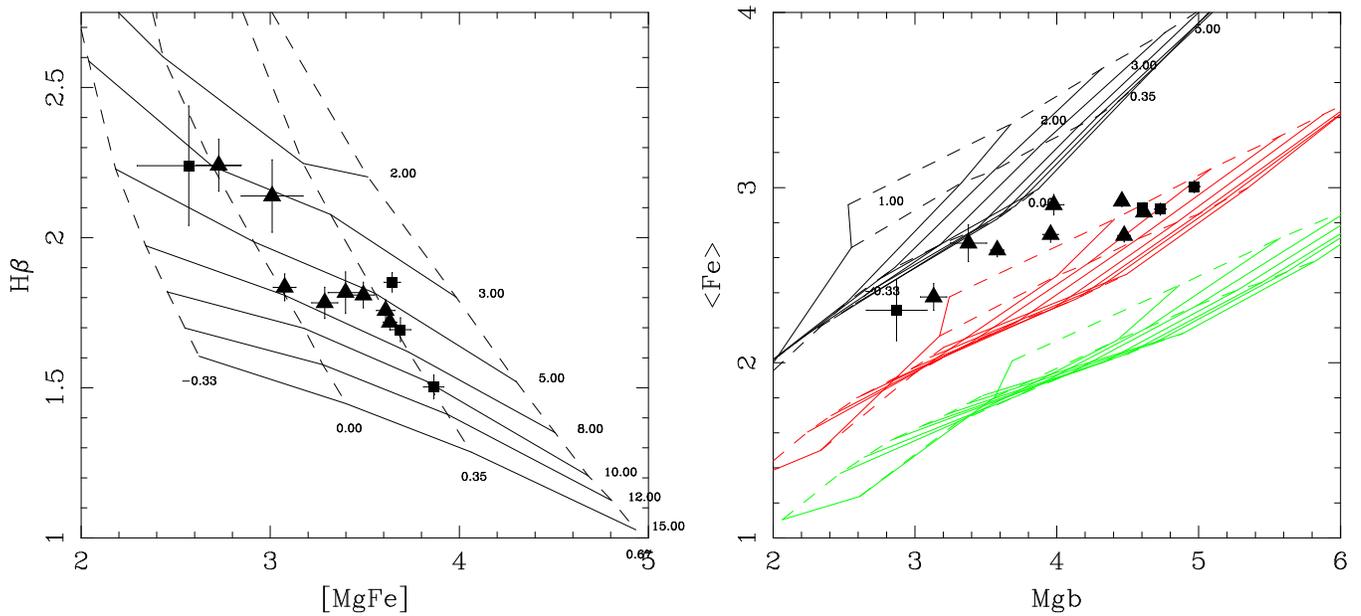}
\caption{As for Figure~\ref{fig:hbmgbfe_lris_w94}, but here the grids
  are those of TMB03, as modified using the \afe\ responses of
  \citet{KMT05}.  Although the inferred metallicities and enhancement
  ratios differ in comparison to the models used in the current work,
  the ages inferred from the \mgfe--\hbeta\ diagram (left) are very
  similar, showing that our preferred model is not the source of the
  young average age we find.
  \label{fig:hbmgbfe_lris_tmb}}
\end{figure*}

We show the new models, with our Coma Cluster ETG data superimposed,
in Figure~\ref{fig:hbmgbfe_lris_w94}.  These new $\enh=+0.3$ grids
fall between the $\enh=+0.3$ and $\enh=+0.5$ grids in, say, \mgb\
\emph{versus} \fe, of older models based on \citet{TB95} or
\citet{KMT05} so that our new results tend to have smaller \enh\ at
high \enh\ than previous studies (compare
Fig.~\ref{fig:hbmgbfe_lris_w94} with Fig.~\ref{fig:hbmgbfe_lris_tmb}).
For the LRIS sample, we find
\begin{eqnarray*}
  \logt_{\mathrm{new}}&=&(1.16\pm0.22)\logt_{\mathrm{Paper\ I}}-(0.21\pm0.03)\\
  \z_{\mathrm{new}}&=&(1.24\pm0.10)\z_{\mathrm{Paper\ I}}+(0.04\pm0.02)\\
  \enh_{\mathrm{new}}&=&(0.87\pm0.08)\enh_{\mathrm{Paper\ I}}+(0.00\pm0.01),\\
\end{eqnarray*}
where the Paper I values were computed using the W94 models and the
\citet{TB95} responses.  These relations suggest that, across the
board, our ages are somewhat lower (younger) at all ages,
metallicities are increasingly higher at high \z, and as expected,
enhancement ratios lower at high \enh\ in the new models than in those
presented in Paper I.  We also show the models of TMB03 \citep[as
modified using the \afe\ responses of][]{KMT05} and our Coma Cluster
ETG data in Figure~\ref{fig:hbmgbfe_lris_tmb} to demonstrate that the
newly-modified W94 models predict \emph{nearly the same ages} as the
TMB03 models, even though they predict lower metallicities and
enhancement ratios\footnote{We have not yet computed stellar
population parameters using the TMB03 models.  This is because our
grid-inversion method requires knowing certain internal model
parameters (in particular, the continuum and line fluxes) that are not
available to us in the TMB03 models.  We leave such parameter
estimation to future work.}.  We use our new models for all comparison
with previous studies.  That is, we compute SSP-equivalent parameters
using our present models from the line strengths given in previous
studies when comparisons are made.

Our method is not self-consistent, as we are manipulating the
atmospheric parameters of the stars of interest and not their interior
parameters, as discussed in Paper I.  That is, we are not altering the
isochrones of the \citet{W94} or the \citet{BC03} models to
accommodate changes in \enh\ (or, more generally,
$[X_j/\mathrm{Fe}]$).  \citet{PS02} have examined the methodology of
Paper I in light of $\alpha$-enhanced isochrones from the Padova group
\citep{S00}.  They note that at high metallicity, the \citet{S00}
isochrones are not significantly changed by increasing \enh\ at fixed
\feh; therefore the isochrones appear to depend on \feh, not \z, as we
have assumed.  Proctor \& Sansom therefore choose to enhance
\emph{all} the elements by \enh\ except the Fe-like elements Fe, Ca,
and Cr, which are kept at their original \feh\ level.  This is in
contrast to our method described in Paper I, where we assumed that the
isochrones depend on the total metallicity \z, as discussed in \S3.1.2
and \S5.4 in that paper, and thus some elemental abundances are
enhanced and others decreased to keep \z\ in balance.  The Proctor \&
Sansom method tends to \emph{increase} the ages of the galaxies with
the highest \enh\ compared with our method -- for galaxies with ages
$\logt\ga0.6$, the increase is $\Delta\logt\sim0.25$ (cf.\ their
Fig.~11) -- but barely affects the other stellar population
parameters.  We agree that our assumptions need updating, but we
currently prefer to use our original assumption that isochrone shapes
are governed by \z\ and wait for self-consistent stellar population
models in which indexes and isochrones are corrected for \enh\ in the
same way (see the discussion in TMB03 and attempts by
\citealt*{WPM95}; \citealt{TM03}; \citealt{LW05}; and
\citealt{Schiavon05}).  Note moreover the recent suggestion by
\citet{Weiss06} that the \citet{S00} isochrones are untrustworthy
because of errors in the low-temperature opacities; this will
certainly affect the conclusions of \citet{PS02}, \citet{TM03}, and
\citet{Schiavon05}.

Finally, we have not (yet) corrected the models for the so-called
\afe-bias inherent in the fitting functions (TMB03).  This `bias' is
however only strong ($\mathrm{[\alpha/Fe]_{intrinsic}}>0.1$ dex) when
$\feh<-0.33$ dex, uncommon in ETGs.  Such a low metallicity is not
seen in the ETGs in LRIS sample (the lowest metallicity is that of GMP
3565, which has $\z=-0.25$ dex).

\subsection{Method}
\label{sec:method}

We have also improved the scheme (`grid inversion') by which line
strengths are fit to models and therefore stellar population
parameters and errors are determined.

Previously we created large, finely-spaced grids in (\logt, \z, \enh)
space and searched the corresponding (\hbeta, \mgb, \fe) grids using a
minimal-distance statistic to find the best-fitting stellar population
parameters (Paper I).  Errors were determined by altering each line
strength by $1\sigma$ in turn and searching the grids again to find
the maximum deviation in each stellar population parameter.

Given the ever-improving speed and memory of modern computers, such a
method is no longer necessary.  We now determine stellar population
parameters directly using a non-linear least-squares code based on the
Levenberg-Marquardt algorithm in which the stellar population models
described above are linearly interpolated in (\logt, \z, \enh) on the
fly.  Confidence intervals are computed by taking the dispersion of
stellar population parameters from $10^4$ Monte Carlo trials using the
errors of the observed line strengths (Table~\ref{tbl:linestrengths}),
assuming Gaussian error distributions.  At the same time, we have
extended the method from (\hbeta, \mgb, \fe) distributions to any
combination of indexes; for example, determining stellar populations
when \ctwo\ is substituted for \mgb\ or \hda\ for \hbeta.  In fact, we
now use Fe5270 and Fe5335 in the fitting process separately rather
than \fe.  We display the data in the (\hbeta, \mgfe) and (\mgb, \fe)
planes\footnote{Here $\fe=(\mathrm{Fe5270}+\mathrm{Fe5335})/2$ and
$\mgfe=\sqrt{\mgb\times\fe}$.}, because these planes are respectively
sensitive to age and metallicity (but mostly insensitive to \enh) and
sensitive to \enh\ (e.g., Fig.~\ref{fig:hbmgbfe_lris_w94}; TMB03).  We
do \emph{not} determine stellar population parameters from these
planes.  We also compute \emph{expected} line strengths and optical
through near-infrared colours (and their errors) based on the computed
stellar population parameters.  We have tested this scheme on the
\citet{G93} data presented in Papers I and II and found it to
reproduce very closely the stellar population parameters derived there
when using models similar to those used in those papers.

\subsection{A check of the models and method}

As a sanity check of the above changes to the models and method, we
have determined the age, metallicity and enhancement ratio of the
galactic open cluster M67 using the Lick/IDS indexes given by
\citet*{SchiavonM67}.  We find $t=4.1\pm0.7$ Gyr, $\z=-0.13\pm0.06$
dex, and $\enh=0.01\pm0.03$ dex (when ignoring blue straggler stars),
in excellent agreement with both the colour-magnitude diagram turnoff
age (3.5 Gyr) and spectroscopic abundances ($\z\approx\enh\approx0$
dex) as well as the model ages and abundances ($t=3.5\pm0.5$ Gyr,
$\z=0.0\pm0.1$ dex, $\mathrm{[Mg/Fe]}=-0.05\pm0.05$) determined by
\citet{SchiavonM67}.  We are therefore confident that we can
accurately and precisely recover the stellar population parameters of
intermediate-aged, solar-composition single stellar populations.

\section{The stellar populations of early-type galaxies in the Coma
  Cluster}
\label{sec:results}

We now explore the resulting stellar population parameters of ETGs in
the Coma Cluster.  In the following, except where indicated, the terms
`age' ($t$), `metallicity' (\z), and `enhancement ratio' (\enh) always
refer to the SSP-equivalent parameters. 
We test our three predictions of \S\ref{sec:introduction} using the
stellar population parameters and their correlations with velocity
dispersion and mass.

\subsection{Line-strength distributions}

In Figure~\ref{fig:hbmgbfe_lris_w94} we plot the distribution of
\hbeta, \mgb, \fe\ line strengths of our twelve Coma Cluster galaxies.
Before discussing results based on stellar population parameters
determined from the grid inversion, three major points can be read
directly from this diagram.  First, these objects span a relatively
narrow range in age (less than a factor of 3, or less than 0.5 in
\logt).  At least 8 of the 12 galaxies have \emph{nearly-identical
ages} around 5 Gyr.  Note that these ages from this plot will not
precisely agree with the parameters given in
Table~\ref{tbl:lris_parameters} below due to lower \hbeta\ at fixed
age for larger \enh.  This means that high-\enh\ galaxies will be
slightly younger when using our age-dating method than ages read
directly from the plot.  Second, the galaxies span a large range in
metallicity \z, about 0.5 dex, as can be seen from the left-hand
panel, centred on a value of $\sim1.5$ times the solar value.  Third,
the \enh\ ratios vary between the solar value and $+0.15$ dex or so
for the newly-modified W94 models, as can be seen from the right-hand
panel.  We note here that differences between models cause subtle
\emph{bulk} changes in age and metallicity, but the overall trends are
not grossly affected by the choice of model.

\subsection{Stellar population parameter distributions}
\label{sec:parameters}

\begin{table}
  \caption{SSP-equivalent stellar population parameters of Coma
    Cluster galaxies through the 2\farcs7-diameter aperture using
    (\hbeta, \mgb, Fe5270, Fe5335)}
  \label{tbl:lris_parameters}
  \begin{tabular}{lrrrr}
    \hline
    GMP&\multicolumn{1}{c}{$\log(t/\mathrm{Gyr})$}&\multicolumn{1}{c}{$t$
    (Gyr)}&\multicolumn{1}{c}{\z}&\multicolumn{1}{c}{\enh}\\
    \hline
    3254&$0.82_{-0.14}^{+0.10}$&$6.6_{-1.9}^{+1.7}$&$0.16_{-0.05}^{+0.07}$&$0.05_{-0.01}^{+0.03}$\\
    3269&$0.96_{-0.05}^{+0.05}$&$9.2_{-1.1}^{+1.2}$&$-0.08_{-0.01}^{+0.04}$&$0.03_{-0.01}^{+0.01}$\\
    3291&$0.61_{-0.31}^{+0.17}$&$4.1_{-2.1}^{+2.0}$&$0.07_{-0.10}^{+0.13}$&$0.03_{-0.03}^{+0.04}$\\
    3329&$0.90_{-0.04}^{+0.05}$&$7.9_{-0.7}^{+1.0}$&$0.38_{-0.01}^{+0.04}$&$0.17_{-0.01}^{+0.01}$\\
    3352&$0.68_{-0.02}^{+0.04}$&$4.8_{-0.2}^{+0.4}$&$0.36_{-0.02}^{+0.02}$&$0.18_{-0.01}^{+0.01}$\\
    3367&$0.66_{-0.04}^{+0.04}$&$4.5_{-0.4}^{+0.4}$&$0.32_{-0.04}^{+0.04}$&$0.19_{-0.01}^{+0.01}$\\
    3414&$0.66_{-0.02}^{+0.04}$&$4.5_{-0.2}^{+0.4}$&$0.36_{-0.04}^{+0.04}$&$0.14_{-0.01}^{+0.01}$\\
    3484&$0.89_{-0.05}^{+0.08}$&$7.8_{-0.9}^{+1.6}$&$0.07_{-0.04}^{+0.05}$&$0.08_{-0.01}^{+0.01}$\\
    3534&$0.66_{-0.07}^{+0.07}$&$4.6_{-0.7}^{+0.8}$&$-0.09_{-0.02}^{+0.05}$&$0.06_{-0.03}^{+0.03}$\\
    3565&$0.70_{-0.22}^{+0.16}$&$5.0_{-2.0}^{+2.2}$&$-0.25_{-0.08}^{+0.17}$&$0.00_{-0.04}^{+0.09}$\\
    3639&$0.48_{-0.02}^{+0.04}$&$3.0_{-0.2}^{+0.3}$&$0.54_{-0.01}^{+0.04}$&$0.20_{-0.01}^{+0.01}$\\
    3664&$0.67_{-0.05}^{+0.05}$&$4.7_{-0.5}^{+0.6}$&$0.41_{-0.04}^{+0.05}$&$0.19_{-0.01}^{+0.01}$\\
    \hline
  \end{tabular}

Note. -- Errors are 68 per cent confidence intervals marginalised over
the other parameters.  Errors are determined from observational
uncertainties only and do not take into account systematic
uncertainties.
\end{table}

\begin{figure*}
  \includegraphics[width=178mm]{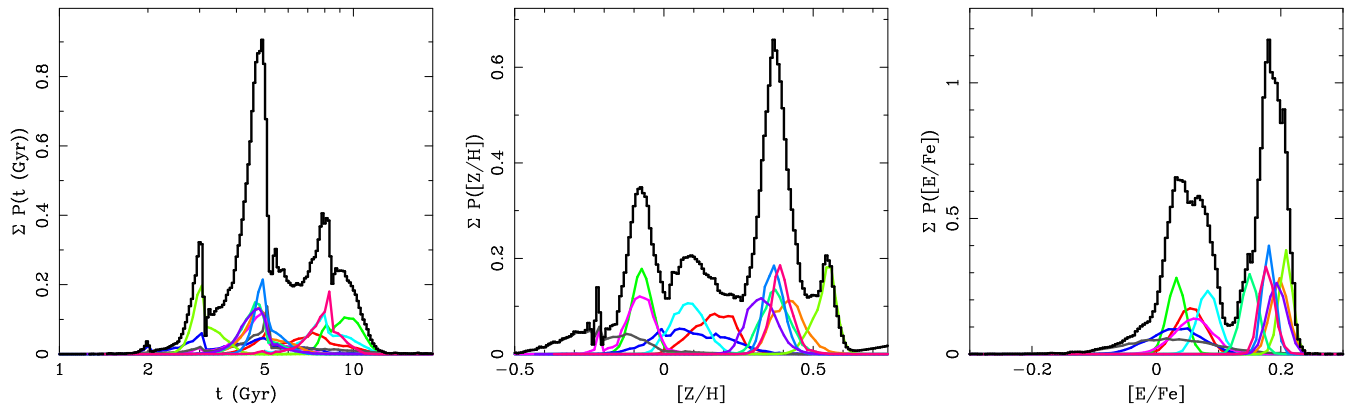}
  \caption{The summed and individual probability distributions of
    stellar population parameters for galaxies in the centre of the
    Coma Cluster, based on the (\hbeta, \mgb, Fe5270, Fe5335) indexes.
    Each galaxy's probability distribution in each parameter is shown
    in a different colour; the (re-binned) sum is shown in black.
    Jagged features in the distributions arise from edge effects in
    the models.  These summed distributions in these panels can be
    thought of as histograms smoothed by the errors in the parameters.
    \label{fig:histograms}}
\end{figure*}

In Table~\ref{tbl:lris_parameters} we present the stellar population
parameters for the twelve Coma Cluster galaxies through the
2\farcs7-diameter synthesised aperture based on the (\hbeta, \mgb,
Fe5270, Fe5335) indexes.  Figure~\ref{fig:histograms} shows the
distribution of stellar population parameters, shown as the
probability distributions of each parameter marginalised over all
other parameters and their sum.  Galaxies in this figure are
distributed as expected from Figure~\ref{fig:hbmgbfe_lris_w94}.

Examining these distributions and Table~\ref{tbl:lris_parameters} in
detail, we find that eight to ten of the twelve ETGs in this sample
have nearly the same age.  Discarding the two most divergent galaxies
-- GMP 3269 and GMP 3639 -- the mean age of the ten remaining ETGs is
$\mu_{\logt}=0.72\pm0.02$ dex ($5.2\pm0.2$ Gyr).  To quantify the age
scatter, we compute a reduced $\chi^2$ for the ETG ages:
\begin{equation}
  \chi^2_{\nu} = \frac{1}{N-1}\sum_{i=1}^{N}\left(\frac{\logt_i -
  \langle\logt\rangle}{\sigma_i}\right),
\end{equation}
where $\langle\logt\rangle=\mu$ is the weighted mean (logarithmic) age
for the $N=10$ galaxies being considered and the $N-1$ term in the
denominator arises from the fact that the we have computed $\mu$ from
the distribution of $\logt$ itself.  We have used the central value
and scale (roughly $\sigma$) of the marginalised age distribution
given by the biweight estimator \citep[see, e.g.,][]{BFG90} to
simplify the calculation.  The biweight ages and best-fitting ages are
nearly identical; the biweight scales closely match the half-width of
the (68 per cent) confidence intervals but are assumed to be symmetric
about the biweight age, unlike the confidence intervals.  We find a
reduced $\chi^2_{\nu}=2.4$ for the age residuals, or a 1 per cent
chance of being consistent with no age spread (although see below).
To determine the amount of permissible internal age scatter, we
compute $\sigma_{\logt}(\mathrm{int})=\sqrt{V-N\sigma^2_{\mu}}$, where
the sample variance $V=\sum_i (x_i-\mu)^2/(N-1)$. The maximum internal
age scatter is then 0.11 dex (1.3 Gyr).  The two deviant ETGs, GMP
3269 and GMP 3639 are notable for having the largest peculiar motions
of the sample.  GMP 3639 has a peculiar motion of
$\approx-2200\,\kms$, more than $2\sigma_{cl}$
[$\sigma_{cl}(\mathrm{Coma})=1021\,\kms$, \citealt{Smith04}] in front
of the cluster, while GMP 3269 has a peculiar motion of
$\approx1000\,\kms$ to the rear of the cluster.  If these ETGs are
assumed to be true cluster members, the mean age decreases negligibly
to $\mu_{\logt}=0.71\pm0.02$ dex ($5.1\pm0.2$ Gyr) and the internal
age spread increases to 0.14 dex (1.7 Gyr).  We conclude therefore
that ten of the twelve ETGs in this sample have \emph{the same age to
within 1 Gyr} and that including the remaining two (at least one of
which may be an interloper) increases the typical age spread to only
1.7 Gyr.

\begin{table*}
  \begin{minipage}{178mm}
  \caption{Kolmogorov-Smirnov probabilities for single-aged
  populations}
  \label{tbl:ksprobs}
  \begin{tabular}{rrrrrrrrr}
    \hline
    \multicolumn{1}{c}{$\logt$}&\multicolumn{1}{c}{Age}&
    \multicolumn{7}{c}{Sample}\\ \cline{3-9}
    \multicolumn{1}{c}{(dex)}&\multicolumn{1}{c}{(Gyr)}&
    \multicolumn{1}{c}{LRIS}&\multicolumn{1}{c}{J99}&\multicolumn{1}{c}{M00}&
    \multicolumn{1}{c}{M02}&\multicolumn{1}{c}{NFPS}&\multicolumn{1}{c}{SB06}&
    \multicolumn{1}{c}{Field}\\
    \hline
    0.1& 1.26&$0.002_{-0.002}^{+0.008}$&&&&&&\\
    0.2& 1.58&&&&&&&\\
    0.3& 2.00&&$0.014_{-0.012}^{+0.032}$&&&&&\\
    0.4& 2.51&&$\mathit{0.253}_{-0.180}^{+0.338}$&&&&&\\
    0.5& 3.16&$0.001_{-0.000}^{+0.002}$&$\mathit{0.499}_{-0.332}^{+0.365}$&$0.002_{-0.002}^{+0.005}$&&&&\\
    0.6& 3.98&$0.011_{-0.008}^{+0.023}$&$0.050_{-0.040}^{+0.116}$&$0.030_{-0.027}^{+0.085}$&&&&\\
    0.7& 5.01&$\mathit{0.614_{-0.365}^{+0.234}}$&$0.001_{-0.001}^{+0.002}$&$\mathit{0.423}_{-0.308}^{+0.260}$&$0.044_{-0.032}^{+0.058}$&&&$0.040_{-0.021}^{+0.037}$\\
    0.8& 6.31&$0.113_{-0.079}^{+0.136}$&&$\mathit{0.652}_{-0.332}^{+0.215}$&$0.006_{-0.005}^{+0.005}$&&$0.001_{-0.001}^{+0.002}$&$0.023_{-0.017}^{+0.026}$\\
    0.9& 7.94&$0.023_{-0.013}^{+0.011}$&&$\mathit{0.399}_{-0.284}^{+0.284}$&&&$0.046_{-0.039}^{+0.069}$&\\
    1.0&10.00&$0.001_{-0.001}^{+0.001}$&&$\mathit{0.126}_{-0.110}^{+0.194}$&&$0.043_{-0.035}^{+0.058}$&$\mathit{0.277}_{-0.162}^{+0.209}$&\\
    1.1&12.59&&&$0.004_{-0.003}^{+0.012}$&&$0.001_{-0.001}^{+0.002}$&$\mathit{0.300}_{-0.185}^{+0.186}$&\\
    1.2&15.85&&&&&&$0.028_{-0.025}^{+0.087}$&\\
    \hline
  \end{tabular}

  Entries in \emph{italics} are those that are consistent with a
  constant age population.  Errors are the extrema of the 68 per cent
  confidence intervals, determined from 100 realisations at the given
  age (see text).  Sample names are defined in
  Table~\ref{tbl:literature}.
  \end{minipage}
\end{table*}

\begin{figure*}
  \includegraphics[width=178mm]{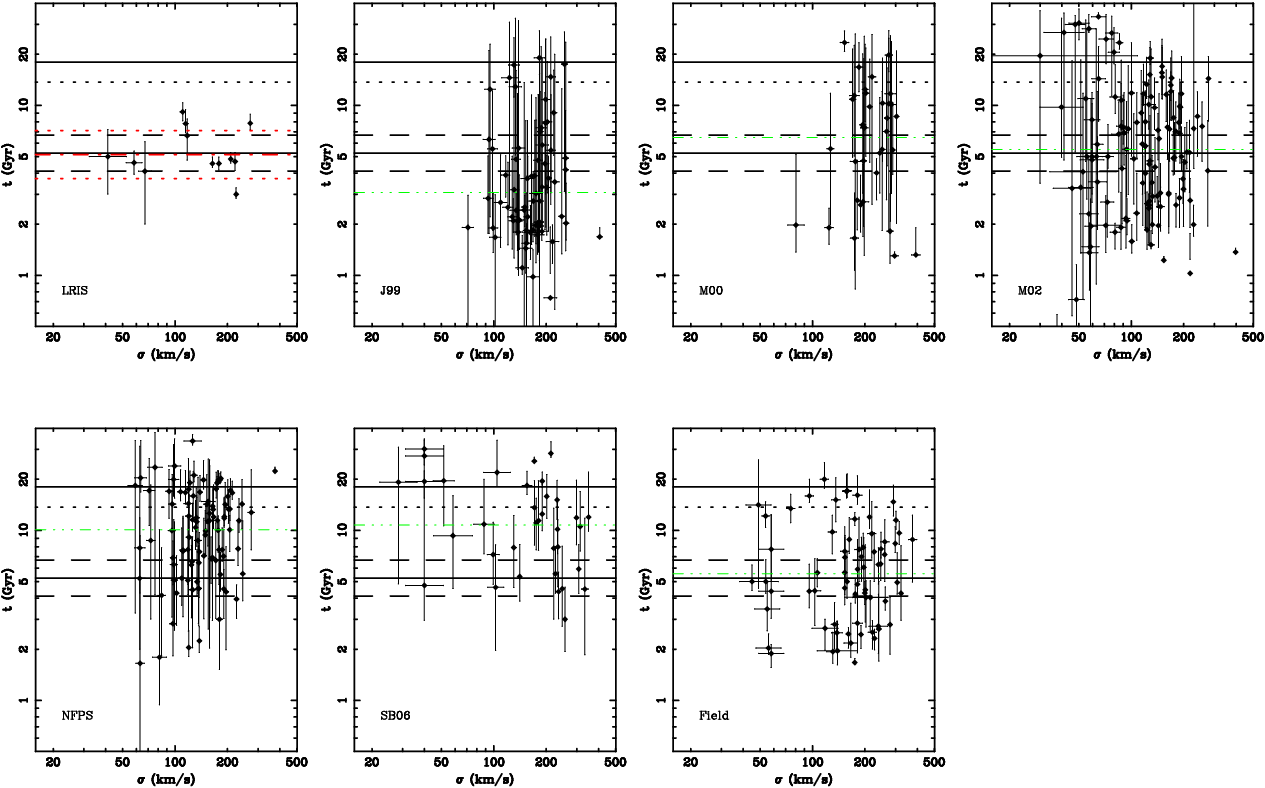}
  \caption{The $\log\sigma$--\logt\ distributions for all Coma Cluster
  ETG and RSG samples.  Top row, from left to right: the LRIS ETG
  sample; the \citet{J99} ETG sample; the \citet{Mehlert00} ETG sample
  and the \citet{M02} ETG sample.  Bottom row, from left to right: the
  \citet{Nelan05} RSG sample (Coma Cluster galaxies only); the
  \citet{SB06a} ETG sample (Coma Cluster galaxies only); and our field
  sample of ETGs.  The solid lines in all panels represent the mean
  age of the Coma Cluster ETGs after removing the two outliers GMP
  3269 and GMP 3639, and the dashed lines represent the maximum
  internal scatter in age permitted by the data.  The (red)
  dashed-dotted line in the LRIS panel is the mean age of the ETGs
  including the outliers, and the (red) dotted lines are the maximum
  scatter permitted by all twelve galaxies.  The (green) dot-dashed
  line in each other panel is the (biweight) mean age of the sample.
  Note that many field ETGs are significantly older than the Coma
  Cluster ETGs \emph{at all velocity dispersions}; this in itself is
  not a contradiction with prediction (ii) \emph{if} the scatter in
  the field galaxies is larger at all velocity dispersion than in
  clusters.  The upper dotted line is the current age of the Universe
  \citep[13.7 Gyr,][]{WMAP3} and the upper solid line is the maximum
  age of the W94 models (18 Gyr).}
  \label{fig:sigmatrels}
\end{figure*}

In order to test this single age hypothesis, we have performed a Monte
Carlo analysis in which we assume a single age for all of the galaxies
in each sample but allow each galaxy in the sample to have its
measured metallicity and enhancement ratio.  We use our models to
predict its line strengths and then perturb these using the
\emph{observed} errors (assuming a normal distribution).  We then
measure its \emph{predicted} stellar population parameters.  We do
this in total one hundred times for each sample for each assumed age,
in steps of $\Delta\logt=0.1$ dex from 0.1--1.2 dex (1.26--15.8 Gyr).
At each age, we use a Kolmogorov-Smirnov (K-S) test to determine
whether the age (and metallicity and enhancement ratio) distributions
of the observed and simulated galaxies are drawn from the same parent
distribution \citep[similar to the approach of][]{Moore01}.  We
compute the K-S probability $P_{\mathrm{KS}}$ for each of the 100
realisations at each age and take the average of the central 68 per
cent of the $P_{\mathrm{KS}}$ distribution; we take the extremes of
this central part of the distribution as the confidence limits.  We
assume that the null hypothesis, that the two populations are drawn
from the same parent distribution, is strongly ruled out when
$P_{\mathrm{KS}}\leq0.05$ and marginally ruled out when
$0.05<P_{\mathrm{KS}}\leq0.10$; otherwise we assume that the null
hypothesis is valid.  Table~\ref{tbl:ksprobs} shows the results of
these tests, and Figure~\ref{fig:sigmatrels} plots the ages as a
function of $\log\sigma$.  The results are as follows:
\begin{itemize}
\item Our LRIS sample is completely consistent with a constant age of
$\logt=0.7$ dex and marginally consistent (within the confidence
limits of the K-S probability distribution) with a constant age of
$\logt=0.8$ dex; other mean ages are strongly ruled out.
\item The \citet{J99} sample is completely consistent with a constant
age of $\logt=0.5$ dex and consistent with a constant age of
$\logt=0.4$ dex.
\item The \citet{Mehlert00} sample is completely consistent with
constant ages of $\logt=0.7$, $0.8$, and $0.9$ dex and marginally
consistent with a constant age of $\logt=1.0$ dex (an age of
$\logt=0.6$ is just on the edge of marginal acceptance).
\item The \citet{M02} sample is marginally consistent with a constant
age of $\logt=0.6$ dex.  This is in agreement with the findings of
\citet{Moore01}, who found that the \citet{M02} ETG sample was
inconsistent with a constant age when considering the the ellipticals
and S0's taken together; taken separately, however, the ellipticals
and S0's were each consistent with a different constant age.  We have
tested this hypothesis and find that \emph{both} the elliptical and S0
galaxies in \citet{M02} are consistent with constant ages of
$\logt=0.7$ or 0.8 dex, and the ellipticals are marginally consistent
with a constant age of $\logt=0.9$ dex.  The K-S probabilities suggest
that the S0's are slightly younger (higher probability at $\logt=0.7$
dex than at 0.8 dex) than the ellipticals (higher probability at
$\logt=0.8$ dex than at 0.7 dex).  Note however that we have ignored
transition morphologies such as E/S0, S0/E, and S0/a, as well as a few
later-type galaxies in these tests.
\item The \citet{Nelan05} sample is at best marginally consistent with a
constant age of $\logt=1.0$ dex.
\item The \citet{SB06a} sample is consistent with constant ages of
$\logt=1.0$ and $1.1$ dex and marginally consistent with with constant
ages of $\logt=0.9$ and $1.2$ dex.  This mean age deviates from all
other Coma ETG samples.  We return to this point in
\S\ref{sec:agescatter} below.
\item Finally, our field sample is marginally consistent with a
constant age of $\logt=0.7$ dex, but only at the extreme end of the 68
per cent confidence interval (as expected from Paper II).  The average
age of this sample is $\mu_{\logt}=0.70\pm0.01$ dex ($5.0\pm0.1$ Gyr),
with a sizable scatter of 0.29 dex (3.3 Gyr) rms.  This is
\emph{identical} within the formal errors to the mean age of the LRIS
galaxies.
\end{itemize}

We have examined the ages of our field sample (\S\ref{sec:literature})
in order to understand our result in the context of prediction (ii),
that ETGs in high-density environments should be older than those in
low-density environments.  The SSP-equivalent ages of the Coma Cluster
and field ETGs and the typical ages and intrinsic age scatter of the
Coma Cluster ETGs are shown as a function of velocity dispersion in
Figure~\ref{fig:sigmatrels}.  This then is our first major result:
Coma ETGs (in our small but extremely high-quality sample) are (i)
(nearly) \emph{coeval} in their SSP-equivalent ages and (ii) are
identical in age to the field ETGs.  In terms of our predictions, Coma
ETGs appear to violate predictions (i), that lower-mass ETGs have
younger stellar populations that high-mass ETGs, and (ii), that ETGs
in high-density environments are older than those in low-density
environments.

\begin{figure}
  \includegraphics[width=89mm]{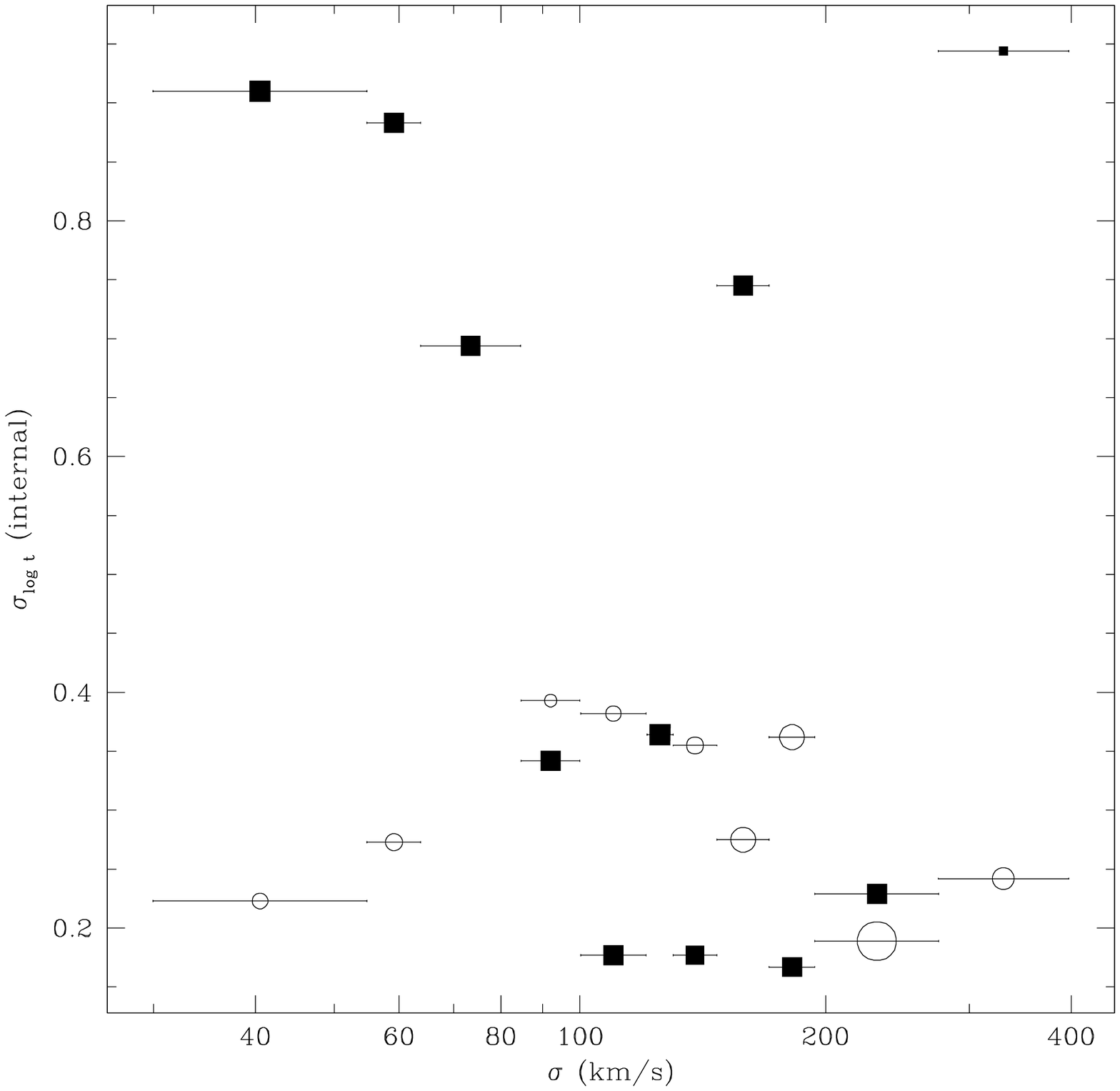}
  \caption{The estimated intrinsic logarithmic scatter as a function
  of velocity dispersion ($\sigma$) in the Coma Cluster ETG sample of
  \citet[solid squares]{M02} and our field galaxy sample (see text;
  open circles).  Galaxies were binned in $\log\sigma$ such that each
  bin had an equal number of galaxies in the \citet{M02} sample (12
  galaxies per bin), except the bin with the highest $\sigma$, which
  had only two galaxies.  Ignoring this bin, it is clear that the
  intrinsic age scatter in this sample increases with decreasing
  velocity dispersion.  The field sample was binned into the same bins
  in $\sigma$ as the Coma sample, and the point size of each bin
  represents the number of field galaxies in that bin.  It is
  important to remember that the field sample is \emph{not} complete
  and is particularly missing galaxies at $\sigma<100\,\kms$, and some
  bins are completely empty.  Even so, it appears that the intrinsic
  age scatter in the field galaxies at high $\sigma$ is typically
  slightly higher than that of the Coma galaxies.}
  \label{fig:agescatter}
\end{figure}

Our LRIS sample is too small to determine the age scatter as a
function of mass, so it is difficult to say whether prediction (iii),
that high-mass ETGs have a smaller age spread than low-mass ETGs, is
violated or not; all we can say is that the intrinsic in our
\emph{entire} sample is small.  However, the \citet{M02} sample is
large enough to make this test, as it contains 121 galaxies with
usable stellar population parameters.  We have binned these galaxies
in velocity dispersion and determined the intrinsic scatter as
described above; the results are plotted in Fig.~\ref{fig:agescatter}.
It is clear that the internal scatter tends to increase with
decreasing velocity dispersion (except in the highest velocity
dispersion bin, where only two galaxies contribute).  Such an increase
in the scatter in stellar population age with decreasing velocity
dispersion has been reported previously by \citet{P01}, although their
data were not as high quality as that of \citet[see
Appendix~\ref{sec:others}]{M02}.  We therefore suggest that the
stellar populations of ETGs in the Coma Cluster are consistent with
prediction (iii), in agreement with previous results.  We find
additionally that the field sample, at least for $\sigma>100\,\kms$,
where this sample may be representative (if not complete) has
typically a slightly larger intrinsic age scatter at a given velocity
dispersion.  This further supports prediction (iii), but the
difference is not large.

\subsubsection{Caveats on stellar population ages}
\label{sec:agecaveats}

We have considered the possibility that our line strengths may be
systematically too high in \hbeta, \mgb, Fe5270, and Fe5335.  We
tested the effects on the inferred stellar population parameters of
offsets of $\Delta\hbeta=-0.056$ \AA, $\Delta\mgb=-0.139$ \AA,
$\Delta\mathrm{Fe5270}=-0.090$ \AA, and $\Delta\mathrm{Fe5335}=-0.091$
\AA\ -- the root-mean-square deviations of calibrations onto the
Lick/IDS system (Table~\ref{tbl:calib}).  These are the maximum
allowable systematic shifts we can reasonably apply to our data, and
are larger than the average differences with respect to other
measurements in the literature (Table~\ref{tbl:comparisons}), except
for Fe5335.  We find that our LRIS galaxies are older by
$\Delta\logt=0.16\pm0.02$ dex and more metal-poor by
$\Delta\z=-0.12\pm0.02$ dex (with negligible change in \enh).  This
age shift translates into a mean age for the entire LRIS sample of
$\mu_{\logt}=0.88\pm0.02$ dex ($7.5\pm0.3$ Gyr).  If we require an
average age of 10 Gyr for this sample, an offset of
$\Delta\hbeta=-0.2$ \AA\ (with no other index changes) is required for
each galaxy, or nearly four times the Lick/IDS calibration
uncertainties.  We believe that this large shift is unlikely, and we
can therefore accept a maximum average age of roughly 7--8 Gyr for
this sample.

As mentioned in \S\ref{sec:calibration} above, we have \emph{not}
applied corrections for emission-line fill-in of \hbeta\ in our LRIS
line strengths.  We warn the reader that this means that our age
estimates are \emph{upper limits}.  Normal weak-lined red-sequence
ellipticals are nearly always LINERS, in which case we expect
$\mathrm{EW(\hbeta)}=0.62\times\mathrm{EW([O\,\textsc{iii}])}$ on
average, with little scatter \citep*[e.g.,][]{HFS97,T00a,Yan06}.
Therefore our detection of [O\textsc{iii}] emission in most of our
sample means that undetected \hbeta\ emission is filling in our
\hbeta\ absorption lines in those galaxies, making them appear older
than they truly are.  We have made a simple attempt to make such a
correction for fill-in using the correction quoted above and find that
the mean age of our twelve galaxies is $\mu_{\logt}=0.618\pm0.018$ dex
($4.1\pm0.1$ Gyr).  This is younger than that inferred above, as
expected.  This would actually make the Coma ETGs \emph{younger} than
the field ETGs, seriously violating prediction (ii).

Stellar population model differences can also affect the determination
of stellar population parameters.  The standard deviation of mean ages
for the vanilla W94, Padova W94, and BC03 models modified as described
in \S\ref{sec:method} is 28 per cent for the current sample, in the
sense that the Padova W94 models give younger ages
($\langle\logt\rangle=0.59\pm0.01$) than the vanilla W94 models
($\langle\logt\rangle=0.71\pm0.02$), which in turn give younger ages
than the BC03 models ($\langle\logt\rangle=0.81\pm0.01$).  Comparison
of Figures~\ref{fig:hbmgbfe_lris_w94} and \ref{fig:hbmgbfe_lris_tmb}
shows that the ages from the vanilla W94 models and TMB03 should in
principle be very similar.  Further, as discussed in
\S\ref{sec:models}, it possible that our models may be underestimating
ages by as much as $\Delta\logt=0.25$ dex for $\logt\ga0.6$ dex due to
incorrect treatment of abundance ratio effects \citep{PS02}, but the
true magnitude of this correction awaits the next generation of
stellar population models.

Calibration, emission fill-in correction, and model differences may
drive differences in the \emph{absolute} stellar population
parameters, but as shown by many previous studies (e.g., Paper I),
\emph{relative} stellar population parameters are nearly insensitive
to changes in the overall calibration, emission corrections, or
stellar population model.  We therefore believe that the
\emph{uniformity of ages} of our Coma ETG sample and their
\emph{similarity in ages} when compared with field ETGs are robust
results.

\subsection{Correlations of stellar population parameters with
  each other and with velocity dispersion and mass}
\label{sec:correlations}

We now ask whether there are trends in the stellar population
parameters as a function of other stellar population parameters or
with other parameters such as velocity dispersion or mass.  The latter
correlations -- if they exist -- are relevant to prediction (i), the
downsizing of the stellar populations of ETGs.

\subsubsection{The $Z$-plane and the \enh--$\sigma$ relation}
\label{sec:zplane}

\begin{table*}
  \begin{minipage}{178mm}
    \caption{$Z$-plane and \enh--$\sigma$ relation parameters for ETGs
    through an aperture of 2\farcs7 projected to the distance of the
    Coma Cluster, using new W94 models}
    \label{tbl:hyperplane}
    \begin{tabular}{llrrrr}
      \hline
      &\multicolumn{1}{c}{$\alpha$}&\multicolumn{1}{c}{$\beta$}&
      \multicolumn{1}{c}{$\gamma$, Zero-point}&
      \multicolumn{1}{c}{$\delta$}&\multicolumn{1}{c}{$\epsilon$, Zero-point}\\
      Data set&
      \multicolumn{1}{c}{$d\z/d\log\sigma$}&\multicolumn{1}{c}{$d\z/d\logt$}&
      \multicolumn{1}{c}{($Z$-plane)}&\multicolumn{1}{c}{$d\enh/d\log\sigma$}&
      \multicolumn{1}{c}{(\enh--$\sigma$)}\\
      \hline
      \multicolumn{6}{l}{Low-density environment ETG samples:}\\
      \quad Paper II$^\mathrm{a}$&$0.76\pm0.13$&$-0.73\pm0.06$&$-0.87\pm0.30$&$0.33\pm0.01$&$-0.58\pm0.01$\\
      \quad Paper II$^\mathrm{b}$&$1.05\pm0.06$&$-0.71\pm0.05$&$-1.51\pm0.15$&$0.25\pm0.02$&$-0.41\pm0.01$\\
      \quad Field$^\mathrm{c}$&$1.19\pm0.07$&$-0.72\pm0.05$&$-1.85\pm0.17$&$0.24\pm0.01$&$-0.40\pm0.01$\\
      \multicolumn{6}{l}{Coma Cluster ETG and RSG samples:}\\
      \quad LRIS&$0.97\pm0.12$&$-0.78\pm0.12$&$-1.26\pm0.31$&$0.35\pm0.03$&$-0.64\pm0.01$\\
      \quad \citet{J99}&$1.38\pm0.21$&$-0.94\pm0.07$&$-2.09\pm0.46$&$0.30\pm0.04$&$-0.51\pm0.01$\\
      \quad \citet{Mehlert00}&$1.39\pm0.31$&$-0.79\pm0.08$&$-2.34\pm0.78$&$0.32\pm0.07$&$-0.57\pm0.01$\\
      \quad \citet{M02}&$1.12\pm0.09$&$-0.81\pm0.04$&$-1.58\pm0.19$&$0.33\pm0.02$&$-0.56\pm0.01$\\
      \quad \citet{SB06a}$^\mathrm{d}$&$0.94\pm0.12$&$-0.88\pm0.12$&$-1.14\pm0.34$&$0.21\pm0.04$&$-0.36\pm0.01$\\
      \quad \citet{Nelan05}$^\mathrm{d}$&$1.23\pm0.14$&$-0.96\pm0.08$&$-1.80\pm0.28$&$0.31\pm0.02$&$-0.56\pm0.01$\\
      \hline
    \end{tabular}

  $^\mathrm{a}$As published in Paper II. These parameters were not
    measured from indexes projected to Coma distance but those in
    \reo{8}-diameter aperture and were also inferred from original
    vanilla W94 models using the \citet{TB95} non-solar abundance
    index response functions.

  $^\mathrm{b}$Using vanilla W94 models with new non-solar abundance index
    response functions, as described in the text.

  $^\mathrm{c}$Galaxies from \citet{G93}, \citet{K00}, and
    \citet{FFI96}, excluding Virgo Cluster galaxies to simulate a
    `low-density environment' sample, as described in
    \S\ref{sec:literature}.

  $^\mathrm{d}$Coma Cluster galaxies only.
  \end{minipage}
\end{table*}

The stellar population parameters \logt, \z, and \enh\ together with
the velocity dispersion $\log\sigma$ form a two-dimensional family in
these four variables, as shown in Paper II for elliptical galaxies in
environments of lower density than Coma (including the Virgo and
Fornax clusters).  The correlation between age and velocity dispersion
in that sample was weak and therefore we associated the two primary
variables in the four-dimensional space with age and velocity
dispersion in Paper II.  This association is tantamount to declaring
that there exists a temporal relation between SSP-equivalent age and
metallicity and also that velocity dispersion plays a role in the
formation of ETGs.  We also associate age and velocity dispersion with
the primary variables in this set of galaxies, as we find no
correlation between age and velocity dispersion in the present sample.
As in Paper II, we find at best a weak anti-correlation between \logt\
and \enh\ (correlation coefficient of $-0.51$ for the LRIS sample), so
we claim again that the variation in stellar population parameters can
be split into an \enh--$\sigma$ relation and a metallicity hyperplane,
the $Z$-plane.  The $Z$-plane has the form
\begin{equation}
  \z=\alpha\log\sigma+\beta\logt+\gamma \label{eq:zplane}
\end{equation}
Coefficients of Eq.~\ref{eq:zplane} are given in the first three
columns of Table~\ref{tbl:hyperplane} for the original sample of Paper
II using the models described therein; the sample of Paper II using
the current vanilla W94 models; a sample consisting of local field E
and S0's from \citet{G93}, \citet{FFI96}, and \citet{K00}, removing
the Virgo Cluster galaxies; the LRIS sample; and five other samples of
Coma Cluster galaxies: \citet{J99}, \citet{Mehlert00}, \citet{M02},
\citet[Coma Cluster galaxies only]{Nelan05}, and \citet[Coma Cluster
galaxies only]{SB06a}.  Coefficients were determined by minimising the
minimum absolute deviations from a plane (after subtracting the mean
values of each quantity), as described in \citet{JFK96} and used in
Paper II.  Uncertainties were determined by making 1000 Monte Carlo
realisations in which the the line strength indexes of the galaxies
were perturbed using their (Gaussian) errors, stellar population
parameters were determined from the new indexes, and new planes were
fit to these parameters.  We find from these realisations that the
slopes $\alpha$ ($=d\z/d\log\sigma$) and $\beta$ ($=d\z/d\logt$) are
nearly uncorrelated with each other, but the zero-point $\gamma$ is
strongly correlated with $\alpha$ and somewhat less with $\beta$.

\begin{figure*}
  \includegraphics[width=178mm]{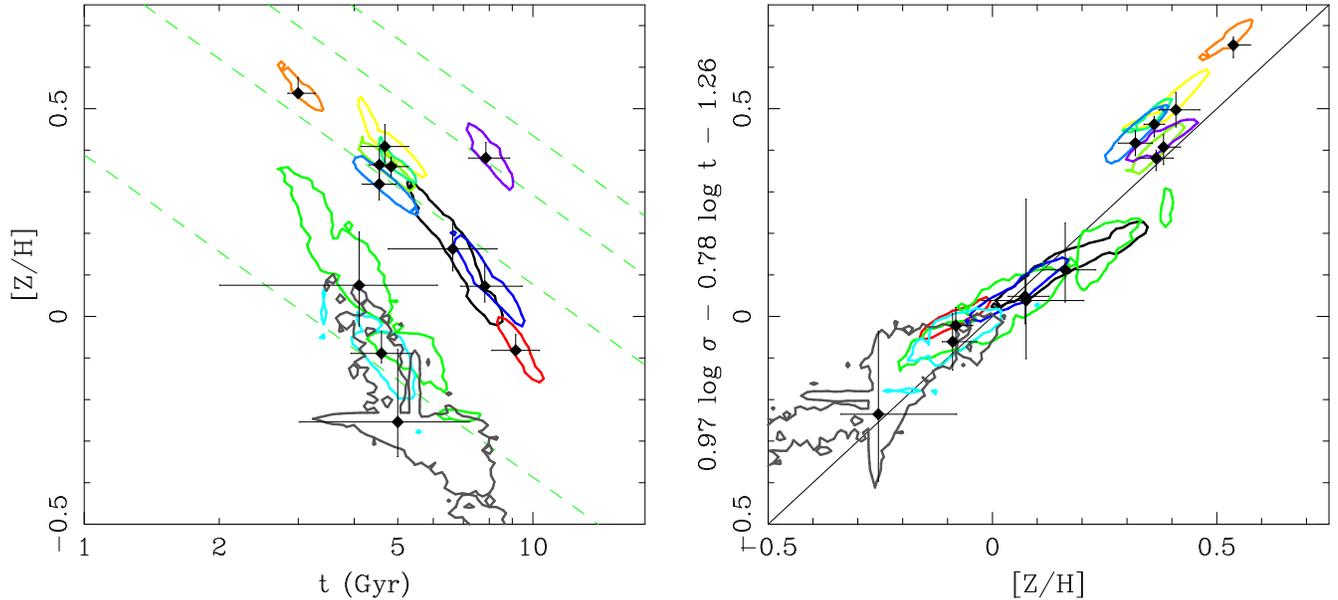}
  \caption{Two views of the $Z$-plane (Paper II) for the LRIS
    galaxies.  Left: the \logt--\z\ projection (roughly face-on).
    Contours are 68 per cent confidence intervals of the stellar
    population parameters, marginalised over \enh.  The solid lines
    are lines of constant velocity dispersion $\sigma$ (from bottom to
    top: 50, 150, 250, 350 \kms).  Right: the (long-) edge-on
    projection, showing the thinness of the plane. \label{fig:zplane}}
\end{figure*}

Figure~\ref{fig:zplane} shows a roughly face-on view of the $Z$-plane
-- the \logt--\z\ projection -- and the long edge-on view.  The
face-on view shows that there exists an \emph{age--metallicity
  relation} for each value of $\sigma$, as shown in Paper II.  We have
argued in Paper II that the age--metallicity relation at fixed
$\sigma$ in field samples is not a result of correlated errors in the
age--metallicity plane, as the variations in ages and metallicities
are many times larger than the (correlated) errors (see, e.g., right
panel of Fig.~\ref{fig:allzplanes} below).  It is possible that
correlated errors may bias the \emph{slope} of the plane, but the
existence of the plane is not driven by the correlated errors.  In the
current dataset, this age--metallicity relation is not strong, as the
dispersion in age is very small for these galaxies, as shown above.
The existence of an age--metallicity relation at fixed $\sigma$ with a
slope $d\z/d\logt\sim-2/3$ means that (optical) colours and metal-line
strengths should be nearly constant at a given velocity dispersion,
following the `Worthey 3/2 rule' \citep{W94}.  This results in thin
Mg--$\sigma$ (as show in Paper II) and colour--magnitude relations.
The thinness of the $Z$-plane (that is, the scatter perpendicular to
the plane) suggests that age and velocity dispersion `conspire' to
preserve the thinness of such relations, which are nearly -- but not
quite \citep[Paper II;][]{TMBO05,Gallazzi06} -- edge-on projections of
the $Z$-plane.

\begin{figure*}
  \includegraphics[width=178mm]{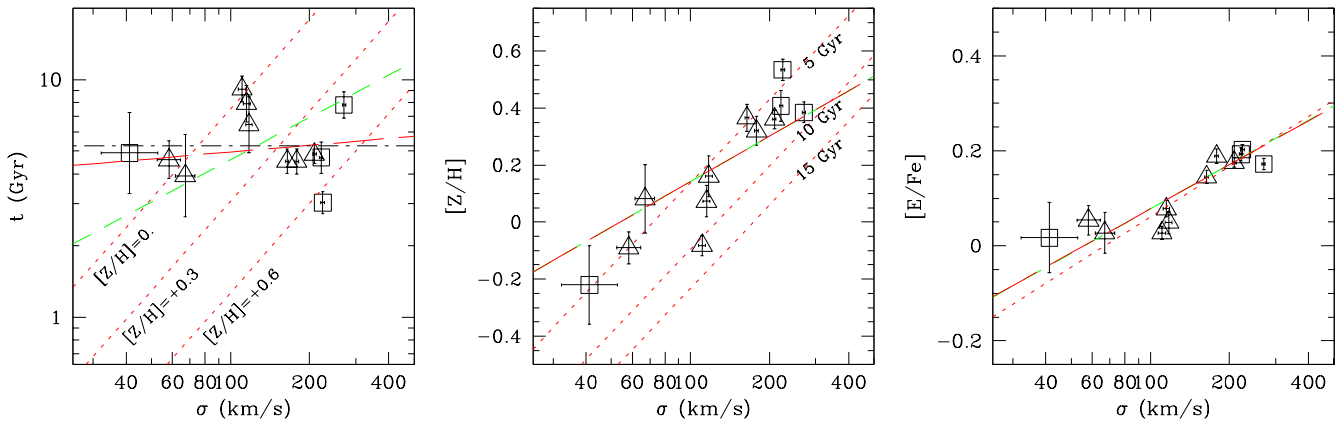}
  \caption{Correlations of stellar population parameters with velocity
    dispersion $\sigma$.  From left to right: $\log\sigma$--\logt;
    $\log\sigma$--\z; $\log\sigma$--\enh.  In all panels, green
    short-dashed lines are the inferred $\log\sigma$--stellar
    population parameter relations of \citet{Nelan05}, zero-pointed to
    the LRIS stellar population parameters, and red long-dashed lines
    are those inferred from the LRIS index strengths following the
    precepts of \citet{Nelan05}.  In the left panel, the red dotted
    lines are the predictions of the $Z$-plane for populations with
    $\z=0$, $+0.3$ (close to the mean metallicity of this sample), and
    $+0.6$.  In the middle panel, the three red dotted lines are the
    predictions of the $Z$-plane for populations of 5, 10, and 15 Gyr
    from top to bottom. \label{fig:sigmarels}}
\end{figure*}

In Figure~\ref{fig:sigmarels} we plot the stellar population
parameters as a function of the velocity dispersion, which are just
projections of the $Z$-plane and the $\enh$--$\sigma$ relation.  We
find both a strong $\log\sigma$--\z\ relation (with a correlation
coefficient of $0.91$; middle panel of Fig.~\ref{fig:sigmarels}) and a
strong $\log\sigma$--\enh\ relation (with a correlation coefficient of
$0.88$; right panel of Fig.~\ref{fig:sigmarels}), but we see \emph{no}
$\log\sigma$--\logt\ correlation (correlation coefficient of $0.01$;
left panel of \ref{fig:sigmarels}), as expected from our discussion in
\S\ref{sec:parameters}.  The latter result is again in contradiction
of our prediction (i) for the stellar populations of ETGs, suggesting
that there is apparently \emph{no} downsizing in Coma Cluster ETGs.

The $\log\sigma$--\z\ correlation is just the mass--metallicity
relation for ETGs \citep{Faber73,Faber77}.  The distribution of
galaxies in the face-on (\logt--\z) projection of the $Z$-plane (left
panel of Fig.~\ref{fig:zplane}) makes it clear why a strong
mass--metallicity relation exists for the LRIS sample of Coma Cluster
galaxies: the galaxies have nearly a single age, so the dispersion in
metallicity \z\ translates into a velocity dispersion--metallicity
sequence (which is related to a mass--metallicity relation through the
virial relation $M\propto\sigma^2r_e$).  This can be seen from the
$\log\sigma$--\z\ relations predicted from the $Z$-plane (dotted line
in the middle panel of Fig.~\ref{fig:sigmarels}).  This is not the
case in samples that have large dispersions in age, like that of Paper
II, because galaxies in these samples have anti-correlated age and
metallicity at fixed velocity dispersion, which erases the observed
mass--metallicity relation\footnote{We note in passing that if a
sample had a very narrow range in metallicity, the $Z$-plane would
require that the galaxies would have a strong age--$\sigma$ relation
if and only if the sample had a strong Mg--$\sigma$ relation (and, of
course, a colour--magnitude relation).}.  That there is such a strong
velocity dispersion--metallicity relation in the LRIS sample is
further evidence that there is at best a weak velocity dispersion--age
relation.

The $\log\sigma$--\enh\ correlation was discovered by \citet{WFG92}
and called the \enh--$\sigma$ relation by Paper II, who found a
relation of the form
\begin{equation}
  \enh=\delta\log\sigma+\epsilon.  \label{eq:enhsigma}
\end{equation}
The last two columns of Table~\ref{tbl:hyperplane} give the
coefficients of Eq.~\ref{eq:enhsigma} for the samples considered here.
A slope of $\alpha=0.41$ is found for the LRIS galaxies.  This value
is roughly consistent with the relations given by Paper II and
\citet{TMBO05}, which were based on models with different
prescriptions for correcting line strengths for \enh.  We note that
the right panels of Figures~\ref{fig:histograms} and
\ref{fig:sigmarels} suggest that the distribution of \enh\ in the LRIS
sample may be bimodal, but this is likely to be an effect of the small
sample size.

We discuss the origin of both of the $Z$-plane and \enh--$\sigma$
relation in \S\ref{sec:zplaneorigin}.

\subsubsection{Velocity dispersion-- and mass--stellar population
  correlations}
\label{sec:msigmaparams}

In Figure~\ref{fig:sigmatrels} we show the distributions of $\logt$ as
a function of $\log\sigma$ for all of the Coma Cluster samples at our
disposal.  We have fit linear relations to these parameters (not
shown) using the routine FITEXY from \citet{NumRec}, which takes into
account errors in both dimensions.  In all samples except the
\citet{Nelan05} RSG sample, we find \emph{negative} correlations
between age and velocity dispersion, violating prediction (i) for the
ages of ETGs in Coma.

Unfortunately, it is difficult to determine the slopes of relations
such as $\log\sigma$--\logt\ for samples with large scatter in the
stellar population parameters from directly fitting the results of
grid inversion, either due to intrinsic scatter or just very uncertain
measurements.  We have therefore also implemented two other methods
for determining the slopes of $\log\sigma$--, $\log M_*$--, and $\log
M_{\mathrm{dyn}}$--stellar population parameter relations.  The first
is the `differential' method described by \citet{Nelan05}.  The second
(`grid inversion') method is very similar to the `Monte Carlo' method
of \citet{TMBO05}, although our implementation is somewhat different:
(a) we use a full non-linear least-squares $\chi^2$-minimisation
routine (Thomas et al.\ fit `by eye'); (b) we do not attempt to
account for extra scatter in the relations; and (c) we do not attempt
to fit two-component (old plus young) population models to outliers.
Our inferred slopes for the \citet{TMBO05} high-density sample match
their results closely, giving us confidence that our method is at
least similar to theirs.  We find no significant positive $\sigma$--
or mass--age relation for any Coma Cluster ETG sample in either
method.  Only the \citet{Nelan05} RSG sample has a significantly
($>2\sigma$) positive slope in this relation.

These relations imply three important results.  (1) RSGs in nearby
clusters -- here represented by the \citet{Nelan05} samples, including
the Coma Cluster itself -- have a strong age--$\sigma$ relation, such
that low-$\sigma$ or low-mass galaxies have younger ages than
high-$\sigma$ or high-mass galaxies, as pointed out by
\citet{Nelan05}.  (2) Taken together, samples of ETGs in the Coma
Cluster show no significant age--$\sigma$ or age--mass relation.  (3)
ETGs in the field show an age--$\sigma$ relation as strong as the Coma
Cluster RSG sample of \citet{Nelan05}.  Results (1) and (2) are
apparently contradictory -- why should RSGs show a strong
age--$\sigma$ relation while ETGs show no such relation?  In advance
of a full discussion in \S\ref{sec:downsizing}, a difference in
emission-line corrections between the \citet{Nelan05} RSG sample and
the ETGs sample is likely to be the cause, \emph{not} a real
age--$\sigma$ relation in the RSGs.  We are therefore again faced with
the conclusion that prediction (i), the downsizing of the stellar
population ages of ETGs, is apparently violated in the Coma Cluster.

\section{Discussion}
\label{sec:discussion}

In \S\ref{sec:introduction} we made three predictions for the stellar
populations of ETGs -- early-type galaxies, galaxies morphologically
classified as elliptical or S0 -- in high-density environments: (i)
low-mass ETGs in all environments are younger than high-mass ETGs (a
prediction that we have called downsizing in this work); (ii) ETGs in
high-density environments are older than those in low-density
environments; and (iii) massive ETGs in high-density environments have
a smaller spread in stellar population age than lower-mass ETGs and
those in lower-density environments.  We recall that our predictions
are based on associating ETGs -- early-type galaxies, galaxies
selected to have elliptical and S0 morphologies -- with RSGs --
red-sequence galaxies, galaxies selected by colour to be on the red
sequence -- and using the results of high-redshift observations and
the predictions of semi-analytic models of galaxy formation.

We found in \S\ref{sec:results} that ETGs in the Coma Cluster have a
mean age of 5--7 Gyrs (including line-strength index calibration
uncertainties but not model uncertainties) and appear to be drawn from
a single-aged population.  Further, the age scatter decreases with
increasing mass.  Finally, while we do find a $Z$-plane for Coma
Cluster ETGs and RSGs, we find no evidence of an age--$\sigma$ or
age--mass relation for the ETGs.  Therefore ETGs in the Coma Cluster
appear to follow prediction (iii) and (perhaps) prediction (ii) but
violate prediction (i).  In this section, we discuss first what we
mean by `age' for old stellar populations, then discuss why we appear
to disagree with previous studies that found downsizing in
high-density environments, what the mean SSP-equivalent age of the
Coma Cluster ETGs implies for their formation, and finally speculate
about the origin of the $Z$-plane, and mass--metallicity and
\enh--$\sigma$ relations.

\subsection{What are we measuring?}
\label{sec:ages}

A worry with stellar population analysis of non-star-forming galaxies
based on their Balmer-line strengths has long been that these lines
reflect not younger (intermediate-aged) main-sequence turn-off stars
but some other hot population, such as blue stragglers
\citep[e.g.,][Paper I]{Rose85,Rose94} or blue horizontal branch stars
\citep[e.g., \citealt{BFGK84}; Paper I;][and references
therein]{MT00,TWFD05}.  Such populations have Balmer-line strengths
comparable or stronger than intermediate-aged main-sequence turn-off
stars and should significantly alter the observed `ages' if present in
large enough (in luminosity-weighted terms) numbers.  \citet{TWFD05}
showed in detail that blue horizontal-branch stars actually affect
inferred \emph{metallicities} more than \emph{ages}, based on
observations of blue absorption lines in the present sample.
Intermediate-aged populations are therefore still required for the
LRIS sample.  Thus we believe that our age estimates are not affected
by hot blue stars that are \emph{not} intermediate-aged main-sequence
turn-off stars.

\begin{figure}
  \includegraphics[width=89mm]{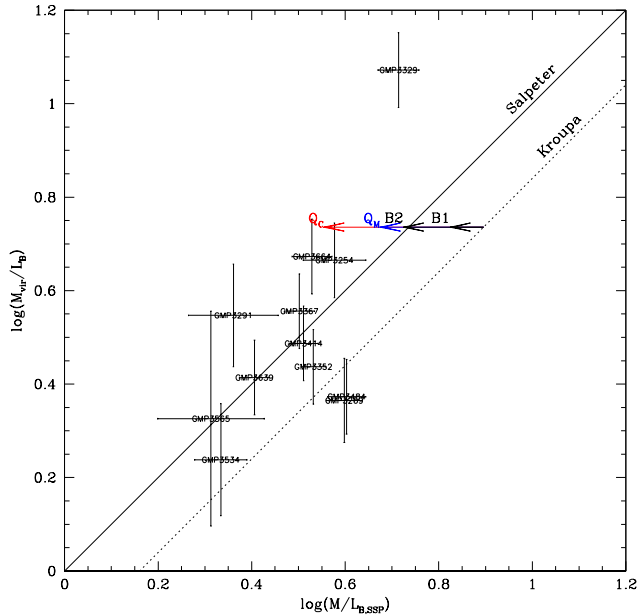}
  \caption{The virial mass-to-light ratios in the $B$-band of our LRIS
  sample ETGs as a function of the stellar mass-to-light ratios as
  determined from the best-fitting SSP models.  The W94 models are
  computed using a \citet{Salpeter55} IMF, represented by the solid
  (one-to-one) line.  Using the \citet{Kroupa01} IMF decreases the SSP
  model mass-to-light ratios by $\sim30$ per cent
  \citep[$\Delta\log(M/L)\sim-0.16$,][]{Cappellari06}, as shown by the
  dotted line.  The arrows represent the effect of different star
  formation histories on the mass-to-light ratios: B1 and B2 are
  bursts occurring 1 and 2 Gyr ago on top of a 12.3 Gyr-old population,
  resulting in $t_SSP=5$ Gyr; Q$_c$ and Q$_m$ are quenching models
  with the same $t_SSP=5$ Gyr.  All of the arrows have the same
  starting location and so have lengths
  $\mathrm{Q}_c>\mathrm{Q}_m>\mathrm{B}1>\mathrm{B}2$.  We note that
  if all the galaxies have a Kroupa IMF and also contain 30 per cent
  of their mass in dark matter within the 2\farcs7 aperture used to
  measure the line-strengths, they should lie on the Salpeter IMF
  line.}
  \label{fig:m2l}
\end{figure}

However, it must always be remembered that the ages, metallicities,
and enhancement ratios we measure with our methods are
\emph{SSP-equivalent} parameters.  We first ask if it is possible that
the galaxies can in fact be the single stellar populations we have
assumed in our modelling.  A simple test of this model is to ask
whether we can reproduce the virial mass-to-light ratios derived in
\S\ref{sec:masses} using SSP models.  We compare the inferred stellar
$M/L$ ratios with the virial $M/L$ ratios in Figure~\ref{fig:m2l}.
Three points can be gleaned from this figure: (a) the
\citet{Salpeter55} IMF appears to be unphysical for these galaxies,
given the presence of many galaxies to the \emph{right} of the
Salpeter IMF line.  Therefore, as in \citet{Cappellari06}, we take a
\citet{Kroupa01} IMF to be a better representation of the (low-mass
star) IMF than the Salpeter IMF; (b) even assuming a small amount (30
per cent) of dark matter within the observed radius in each galaxy
\citep{Cappellari06}\footnote{This might even be a little extreme, as
the \citet{Cappellari06} results are based on $M/L$ ratios within one
$r_e$, while our apertures are in general closer to \reo{2}.}, most of
the galaxies have SSP-equivalent stellar $M/L$ ratios too low for
their virial $M/L$, suggesting that a complex star-formation history
is required in these galaxies; and (c) quenching (arrow Q$_c$ and
Q$_m$) appears to be too extreme for most of the galaxies.

We have further examined the GALEX \citep{GALEX} photometry of
galaxies in the LRIS sample as a probe of young, hot stars.  Only
three of our galaxies -- GMP 3414, GMP 3565, and GMP3664 -- have GALEX
photometry publicly available in GR3.  (Unfortunately, the bright star
HD 112887 prevents GALEX from observing the are directly around the cD
galaxy GMP 3329=NGC 4874.)  Of these three, only GMP 3565 is
`UV-strong' in the notation of \citet{Yi05} -- $(FUV-r)<5$ and
$(NUV-r)<4$ mag -- indicating very young ($t\sim0.1$ Gyr) stars.  A
total of five galaxies in all of the ETGs with line strengths
considered in this study (from all sources) are in this `UV-strong'
class, and their \hbeta-strengths and ages are uncorrelated with their
UV--optical colours.  Using the more generous `young' galaxy criterion
of \citet{Kaviraj07a} -- $(NUV-r)<5.5$ mag -- 51 galaxies in the total
sample are `young' (out of 109 with NUV photometry), although only 17
have $(NUV-r)<5$ mag and only eight (including GMP 3565 in the LRIS
sample) lie significantly off of the $NUV-r$ `red sequence'.  This may
suggest that very young populations are not significantly
contaminating our age estimates.

Clearly therefore the populations of ETGs are more complicated than
single-burst populations \citep[e.g.][]{FY04,deLucia06}.  \citet{ST06}
have explored two-burst `frosting' models and Trager \& Somerville (in
prep.) explore more complicated star-formation histories using
semi-analytic galaxy formation models.  Taken together these studies
find that SSP-equivalent \z\ and \enh\ represent their
luminosity-weighted quantities.  SSP-equivalent age, however,
represents a degenerate mixture of recent star-formation age and burst
strength, as suggested in Paper II.  Moreover young and
intermediate-aged populations contribute \emph{much} more to the
age-sensitive line strengths than is suggested by the phrase
`light-weighted', because younger populations have much higher
mass-to-light ratios \emph{in the Balmer lines} than old populations.
This is why small `frostings' of recent star formation \citep[Paper
II;][]{Gebhardt03} or recent truncation of `quenching' of
previously-on-going star formation \citep[e.g.,][and many
others]{CS87,Bell04,Harker06} lead to much younger SSP-equivalent
ages.

As a simple example, a two-burst model with 98 per cent of the mass in
an 12 Gyr-old population (a formation redshift of $z_f=4$) and the
remaining 2 per cent of the mass in 1 Gyr population (a burst redshift
of $z_b=0.08$) results in an SSP-equivalent age of 5 Gyr.  Note that
as the young population becomes older, much more mass is required: for
a 2 Gyr old burst (a burst redshift of $z_b=0.16$), 12 per cent of the
stellar mass needs to be in the younger population for this population
to also have an age of 5 Gyr.  The effect of these two-burst models on
the stellar $M/L$ ratios are shown in Figure~\ref{fig:m2l}, clearly
reducing the stellar $M/L$ ratios by the addition of much brighter,
slightly more massive stars.

\begin{figure}
  \includegraphics[width=89mm]{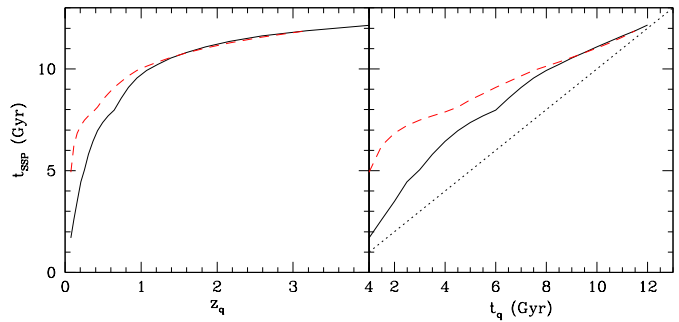}
  \caption{The relation between present-day SSP-equivalent age
  $t_{\mathrm{SSP}}$ and (left panel) quenching redshift $z_q$ and
  (right panel) quenching time $t_q$.  In each panel, the solid line
  represents quenching models with constant star formation from $z=5$
  to $z_q$, while the (red) dashed line represents quenching models
  with star formation that follows the `Madau plot' \citep[as
  parametrised by][see text]{Hopkins06}.  The dashed line in the right
  panel is equivalence between $t_{\mathrm{SSP}}$ and $t_q$.}
  \label{fig:quenching}
\end{figure}

\begin{table}
  \caption{SSP-equivalent, mass-weighted, and $B$-band light-weighted
  ages of quenched galaxies}
  \label{tbl:quenching}
  \begin{tabular}{rrrrrrrr}
    \hline
    \multicolumn{1}{c}{$t_q$}&&\multicolumn{1}{c}{$t^c_\mathrm{SSP}$}&
    \multicolumn{1}{c}{$t^c_M$}&\multicolumn{1}{c}{$t^c_B$}&
    \multicolumn{1}{c}{$t^m_\mathrm{SSP}$}&
    \multicolumn{1}{c}{$t^m_M$}&\multicolumn{1}{c}{$t^m_B$}\\
    \multicolumn{1}{c}{(Gyr)}&\multicolumn{1}{c}{$z_q$}&
    \multicolumn{1}{c}{(Gyr)}&\multicolumn{1}{c}{(Gyr)}&
    \multicolumn{1}{c}{(Gyr)}&\multicolumn{1}{c}{(Gyr)}&
    \multicolumn{1}{c}{(Gyr)}&\multicolumn{1}{c}{(Gyr)}\\
    \hline
     1.0&0.075& 1.84& 4.60& 3.33& 5.23& 8.41& 7.32\\
     1.5&0.116& 2.60& 5.22& 3.90& 6.21& 8.48& 7.48\\
     2.0&0.160& 3.50& 5.74& 4.57& 6.83& 8.56& 7.68\\
     2.5&0.206& 4.45& 6.21& 5.19& 7.20& 8.65& 7.89\\
     3.0&0.256& 5.04& 6.65& 5.77& 7.47& 8.75& 8.09\\
     3.5&0.308& 5.81& 7.05& 6.32& 7.68& 8.85& 8.30\\
     4.0&0.365& 6.45& 7.43& 6.82& 7.87& 8.97& 8.50\\
     4.5&0.427& 6.97& 7.79& 7.28& 8.09& 9.09& 8.69\\
     5.0&0.493& 7.37& 8.14& 7.71& 8.44& 9.22& 8.88\\
     5.5&0.566& 7.69& 8.48& 8.11& 8.75& 9.36& 9.07\\
     6.0&0.646& 7.98& 8.80& 8.50& 9.05& 9.51& 9.27\\
     6.5&0.735& 8.56& 9.12& 8.86& 9.34& 9.66& 9.46\\
     7.0&0.833& 9.09& 9.43& 9.21& 9.63& 9.83& 9.66\\
     7.5&0.945& 9.55& 9.72& 9.55& 9.88&10.00& 9.86\\
     8.0&1.072& 9.92&10.02& 9.88&10.11&10.18&10.07\\
     8.5&1.218&10.23&10.30&10.20&10.32&10.37&10.29\\
     9.0&1.390&10.53&10.58&10.51&10.53&10.57&10.51\\
     9.5&1.596&10.81&10.86&10.80&10.75&10.78&10.74\\
    10.0&1.848&11.09&11.13&11.09&10.99&11.01&10.98\\
    10.5&2.166&11.36&11.39&11.37&11.24&11.26&11.24\\
    11.0&2.587&11.63&11.65&11.64&11.52&11.53&11.52\\
    11.5&3.174&11.88&11.91&11.91&11.81&11.83&11.83\\
    \hline
  \end{tabular}

  Model galaxies are assumed to begin star formation at $z=5$
  (lookback time of 12.3 Gyr). Columns.-- (1) Quenching time. (2)
  Quenching redshift.  (3) Present-day SSP-equivalent age of composite
  stellar population for constant star formation model ($c$).  (4)
  Present-day mass-weighted age of composite stellar population for
  constant star formation model.  (5) Present-day $B$-band
  light-weighted age of composite stellar population for constant star
  formation model.  (6)--(9) As in columns (4)--(6) for Madau-curve
  model ($m$).  See text for details.
\end{table}

As slightly more complex examples, we construct two simple `quenching'
models.  In this sort of model, a galaxy forms stars -- perhaps with a
constant star formation rate, or with a declining rate -- until star
formation is suddenly truncated \citep[e.g.,][]{Bell04,Faber05}.
We assume that a galaxy starts forming stars at $z=5$ (a lookback time
of 12.3 Gyr in our assumed cosmology) and ceases forming stars at some
`quenching redshift' $z_q$ corresponding to a `quenching age'
(lookback time) of $t_q$.  We then ask what its SSP-equivalent age
$t_{\mathrm{SSP}}$ is today.  In the first model, we assume that the
galaxy forms stars at a constant rate from $z=5$ to $z_q$; this is
model $c$ (for constant star formation), typical of the star-formation
histories of Sc disc galaxies \citep{Sandage86,Kennicutt98}.  In the
second model, we assume that the galaxy forms stars at rate that
follows the star formation history of the Universe -- the `Madau
plot', after \citet{Madau96} -- as parametrised by \citet{Hopkins06};
this is model $m$ (for `Madau'), and is similar to the star-formation
histories of early-type (Sa--Sb) spirals
\citep{Sandage86,Kennicutt98}.  In both models we assume star
formation is stopped completely at $z_q$ with no associated burst.  We
further assume no chemical evolution; rather, we assume that
$\z=\enh=0$ dex at all times (an unrealistic assumption!).  The
results are plotted in Figure~\ref{fig:quenching} and tabulated in
Table~\ref{tbl:quenching}.  We see that the SSP-equivalent age
$t_{\mathrm{SSP}}$ is a good tracer of the quenching time $t_q$ or
redshift $z_q$, although $t_{\mathrm{SSP}}\geq t_q$ at all ages in
these models.  This is due to the composite nature of
$t_{\mathrm{SSP}}$, in which stars of all ages contribute to the age
indicators (here \hbeta).  However, in nearly all cases,
$t_{\mathrm{SSP}}<t_{M,B}$, where $t_M$ and $t_B$ are the
mass-weighted and $B$-band-luminosity-weighted ages, because the
youngest populations contribute most to $t_{\mathrm{SSP}}$ due to
their low mass-to-light (high light-to-mass) ratios.  We show the
effect of these models on the stellar $M/L$ ratios in
Figure~\ref{fig:m2l}; model $c$ appears to be too extreme if dark
matter is present within the observed apertures of these galaxies, but
model $m$ is possibly consistent with the observed trend for most
galaxies.  

\begin{figure}
  \includegraphics[width=89mm]{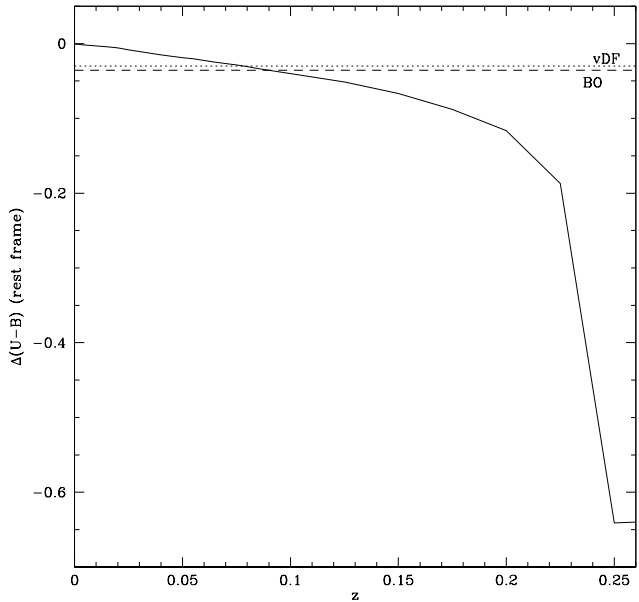}
  \caption{The deviation of rest-frame $U-B$ colour from a typical red
  sequence galaxy as a function of redshift $z$ for a 12.3 Gyr-old
  galaxy quenched at $z_q=0.25$, having formed stars at a constant
  rate before that.  The red sequence galaxy is assumed to have formed
  6.5 Gyr ago ($z_f=0.74$) and so has the same colour as the quenched
  galaxy at $z=0$.  The dotted line at $\Delta(U-B)=-0.030$ is RMS
  dispersion of cluster red sequences from \citet{vDF01} and the
  dashed line at $\Delta(U-B)=-0.036$ corresponds to the \citet{BO84}
  division between red and blue galaxies at $\Delta(B-V)=-0.2$.  Note
  that the colour difference becomes larger than both of these
  divisions at $z>0.09$.}
  \label{fig:quenchingcolor}
\end{figure}

These simple models point out that recent `quenching' can produce
significantly younger populations, as measured by the line strengths,
than might be expected from a simple mass- or light-weighted estimate
\citep[cf.][]{Harker06}.  An advantage of quenching models is that
relations like the mass--metallicity and the \enh--$\sigma$ relation
are generated naturally from the progenitors, which already possess
these relations (\S\ref{sec:zplaneorigin} below).  There is a problem
with such simple quenching models, however.  If the galaxies have
continuous (if not constant) star formation before quenching, they are
quite blue for a significant period \emph{after} quenching.  We
compare the rest-frame $U-B$ colour evolution of a model $c$ galaxy
quenched $z_q=0.2$ ($t_q=3$ Gyr) with an SSP galaxy with the same
colour at $z=0$, which has $t_{\mathrm{SSP}}=6.5$, in
Figure~\ref{fig:quenchingcolor}.  We show also the typical scatter in
rest-frame $U-B$ in the red sequences clusters at $z\la0.8$,
$\sigma_{U-B}\approx0.03$ \citep{vDF01}, and the Butcher-Oemler colour
division between red and blue galaxies, $\Delta(B-V)=-0.2$
\citep{BO84}, corresponding to $\Delta(U-B)=-0.036$ -- strikingly
similar to the typical scatter in the red sequence \citep[as
desired][]{BO78}.  The simple quenching model remains a `blue'
Butcher-Oemler galaxy until as late as $z=0.09$, significantly below
$z_q=0.25$ (a total time of 1.8 Gyr).  Therefore, as suggested by
\citet{vDF01}, quenched galaxies must \emph{continually} join the red
sequence at all redshifts to preserve the observed tight red sequences
in clusters.

\subsection{The mean age of Coma Cluster ETGs}
\label{sec:agescatter}

In \S\ref{sec:parameters} we found a mean age of $\logt=0.72\pm0.02$
dex, or $5.2\pm0.2$ Gyr, for the high-precision and high-accuracy LRIS
ETG sample, and that we could accept a mean age as old as 7.5 Gyr.
Such a young mean age of the Coma Cluster ETGs -- 5--7 Gyr, including
calibration uncertainties -- is surprising.  We must ask whether we
see other `young' ETGs in other clusters at the same masses?  We
certainly see signs of recent star formation and accretion activity in
massive galaxies at the centres of clusters: the young globular
clusters in NGC 1275 \citep[Pegasus A: see,
e.g.,][]{Holtz92,Carlson98}; the multiple nuclei of NGC 6166
\citep[the cD of Abell 2199: see,
e.g.,][]{Minkowski61,Tonry84,Lauer86} and indeed of many other cD
galaxies, more than half of which are likely gravitationally bound
\citep{Tonry85}; and the depressed \mgtwo\ and $D_{4000}$ indexes
found in the central galaxies of cool-core clusters, indicative of
recent star formation \citep*{Cardiel95,Cardiel98}.  These galaxies
appear to have recently-formed stars accreted from smaller objects.
On the other hand, the presence of young low-mass ETGs has been noted
for some time \citep[e.g.,][just to name a few
studies]{Rose85,Rose94,G93,TFGW93,T00b,CRC03,TMBO05,Nelan05,Bernardi06},
and these galaxies may have formed their stars \emph{in situ}
\citep[e.g.,][]{TMBO05}.  But the striking result here is that the
\emph{massive but not central} ETGs in the Coma Cluster have (on
average) young SSP-equivalent ages.

If we apply to the `quenching' models described in \S\ref{sec:ages}
above, we find that model $c$, constant star formation followed by
sudden quenching, predicts a quenching redshift of
$z_q\approx0.25$--0.43 (Table~\ref{tbl:quenching}); model $m$ predicts
a much more recent quenching epoch, $z_q\approx0.08$--0.2.  It appears
from these models that Coma Cluster ETGs have recently been quenched
by some process.  For either model, such a recent quenching epoch
suggests that the galaxies either \emph{just} arrived on (model $c$)
or should still be too blue for (model $m$) the red sequence, and that
there will be no red sequence in the Coma Cluster at $z\ga0.2$
\emph{if all} of the ETGs quenched at the same, very recent time.  We
certainly do not see a \emph{large} population of `young' or blue ETGs
in intermediate-redshift clusters, at least at moderate-to-high ETG
masses, as judged from studies of the evolution of galaxy colours
\citep[e.g.,][]{BO78,BO84,Morphs97,SED98}, the Fundamental Plane
\citep[e.g.,][]{vDF96,vD98,vD99,vdW04,Treu05}, mass-to-light ratios
\citep{vdM06b}, and absorption-line strengths
\citep{Jorgensen05,KIFvD06}.  The majority of the massive galaxies in
intermediate-redshift clusters are quite red
\citep[e.g.,][]{BO78,BO84,Morphs97,Yee05}, with very few, if any, blue
galaxies among the bright ($L>2L_{\ast}$) population.  We therefore
consider such extreme quenching models ruled out.

If we adopt instead a two-burst model of star formation in Coma
Cluster ETGs and assume an mean age of 5 Gyr, we require that 2 per
cent of the mass (in our 2\farcs7 aperture) in each galaxy was formed
at $z=0.08$ or 12 per cent of the mass at $z=0.16$, while the rest of
the mass formed at $z_f=4$ (\S\ref{sec:ages}).  This scenario allows
for most of the mass to be formed at high redshifts while requiring
only small bursts of recent star formation.  Moreover, \citet{Yi05}
have shown that the FUV- and NUV-optical colours of massive early-type
galaxies suggest that 15\% of these objects have had recent star
formation.  \citet{Kaviraj07b} have shown further that truly passive
evolution of ETGs is in conflict with the evolution of their
rest-frame UV-optical colours, such that 5--13 per cent of the entire
mass in ETGs at $0.5<z<1$ resulted from star formation events less
than 1 Gyr previous to the epoch of observation, although this number
decreases by a factor of two by $z=0$.  They suggest that massive ETGs
have formed 10--15 per cent of their total mass since $z=1$, while
low-mass ETGs have formed as much as 60 per cent of their mass in that
time.  We note however that their sample considered is a field sample,
unlikely to contain a significant number of cluster galaxies.

Simplistically, in the two-burst case, we require that \emph{most}
ETGs in the Coma Cluster suffered an event that either triggered star
formation \emph{simultaneously} at redshifts in the range
$z\sim0.1$--0.2.  This agrees well with the observation by
\citet{Gerhard07} that `perhaps 30 per cent' of galaxies in the core
Coma Cluster are involved in an on-going subcluster merger, suggesting
that `Coma is forming now!'  (their emphasis).  Our results support
the view that the Coma Cluster is a very active region, with a large
fraction of the ETGs within $r_{vir}/3$ having suffered star formation
recently, at redshifts around $z\sim0.1$--0.2.  However, this scenario
also requires there to be a significant population of blue galaxies at
\emph{all masses} in the Coma Cluster at those redshifts -- which we
have said above is unlikely, given the relatively tight red sequences
in intermediate-redshift clusters.

There is also the possibility that we have been unlucky with our
sample selection.  The mean ages of the \citet{SB06a} Coma sample
deviate from the other Coma ETG samples, as is clear from
Table~\ref{tbl:ksprobs} (ignoring for present the red-sequence sample
of \citealt{Nelan05}).  However we have found that the ages of four of
the five galaxies in common (GMP 3254, 3269, 3639, and 3664) are the
same within $1\sigma$, and the fifth, GMP 3329 (=NGC 4874), has a
younger age but a higher metallicity from our data as a result of a
higher \mgb\ -- but nearly identical \hbeta\ -- strength in the LRIS
data.  It is notable that the four galaxies in common with the same
ages in both samples are among the youngest in the \citet{SB06a}
sample, and thus we may have been unlucky to select an
unrepresentative sample of galaxies in the cluster.  On the other
hand, we note here that 20 per cent of the \citet{SB06a} sample (7/35
galaxies) have ages that are more than $1\sigma$ older than 14 Gyr --
and therefore older than current estimates of the age of the Universe
\citep{WMAP3} -- using the vanilla W94 models.  This suggests that the
\citet{SB06a} galaxies may be on average too old, and that this is
likely due to uncorrected emission.  We therefore consider that their
old mean age of 12.3 Gyr (after correcting for a reasonable amount of
intrinsic scatter; the weighted mean without this correction is
$>18\,\mathrm{Gyr}$, older than the oldest models and significantly
older than the present age of the Universe) may be unreliable.

We are left with a conundrum: we either were very unlucky in our
sample selection or we require Coma Cluster galaxies to form stars
over an extended time in such a way as to `conspire' to have the same
$t_{\mathrm{SSP}}$ today but not produce too many blue galaxies at
relatively recent lookback times.  As we noted above, massive, central
galaxies have young stars apparently acquired through accretion, while
low-mass galaxies may have just shut down their internal star
formation; perhaps these process have gone on independently and we
have just chanced upon the right time to see them all have the same
age.  The increased scatter in the ages of low-mass Coma ETGs
(Fig.~\ref{fig:agescatter}) suggests that the process of shutting down
star formation in the low-mass galaxies is an extended process, and we
may have just gotten lucky in finding the ages well-synchronised.

Finally, we find a mean age of $\logt=0.70\pm0.01$ dex, or $5.0\pm0.1$
Gyr, for the field sample of \citet{G93}, \citet{FFI96}, and
\citet{K00}, completely consistent with the age of the LRIS ETG
sample, and consistent with the typical ages of nearly all of the Coma
Cluster ETG samples (except \citealt{SB06a}; see
Table~\ref{tbl:ksprobs}).  Thus, unlike \citet{TMBO05},
\citet{Bernardi06} and \citet{SB06b}, we find no significant
difference between field and cluster ETGs, although this is strictly
true only for the Coma Cluster.

\subsection{Downsizing in the Coma Cluster or not?}
\label{sec:downsizing}

One of our three robust predictions for the stellar populations of
local RSGs -- and by assumption, local ETGs -- is that low-mass RSGs
and ETGs are younger than high-mass RSGs and ETGs.  In
\S\ref{sec:introduction} we called this phenomenon downsizing by
analogy with the decrease in specific star formation rate with
decreasing redshift.  In \S\ref{sec:results} we find \emph{no}
evidence of an age--mass or age--$\sigma$ relation at the $>1.5\sigma$
level (ignoring model variations) in any of the Coma Cluster ETG
samples.  However, we do find significant age--$\sigma$ and age--mass
relations for the \citet{Nelan05} Coma Cluster RSG sample and a
significant age--$\sigma$ relation for the entire \citet{Nelan05}
cluster RSG sample and in the \citet{TMBO05} high-density ETG sample.

\subsubsection{Why does the Coma Cluster not show downsizing?}

One possibility is that the Coma Cluster is somehow special, being a
\emph{very} rich cluster.  In the \citet{Nelan05} sample, it has the
twelfth-highest cluster velocity dispersion and is the fifth
most-X-ray-luminous cluster in the full sample, and it is the X-ray
brightest and most massive cluster at $cz_{hel}<10000\,\kms$.  Because
of its richness and velocity dispersion, it might be expected to
contain old galaxies with little recent star formation.  We have
examined the age--$\sigma$ relations for both the \emph{full}
\citet{Nelan05} sample, containing nearly 3500 RSGs (after removing
galaxies contaminated by emission) in 93 clusters, and that sample
restricted to just the Coma Cluster (97 RSGs).  We find a significant
age--$\sigma$ relation for both the full \citet{Nelan05} sample --
using the `differential' method described by \citet{Nelan05} and using
the W94 models, we find $t\propto\sigma^{0.58\pm0.15}$ -- and for the
restricted Coma Cluster sample -- $t\propto\sigma^{0.39\pm0.12}$. The
relation for the Coma Cluster is only marginally shallower than that
found for the entire \citet{Nelan05} sample: a slope difference of
$0.19\pm0.19$.  We suggest below that the significant age--$\sigma$
slope for the \citet{Nelan05} Coma Cluster sample may be due to a lack
of emission-line correction in the Balmer line strengths of that
sample, which is also true for the entire sample.  If the Coma Cluster
RSGs truly possess an age--$\sigma$ relation, the results of
\citet{Nelan05} and our analysis suggest that its slope is cannot be
much shallower than that of RSGs in typical high-density regions.
This suggests that the lack of an age--$\sigma$ relation for the ETG
samples is not due \emph{solely} to the overall richness of the Coma
Cluster.

Another possibility is that galaxies in the centre of the Coma Cluster
are preferentially younger than the cluster as a whole.  Studies of
the diffuse light in the centre of the Coma Cluster
\citep[e.g.,][]{TK77,GW98,Adami05a,Adami05a} and intracluster
planetary nebulae \citep{Gerhard07} suggest that the centre of the
Coma Cluster is a violent place, with a massive on-going merger of a
subcluster \citep{Gerhard07}.  The \citet{M02} sample however covers
the inner $1^{\circ}$ of the cluster, which corresponds to a radius of
$r_{vir}/3$ \citep{LM03}.  We find no age--$\sigma$, age--$\log M_*$
nor age--$\log M_{\mathrm{dyn}}$ relation in this sample, and so a
seriously different age of the centre -- older or younger -- is
unlikely.

Finally, we note that we are not the first to find a flat
age--$\sigma$ relation in cluster ETGs, nor even in Coma Cluster ETGs.
\citet{SB06a,SB06b} have claimed that there is \emph{no} age--$\sigma$
relation in cluster ETGs, although there is a significant dispersion
(and many of their galaxies appear to be too old, as discussed in
\S\ref{sec:agescatter} above).  Their cluster ETG sample is dominated
by Coma galaxies (with a non-negligible minority of Virgo galaxies as
well) and therefore is a similar result, with a different mean age, to
ours.  As mentioned in \S\ref{sec:introduction}, \citet{KIFvD06} have
also recently shown that the age--$\sigma$ relation for ETGs in the
cluster CL1358+62 at $z=0.33$ is flat.  Although they suggest that
this is due in part to a different method for correcting the
line-strength indexes for the effects of velocity dispersion (see
Appendix~\ref{sec:indexcorrections}), their Figure 10 shows that this
correction is a minor effect and that the ETGs in that cluster do not
show a significant age--$\sigma$ relation.  Some amount of caution
must be taken here, though, as \citet{KIFvD06} came to this conclusion
using only the blue indexes (H$\delta$--\ctwo) due to the redshift of
the cluster.  Further, \citet{vdM06b} have used resolved internal
kinematics of ETGs in clusters at $z\approx0.5$ to probe the evolution
of \emph{rotation-corrected} dynamical mass-to-light ratios.  They
find no evidence for change in mass-to-light ratio with velocity
dispersion as a function of redshift.  This suggests that age and
velocity dispersion are not correlated in that sample, as
mass-to-light ratios are more sensitive to age than to metallicity
\citep{G93,W94}.

\subsubsection{Why do we disagree with \citet{TMBO05}?}

\begin{figure}
  \includegraphics[width=89mm]{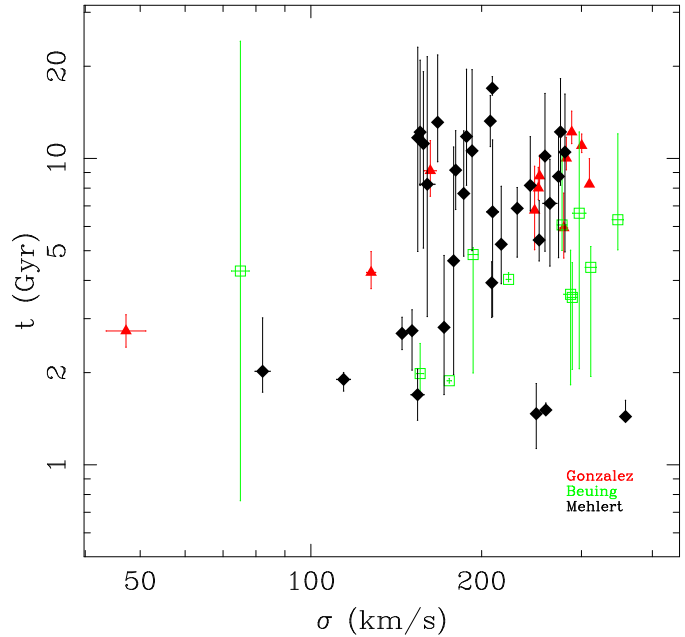}
  \caption{The age--$\sigma$ relation for the high-density sample of
  \citet{TMBO05}.  Diamonds are Coma Cluster ETGs from
  \citet{Mehlert03}, based on the sample of \citet{Mehlert00}, open
  squares are cluster ETGs from \citet{Beuing02}, and triangles are
  cluster galaxies from \citet[mostly Virgo cluster galaxies]{G93}.}
  \label{fig:sigmat_tmbo05}
\end{figure}

We now ask why \citet{TMBO05} find an apparently significant slope in
the age--$\sigma$ relation for ETGs in high-density regions while we
do not find one for ETGs in the Coma Cluster.  We note that their
high-density ETG sample contains Coma Cluster ETGs from \citet[a
compilation of aperture-corrected data from
\citealt{Mehlert00}]{Mehlert03}, Virgo\footnote{Note that the field
galaxies NGC4261 and NGC 4697 are included in the high-density sample
of \citet{TMBO05}, apparently mistaken as Virgo Cluster galaxies, and
the galaxy NGC 636 appears twice in their low-density ETG sample,
taken once each from \citet{G93} and \citet{Beuing02}.} and Pegasus
Cluster ETGs from \citet{G93}, and a collection of mostly compact
group galaxies from \citet{Beuing02}.  We show the age--$\sigma$ data
from \citet{TMBO05} in Figure~\ref{fig:sigmat_tmbo05}.  If we consider
only the Coma Cluster galaxies in their sample -- the ETG sample of
\citet{Mehlert00} -- we do not find an age--$\sigma$ relation.  As
\citet{TMBO05} did not publish error bars or confidence levels on
their age--$\sigma$ relation, it is difficult to infer the robustness
of their result.  We therefore cannot say with confidence whether our
conclusion truly disagrees with their findings, but we suggest that
the \citet{Mehlert00} data do not by themselves support downsizing in
the Coma Cluster.

\subsubsection{Why do we disagree with the Coma Cluster RSGs of
  \citet{Nelan05}?}

\begin{figure*}
  \includegraphics[width=178mm]{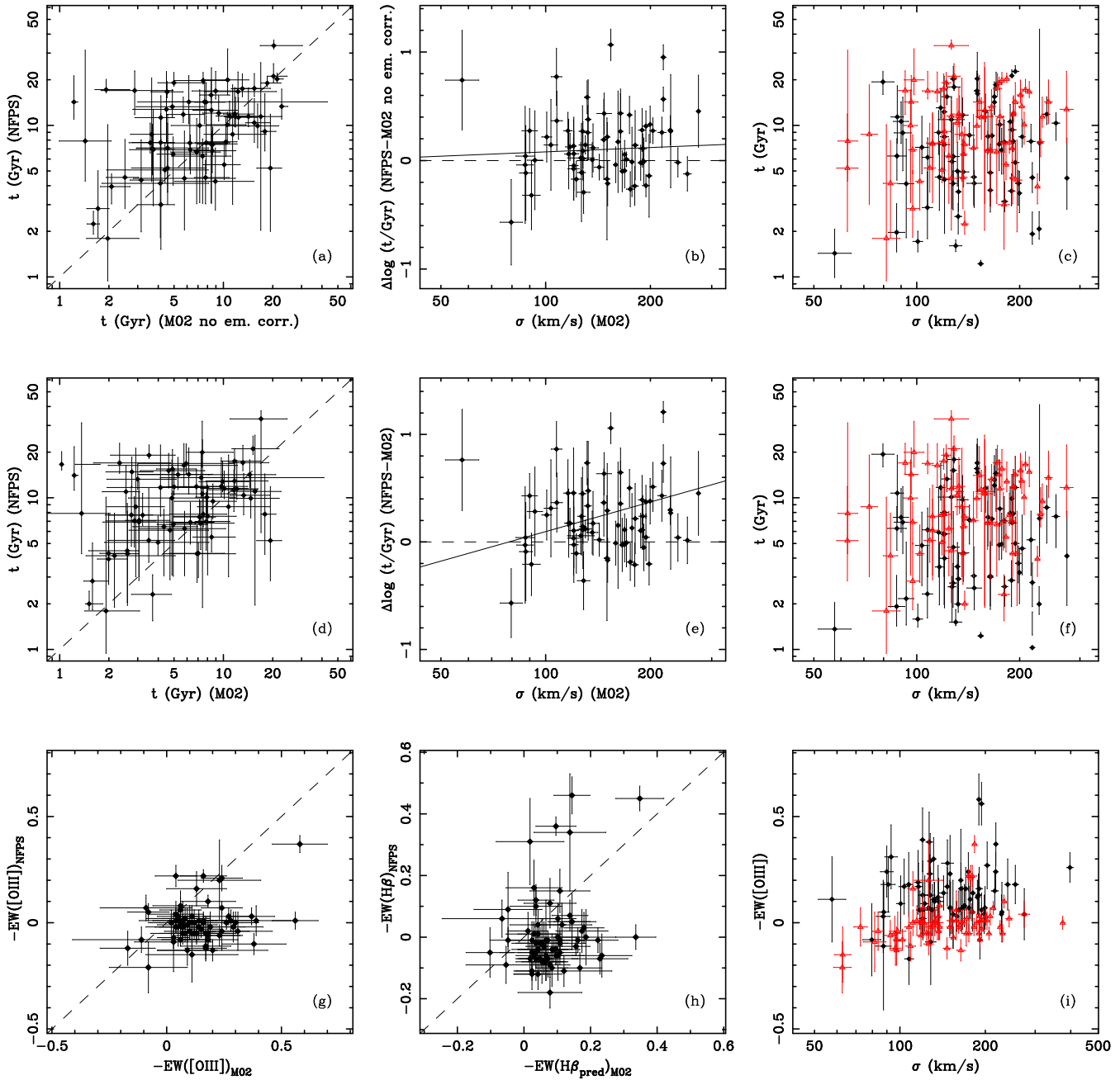}
  \caption{A comparison of ages and emission-line strengths of the
  \citet{M02} ETG and \citet{Nelan05} RSG samples for galaxies in
  common.  Panels (a)--(f): A comparison of galaxy ages in the
  samples.  In these panels, the strong outlier GMP 2921 (=NGC 4889)
  has been removed: its age inferred from the \citet{Nelan05} data is
  nearly ten times higher than that inferred from the \citet{M02}
  sample, with very small formal errors in each sample.  The dashed
  lines represent equality in the ages.  The solid lines in panels (b)
  and (e) are fits to the age differences as a function of
  $\log\sigma$, accounting for errors along both axes; the slope of
  the fit in panel (e) is significant, but that in panel (b) is not.
  In panels (a)--(c) (top row), the \hbeta\ strengths of the
  \citet{M02} galaxies have \emph{not} been corrected for emission,
  while such a correction has been made in panels (d)--(f) (middle
  row).  Note that the SSP-equivalent ages of the \citet{Nelan05}
  sample are on average \emph{older} than those of the \citet{M02} for
  galaxies in common, even without the emission-line correction of
  \hbeta.  Panels (c) and (f) show the age--$\sigma$ relations for the
  two samples for galaxies in common to both samples.  A comparison of
  panels (b) and (e) show that neglecting the emission-line correction
  can impose an age--$\sigma$ relation on the \citet{Nelan05} RSG
  sample.  Black diamonds: \citet{M02}; red triangles:
  \citet{Nelan05}.  Panels (g)--(i): A comparison of emission-line
  strengths in the samples.  (GMP 2921 is included in these panels.)
  Panels (g) and (h) compare the emission line strengths of the two
  samples.  The predicted \hbeta\ emission-line strength of the
  \citet{M02} sample,
  $-\mathrm{EW(\hbeta)}=-0.6\times\mathrm{EW([O\,\textsc{iii}])}$
  \citep{T00a}, is plotted as a function of the measured \hbeta\
  emission-line strength in panel (h).  The correlation between the
  samples is stronger in panel (h) but only marginally significant
  there (4 per cent probability of being uncorrelated).  Panel (i)
  shows that $\mathrm{EW([O\,\textsc{iii}])}$ is strongly correlated
  with $\sigma$ in the \citet{Nelan05} RSG sample, suggesting again
  that neglecting emission corrections may result in a false detection
  of an age--$\sigma$ relation.}
  \label{fig:moorenfps}
\end{figure*}

We next ask why we find a significant age--$\sigma$ relation for the
\citet{Nelan05} Coma Cluster RSG sample but not for the any of the
Coma Cluster ETG samples.  We compare the large ETG sample of
\citet{M02} with the \citet{Nelan05} sample in
Figure~\ref{fig:moorenfps} for the 71 galaxies in common.  In the top
and middle rows, we compare the inferred ages of the two samples.  In
the top row, we compare the ages of the \citet{M02} sample,
\emph{uncorrected} for emission-line fill-in of \hbeta, with those of
the (uncorrected) \citet{Nelan05} sample.  Apart from a few outliers
[and neglecting the strongly deviant galaxy GMP 2921=NGC 4889, which
has been removed in panels (a)-(f)], the ages of the two samples are
very comparable: the middle panel shows the difference in ages in the
samples as a function of velocity dispersion.  We do not find a
significant \emph{slope} difference between the samples, merely a
small offset, such that the \citet{Nelan05} ages are
$\Delta\logt=0.18\pm0.13$ dex ($66\pm31$ per cent) older than the
uncorrected \citet{M02} ages.  It is important to note that
\citet{Nelan05} rejected galaxies with $\mathrm{EW(\hbeta)}<-0.6$ \AA\
(and $\mathrm{EW([O\,\textsc{iii}])}<-0.8$ \AA) from their sample.  No
galaxy in the Coma Cluster has such strong emission, but certainly
small amounts of emission are detected in both the LRIS and
\citet{M02} samples.  In fact, ten galaxies (out of 97) in the
\citet{Nelan05} data set have detectable emission with
$\mathrm{EW(\hbeta)}\leq-0.2$ \AA, sufficient to make these galaxies
have older SSP-equivalent ages than if their \hbeta\ strengths had
been corrected for this emission.  In the middle row of of
Figure~\ref{fig:moorenfps}, we compare the ages of the \citet{M02}
sample, corrected for emission-line fill-in of \hbeta\ using the
precepts of \citet[see \S\ref{sec:literature}]{T00a}, with those of
the (uncorrected) \citet{Nelan05} sample.  In panel (e) we find a
strong discrepancy in ages which grows stronger with increasing
velocity dispersion.  As we believe that an emission correction to
\hbeta\ \emph{should} be applied, we suggest that the age--$\sigma$
relation seen in the \citet{Nelan05} RSG sample results from their
lack of emission-line correction and is not intrinsic to their sample.

Finally, in the bottom row of Figure~\ref{fig:moorenfps}, we compare
the emission-line strengths of [O\textsc{iii}] of the two samples
(panel g), the predicted \hbeta\ emission-line strengths of the
\citet{M02} sample with the measured \hbeta\ emission-line strengths
of the \citet{Nelan05} sample (panel h), and the variation in
[O\textsc{iii}] strength as a function of velocity dispersion (panel
i).  The [O\textsc{iii}] strength of the \citet{Nelan05} sample is
correlated with velocity dispersion, reinforcing our suggestion that
the age--$\sigma$ relation found in that sample is an artefact of
ignoring the (necessary) emission correction.  Clearly, larger samples
of high-signal-to-noise spectra with careful emission-line correction
in the Balmer lines \citep[using, say, the techniques of][]{Sarzi06}
will be required to resolve this discrepancy completely -- but even
those techniques are imperfect, as shown by the fact that we detect
[O\textsc{iii}] but \emph{not} \hbeta\ emission in our galaxies (which
we claim we should have, as it is nearly impossible to have
[O\textsc{iii}] but not \hbeta\ emission: \citealt{Yan06}) using the
\citet{Sarzi06} method (Appendix~\ref{sec:emission}).

\begin{figure}
  \includegraphics[width=89mm]{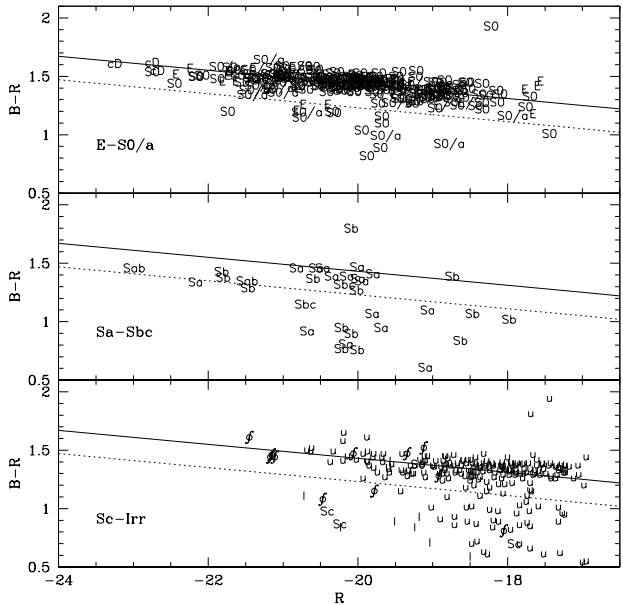}
  \caption{The colour--magnitude relation of Coma Cluster galaxies
    coded by morphological type.  Colours and magnitudes are taken
    from Beijersbergen (2002) and morphologies from NED.  Solid lines
    are a fit to the colour--magnitude relation; dashed lines are 0.2
    magnitudes bluer.  Galaxies with types earlier than Sd and Irr
    galaxies are labelled with their morphological type (I=Irr); Sd
    and spiral galaxies without specific type are labelled as $\oint$.
    Galaxies without morphological type in NED are labelled as "u".
    Top panel: E--S0/a galaxies.  Middle panel: Sa--Sbc galaxies.
    Bottom panel: Sc--Irr galaxies and galaxies with unknown
    morphological types.}
  \label{fig:cmdmorphs}
\end{figure}

\begin{table}
  \caption{Deviations from red sequence by morphological type}
  \label{tbl:rsdev}
  \begin{tabular}{lrrr}
    \hline
    \multicolumn{1}{c}{Morphological type}&
    \multicolumn{1}{c}{$\langle\Delta(B-R)\rangle$}&
    \multicolumn{1}{c}{$\sigma_{\langle\Delta(B-R)\rangle}$}&
    \multicolumn{1}{c}{$N_{\mathrm{gal}}$}\\
    \hline
    cD&$-0.011$&0.024&3\\
    E&0.018&0.007&59\\
    E/S0&0.010&0.014&16\\
    S0&0.006&0.005&146\\
    S0/a&$-0.021$&0.015&27\\
    Sa&$-0.027$&0.014&9\\
    Sab&$-0.160$&0.014&3\\
    Sb--Irr&$-0.029$&0.029&18\\
    \hline
  \end{tabular}
  
  Only galaxies with $R<18$ and $\Delta(B-R)>-0.2$ included.
\end{table}

Although we suspect that emission corrections are the primary cause of
the discrepancy between the age--$\sigma$ slopes -- and thus the
detection of downsizing -- of \emph{all} of the Coma Cluster ETG
samples and the age--$\sigma$ slope of the \citet{Nelan05} sample, it
is possible that target selection could drive the difference.  That
is, are the stellar populations of RSGs intrinsically different than
those of ETGs?  Do colour and morphology drive the presence or lack of
an age--$\sigma$ relation?  The significant difference between the
\citet{Nelan05} sample and the \citet{J99}, \citet{Mehlert00},
\citet{M02}, \citet{SB06a}, and LRIS samples is the colour selection
of the NFPS galaxies and the morphological selection of all of the
other samples.  We note that the red sequence contains not only
elliptical and S0 galaxies but also disk-dominated early-type spiral
galaxies (Fig.~\ref{fig:cmdmorphs}).  We suggest here that a possible
solution is that the presence of disc-dominated galaxies in the
colour-selected samples could be the cause of the age--$\sigma$
relation found by \citet[and by extension,
\citealt{Smith06}]{Nelan05}.  In Table~\ref{tbl:rsdev} we examine the
deviation from the colour--magnitude relation of Coma Cluster galaxies
as a function of morphological type for galaxies that qualify as
`red-sequence galaxies' under the criteria of \citet{Smith04}: redder
than $-0.2$ magnitudes bluer than the mean colour--magnitude relation
in $B-R$.  We find that mean deviations from the red sequence become
bluer as morphological type becomes later, as might be expected,
although the numbers are small.  However, \citet{Smith06} have
examined the influence of morphology for a subset of NFPS galaxies by
taking only those galaxies with quantitative morphologies and with
$B/T>0.5$ (about 35 per cent of the total NFPS sample) and recomputing
the $\log\sigma$--parameter relations.  They find that a shift of one
unit in $B/T$ -- i.e., going from pure disc to pure bulge -- increases
\logt\ by $0.176\pm0.026$, which is not enough to erase the
age--$\sigma$ relation.  Moreover, virtually all of the
\citet{Nelan05} Coma Cluster RSGs are ETGs (only two are typed as Sa
in NED).  We therefore come to the conclusion that the lack of
emission-line corrections to the Balmer lines in the \citet{Nelan05}
sample is likely to be the largest contributor to the difference
between that sample and all the others, and that sample selection --
RSGs \emph{versus} ETGs -- is unlikely to play a significant r{\^o}le
in that difference.

To summarise this section, we find no evidence for an age--$\sigma$ or
age--mass relation in ETGs in the Coma Cluster.  We suggest further
that such a relation may not even hold for RSGs in the Coma Cluster,
but this requires further high-quality data.  We have referred to a
significant age--$\sigma$ relation with a positive slope as downsizing
of the stellar populations of local ETGs.  We do not see significant
evidence for such downsizing in Coma Cluster ETGs, and this is not the
only environment where this seems to be the case \citep{KIFvD06}.  We
therefore come to the conclusion that our prediction (i) for the
stellar populations of local ETGs in \S\ref{sec:introduction} is
violated in the Coma Cluster.  But we are still left with the question
of where the \emph{old} galaxies are.  Have we just missed them, or
are they not there, because \emph{all} early-type galaxies have formed
stars recently enough that we see `young' galaxies, as predicted by
\citet{Kaviraj07b}?

\subsection{The $Z$-plane and the \enh-$\sigma$ relation in the Coma
  Cluster}
\label{sec:zplaneorigin}

\begin{figure*}
  \includegraphics[width=178mm]{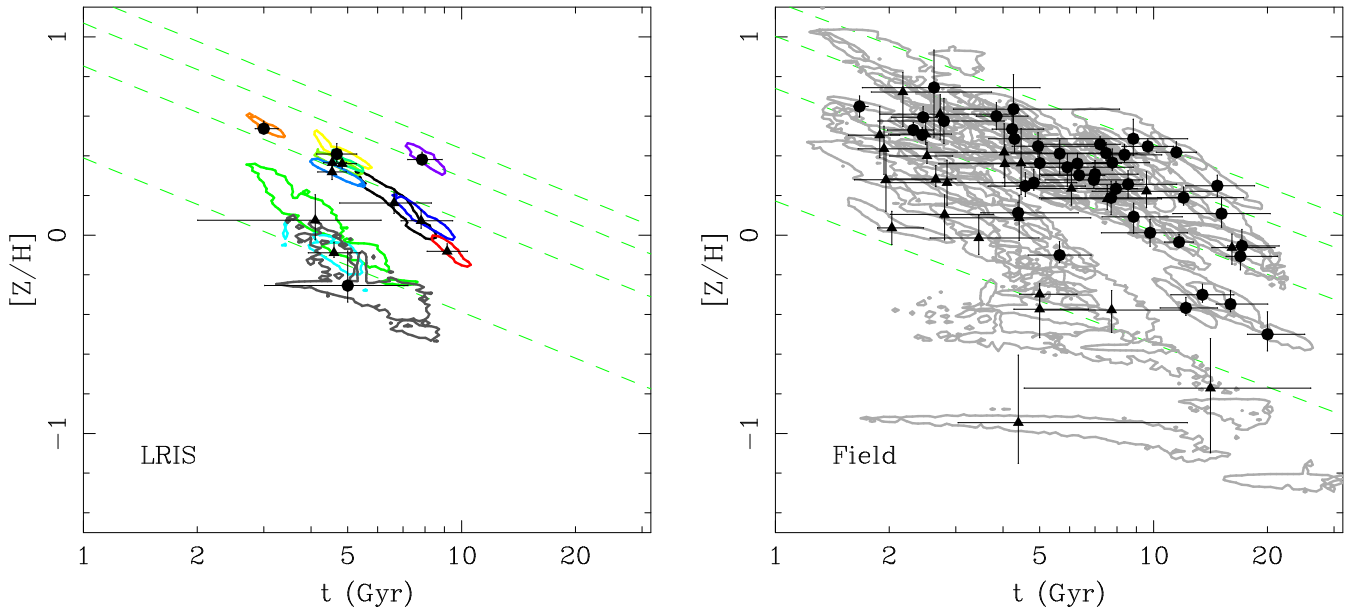}
  \caption{The $Z$-plane for the LRIS Coma Cluster ETG sample (left
  panel) and our field ETG sample (right panel).  Lines of constant
  $\sigma$, inferred from the $Z$-planes given in
  Table~\ref{tbl:hyperplane}, are shown as dashed lines (bottom to
  top: 50, 150, 250, 350 \kms).}
  \label{fig:allzplanes}
\end{figure*}

Finally, we turn to the two relations explored in detail in Paper II:
the $Z$-plane and the \enh--$\sigma$ relation.  The $Z$-plane, as
discussed above and in Paper II, says that there exists an
age--metallicity anti-correlation at each value of $\sigma$, with
metallicity increasing with increasing $\sigma$.  Note that the
$Z$-plane specifically decouples age and $\sigma$, as required from
our discussion of the age--$\sigma$ relation above.  We plot a nearly
face-on projection of the $Z$-plane -- the age--metallicity plane --
for our LRIS Coma Cluster and field ETG samples in
Figure~\ref{fig:allzplanes}.  The fact that the $Z$-plane and
\enh--$\sigma$ relation are seen in \emph{both} field and cluster
populations, as found in \S\ref{sec:zplane}, suggests that they are a
general feature of the stellar populations of ETGs and should
therefore be understood in the context of galaxy formation models.

What are the origins of $Z$-plane and \enh--$\sigma$ relations?  We
first consider a two-burst model, in which the majority of the mass of
ETGs form at high redshift, followed by small bursts of star formation
at $z\sim0.1$--0.3, as discussed above.  It is important to recall
here that \citet{ST06} have shown that the SSP-equivalent \z\ and
\enh\ values are very nearly equivalent to their mass-weighted
quantities.  This suggests that the $Z$-plane (and \enh--$\sigma$)
relations in the Coma Cluster ETGs were put in place during the
initial star formation phases at high redshift and were only mildly
perturbed in the secondary star formation events, \emph{as long as
these secondary events involve only small mass fractions}.  That is,
the secondary bursts must have occurred very recently in order to keep
the mass--metallicity and \enh--$\sigma$ relationships of Coma Cluster
ETGs as tight as is found in Figure~\ref{fig:sigmarels}.  We discussed
the origin of these relations extensively in Paper II.  Here we remind
the reader that apparently the only available scenarios are (a) early,
metal-enriched winds that grow stronger with decreasing ETG velocity
dispersion and (b) an IMF slope that becomes flatter with increasing
ETG velocity dispersion.

We have shown in \S\ref{sec:agescatter} above that all of the ETGs in
the LRIS sample might be assumed to have quenched at $z\approx0.2$
(although we have ruled this scenario out).  Therefore they form a
narrow strip in the age--metallicity plane, because they have nearly
the same age.  Then the question becomes why do they exhibit both a
mass--metallicity ($\sigma$--\z) relation and a \enh--$\sigma$
relation?  In the context of the quenching model, this is because they
came from \emph{blue, star-forming galaxies that already exhibited
these relations} \citep{Faber05}.  We therefore speculate that the
trends found in Paper II for field ETGs and by \citet{TMBO05} and
\citet{Bernardi06} for both low- and high-density ETGs --
high-$\sigma$ galaxies are older, more metal-rich, and have higher
\enh\ -- were also exhibited by their blue, star-forming progenitors.
We already have evidence of that two of these relations are true for
star-forming galaxies: the larger a disc galaxy is, the redder it is
\citep{RM94} -- which means the stars formed earlier, as shown by
\citep{MacArthur04} -- and the more metal-rich it is \citep[from SDSS
emission-line spectra]{Tremonti}.  We also know that the bulges of
large spirals follow the \enh--$\sigma$ relation \citep{PS02}.
Therefore there is already enough evidence to assert that the
compositions of ETGs are essentially embedded in their spiral galaxy
precursors.  If this is the case, that what we are seeing in the LRIS
sample is a set of objects of different masses that all got quenched
at about the same time.  Their chemical compositions follow naturally
from their velocity dispersions.  In order to fill out the $Z$-plane,
then, one needs galaxies that quenched at different times, both
earlier and later than our Coma Cluster ETGs -- such as the field
sample or the sample of Paper II -- as seen in
Figure~\ref{fig:allzplanes}.  This appears to be a more
straight-forward explanation of our results than the two-burst model,
because the progenitors are clearly identified as blue, star-forming
galaxies which we know have the correct scaling relations.  Moreover,
quenching models of this sort also explain the evolution of the
morphology--density relation in clusters \citep{Dressler97}.  But we
point out again that massive cluster ETGs are generally old at
intermediate redshifts, as discussed above.  The quenching model we
consider here predicts rather that for the Coma Cluster, most of the
ETGs were blue, star-forming galaxies very recently, which we have
already rejected in \S\ref{sec:agescatter} above.

We are left in the position of having a reasonable explanation for the
origin of the $Z$-plane -- that is, that disc galaxies that already
possess mass--metallicity and \enh--$\sigma$ relations are quenched
simultaneously -- that is ruled out by observations of
intermediate-redshift clusters.  We have begun to explore whether
hierarchical galaxy formation models with detailed chemical evolution
can predict these relations (Trager \& Somerville, in prep.; Arrigoni
et al., in prep.).

\section{Summary and conclusions}
\label{sec:conclusions}

In \S\ref{sec:introduction} we made three predictions for the stellar
populations of local ETGs based on observations of RSGs at high
redshifts and the results of models of hierarchical galaxy formation:
\begin{enumerate}
  \item lower-mass ETGs in all environments have younger stellar
  population ages than high-mass ETGs;
  \item ETGs in high-density environments are older than those in
  low-density environments; and
  \item massive ETGs in high-density environments have a small
  stellar population age spread compared with lower-mass ETGs and
  those in lower-density environments.
\end{enumerate}
We have tested these predictions using very high signal-to-noise
spectra of twelve ETGs spanning a wide range in mass in the Coma
Cluster surrounding and including the cD galaxy NGC 4874.  Because of
the small size of this sample, we have augmented it with larger but
less precise samples of ETGs and RSGs in the Coma Cluster.

We find the following results.
\begin{enumerate}
  \item Coma Cluster ETGs in the LRIS sample are consistent with a
  uniform SSP-equivalent age of $5.2\pm0.2$ Gyr (with a possible
  systematic upper limit of 7.5 Gyr using the \citealt{W94} models),
  which is identical within the formal errors to the average
  SSP-equivalent age of a sample of field ETGs drawn from the samples
  of \citet{G93}, \citet{FFI96}, and \citet{K00}.  All Coma Cluster
  ETG samples are consistent with a single-age population of galaxies,
  with the exception of the \citet{M02} sample, in which the
  elliptical and S0 galaxies are each consistent with a single-age
  population.  Differences in calibration onto the Lick/IDS index
  system and the treatment of possible emission-line corrections of
  the Balmer lines are primarily responsible for differences in the
  mean ages between samples.  However, the \citet{Nelan05} RSG sample
  is \emph{inconsistent} with a single-age population of galaxies.
  \item All Coma Cluster ETG samples are consistent with having
  \emph{no} SSP-equivalent age--$\sigma$ or age--mass relation.  That
  is, we see no sign of downsizing in Coma Cluster ETGs.  This is not
  the case in the \citet{Nelan05} Coma Cluster RSG sample; however, we
  have shown that this due to neglect of emission-line corrections to
  the Balmer-line indexes in their sample.
  \item The large Coma Cluster ETG sample of \citet{M02} is consistent
  with the dispersion of SSP-equivalent ages decreasing with
  increasing velocity dispersion.  These age dispersions are typically
  smaller than those of our field ETG sample at the same velocity
  dispersion.
  \item Field ETGs and all Coma Cluster ETG and RSG samples show both
  a $Z$-plane and an \enh--$\sigma$ relation.
\end{enumerate}
Taken together, findings (i)--(iii) mean that predictions (i) and (ii)
above does not hold for the stellar populations of Coma Cluster ETGs;
only prediction (iii) holds.

We have explored two galaxy formation scenarios to explain these
results: (1) one in which old ETGs have recent burst of star formation
triggered by an as-yet unidentified process and (2) one in which the
on-going star formation in blue galaxies is suddenly shutdown and
followed by passive evolution of these galaxies to become the ETGs we
see today.  We have ruled out the second, `rapid quenching' model on
the basis that intermediate-redshift clusters do not have large
populations of the \emph{massive} blue galaxies implied by this model
\citep[as previously remarked on by, e.g.,][]{Bell04}.  We therefore
consider recent star formation on top of old stellar populations as
being the preferred (but not ideal) model.  This star formation either
happened at $z\sim0.2$ for most ETGs in the Coma Cluster \emph{or} the
star formation histories of the ETGs were more complex but `conspire'
to appear simultaneous using our line-strength dating technique at the
present epoch.  An open question is, where are the \emph{old} Coma
Cluster ETGs that did \emph{not} suffer recent star formation?  We do
find a few galaxies in our sample (GMP 3269 and GMP 3484) whose 68 per
cent upper limits on their SSP-equivalent ages approach or exceed 10
Gyr with the W94 models, but the average age at all masses is, again,
5--7 Gyr.

We however must pause and ask whether we have \emph{really} ruled out
downsizing in the Coma Cluster ETG population if all we are detecting
is a `frosting' \citep[to use the phrase of][]{T00b,Gebhardt03} of a
few percent by mass of young stars on top of a massive population that
formed at high redshift.  Taking our very simple two-burst models -- a
young population on top of a 12-Gyr-old population -- at face value,
we could say yes, that most of the stars formed at an early epoch
regardless of their mass.  This is contrary to our definition of
downsizing.  However, we could certainly imagine more complicated
`frosting' scenarios in which low-mass galaxies formed the bulk of
their stars later than high-mass galaxies -- at, say, 8 Gyr rather
than 12 Gyr -- and then all (or at least most) of the galaxies had a
later, small star-formation episode.  We admit that it is difficult to
test these models with the observations we have presented here.
However, others -- for example \citet{TMBO05} and \citet{Nelan05} --
have claimed that they observe downsizing of ETGs directly from their
present-day line strengths.  It is this precise claim of downsizing
that we believe we have falsified, at least in the Coma Cluster.  Even
if these studies did show `downsizing', the `frosting' scenario calls
into question whether this is the same `downsizing' that is seen in
lookback studies, as it may only involve a small fraction of the mass.
We suggest that at present perhaps only lookback studies (like those
mentioned in \S\ref{sec:introduction}) can detect downsizing in the
stellar populations of ETGs.

Are stellar population studies of ETGs therefore not useful?  We
believe that they are, even if they only address a small fraction of
the mass of the population.  Our results suggest that \emph{something
interesting} has happened in the Coma Cluster ETGs that appears not to
be reproduced by current galaxy formation models or expectations from
observations of high redshift galaxies.  However, galaxy formation
models have not (yet) examined the ages of ETGs in the same way as we
determine them locally -- i.e., they do not attempt to model
$t_{\mathrm{SSP}}$.  We (Trager \& Somerville, in prep.) are modelling
line strengths and SSP-equivalent stellar population parameters to see
if our predictions above are still valid when considering the
observational quantities presented in this paper.  By comparing the
results of stellar population analysis of real galaxies to the stellar
populations of model galaxies, we will be able to test the validity of
our galaxy formation models, helping us to understand the formation
processes in real ETGs.

The formation processes of ETGs -- those in clusters or in the field
-- are clearly more complicated than simple, rapid quenching of star
formation leading to downsizing.  Our results show that we can place
new constraints on models of these processes.  Of course, considering
the ETGs in just one local cluster is a necessary but not sufficient
step forward in understanding their formation and evolution.  Further
clusters must be tested with data of the same quality that (or better
than) we have presented here.

\section*{Acknowledgements}

The authors wish to recognise and acknowledge the very significant
cultural role and reverence that the summit of Mauna Kea has always
had within the indigenous Hawaiian community.  We are most fortunate
to have had the opportunity to conduct observations from this
mountain.

It is a pleasure to thank E.~Bell, J.~Gorgas, J.~van Gorkom, A.~Helmi,
D.~Kelson, S.~Khochfar, C.~Maraston, D.~Mehlert, B.~Poggianti,
J.~Rose, P.~S\'anchez-Bl\'azquez, R.~Schiavon, P.~Serra, R.~Shipman,
R.~Smith, R.~Somerville, D.~Thomas, and G.~Worthey for helpful
discussions.  We also thank D.~Kelson for much very useful software;
A.~Phillips for slitmask design software; G.~Worthey for providing
stellar population models in advance of publication; J.M.~van der
Hulst for access to the images of \citet{Beijersbergen} used to
compute surface brightness parameters of GMP 3565; C.~Peng for helpful
advice on the use of GALFIT; P.~Serra for help with GANDALF, and
M.~Sarzi, J.~Falcon-Barroso \& R.~Peletier for writing GANDALF and
making it available; B.~Poggianti for an electronic copy of the
\citet{P01} data for galaxies in common with our sample and a careful
reading of an early draft of the manuscript; J.~B.~Oke and J.~Cohen
for designing and building the LRIS spectrograph; J.~Nelson and the
entire CARA staff past and present for designing, building, and
maintaining the Keck Telescopes; the directors of Lick/UCO and Keck
Observatories for the generous allocation of observing time; and
finally Pele for bringing us clouds but not humidity on 7 April 1997.
Support for this work was provided by NASA through Hubble Fellowship
grant HF-01125.01-99A to SCT awarded by the Space Telescope Science
Institute, which is operated by the Association of Universities for
Research in Astronomy, Inc., for NASA under contract NAS 5-26555; by a
Carnegie Starr Fellowship to SCT; by NSF grant AST-9529098 to SMF; and
by NASA contract NAS5-1661 to the WF/PC-I IDT.  This research has made
use of the NASA/IPAC Extragalactic Database (NED) which is operated by
the Jet Propulsion Laboratory, California Institute of Technology,
under contract with the National Aeronautics and Space Administration.
This research has also made use of the Sloan Digital Sky Survey
(SDSS).  Funding for the SDSS and SDSS-II has been provided by the
Alfred P. Sloan Foundation, the Participating Institutions, the
National Science Foundation, the U.S. Department of Energy, the
National Aeronautics and Space Administration, the Japanese
Monbukagakusho, the Max Planck Society, and the Higher Education
Funding Council for England. The SDSS Web Site is
http://www.sdss.org/.

\begin{appendix}

\section{Calibrating onto the Lick/IDS system}
\label{sec:appcal}

\subsection{Initial calibration}

As described below, the wavelengths of the Lick/IDS system bandpasses
are defined relative to a few template stars.  Moreover, the Lick/IDS
system is defined at a resolution that varies from about 8 \AA\ at
5000 \AA\ to 10--12 \AA\ at the extreme blue (4000 \AA) and red (6400
\AA) ends of the system \citep{WO97}.  As a first step, we choose a
template star on which to define the wavelength system.  The K1 giant
HR 6018 is the template for G and K stars and most galaxies on the
Lick/IDS system; we observed this star as well
(\S\ref{sec:selection}).  Next we determine the intrinsic resolution
of the template $\sigma_{int}$, which is done by fitting (using LOSVD)
the spectrum of the template to a digital echellogram of Arcturus
\citep{HWVH00}.  The template spectrum is then smoothed to the
Lick/IDS resolution using a variable-width Gaussian filter with an
intrinsic dispersion of
\begin{equation}
\sigma_{\mathrm{Lick/IDS}}=3492.88-1.30364\,\lambda+0.000128619\,\lambda^2\;
\ \kms\label{eq:idsres}
\end{equation}
(with $\lambda$ in \AA), determined from fitting our spectrum of HR
6018 to the Lick/IDS spectrum of this star.  This quadratic fit to the
resolution data is very nearly that given by \citet{WO97}.  The net
smoothing kernel has a width
$\sigma_b=(\sigma_{\mathrm{Lick/IDS}}^2-\sigma_{int}^2)^{1/2}$
\citep[cf.][]{PS02}.

We next place the wavelengths of the usable Lick/IDS bandpasses on the
smoothed template spectrum.  Due to the observational material from
which it was defined, the Lick/IDS system is not simple to reproduce
\citep[see e.g.,][just to name a few descriptions of the steps
required]{WO97,K00}.  One particular issue is the wavelength scale of
the Lick/IDS system.  As described by \citet{WFGB94} and
\citet{TWFBG98}, the zero-point and scale of the IDS spectra could
change between observing runs and even between consecutive exposures
as the local magnetic field changed and altered the channel the
incoming electrons hit on the IDS detector.  Each IDS stellar spectrum
was therefore adjusted to have zero redshift and a fixed
\emph{average} wavelength scale, set in AUTOINDEX
\citep{WFGB94,TWFBG98} by fixing the wavelengths of the strongest two
features at (roughly) either end of the spectrum.  For cool giants and
dwarfs, the (blended head of the) G band and the (blended) Na D
doublet were used, defined to have wavelengths of 4306.000 \AA\ and
5894.875 \AA\ respectively; for hot dwarfs, H$\gamma$ was used in the
blue; for very cool stars, \mbox{Ca\,\textsc{i}} was used in the blue.

However, small-scale fluctuations in the wavelength scale still
persisted.  To overcome this difficulty, AUTOINDEX implemented an
index centring scheme that used a high-quality template star to place
the bandpasses on each index \citep{WFGB94,TWFBG98}.  There were three
templates: HR 6018 for G-K stars, HR 8430 for mid-F and earlier stars,
and HR 6815 for early- to mid-M stars.  The bandpasses on each
template were carefully placed to best reproduce the `eye' system of
\citet{BFGK84}.  Therefore, the \emph{true} wavelength definitions of
the Lick/IDS passbands can be traced to the wavelength scales of these
three stars.  The passband definitions of \citet{WFGB94} and
\citet{TWFBG98} were based on a comparison of the bandpasses given by
the Lick/IDS template stars to the wavelength scales of modern CCD
spectra taken by G. Worthey and J. Gonz\'alez \citep[see][]{G93}.

In the current study, we use a scheme (SPINDEX2; see below) very
similar to AUTOINDEX, in which a template star is defined to have the
`correct' passband definitions and then is used to centre the indexes
on each spectrum of interest.  This was necessary in part because it
is difficult to calibrate the wavelength scale of LRIS-R spectra in
the blue region that concerns us here.  The template star \emph{must}
be `on' the Lick/IDS system in order to calibrate the line strengths
of the individual objects onto that system.  Fortunately, one of our
comparison stars is the Lick/IDS K giant standard HR 6018
(\S\ref{sec:selection}), also the template for cool stars and almost
all galaxies in the Lick/IDS system.  Using SPINDEX2, we smoothed our
spectrum of HR 6018 to the resolution of the Lick/IDS system, after
correcting this spectrum to zero velocity as described above.  We then
shifted the bandpasses given in the original AUTOINDEX template file
for HR 6018 by measuring the velocity shifts of each index in the IDS
spectrum of HR 6018 (observation 550010, that used as a template for
the Lick/IDS system) with respect to the smoothed LRIS spectrum of HR
6018.  The wavelength shifts are generally no more than 1.25 \AA\ and
typically $\pm0.125$ to $\pm0.375$ \AA\ for most indexes in the
observed range of the current data (from \cnone\ to Fe5406).

\subsection{Emission corrections}
\label{sec:emission}

As discussed in previous works \citep[e.g.,][]{G93,GE96,K00}, nebular
emission lines due to, e.g., low-luminosity AGN \citep{HFS97} are
common in ETGs.  Emission lines of atomic hydrogen, oxygen, and
nitrogen can pollute the absorption-line indexes and distort the age
and metallicity estimates.  The hydrogen Balmer line indexes
(H$\delta_{A,F}$, H$\gamma_{A,F}$, \hbeta), Fe5015 (which contains
both [\mbox{O\,\textsc{iii}}]$\lambda4959$ \AA\ and
[\mbox{O\,\textsc{iii}}]$\lambda5007$ \AA) and \mgb\ (which contains
[\mbox{N\,\textsc{i}}] in its red sideband, \citealt{GE96}) are all
susceptible to emission-line contamination.

\begin{figure}
\includegraphics[width=89mm]{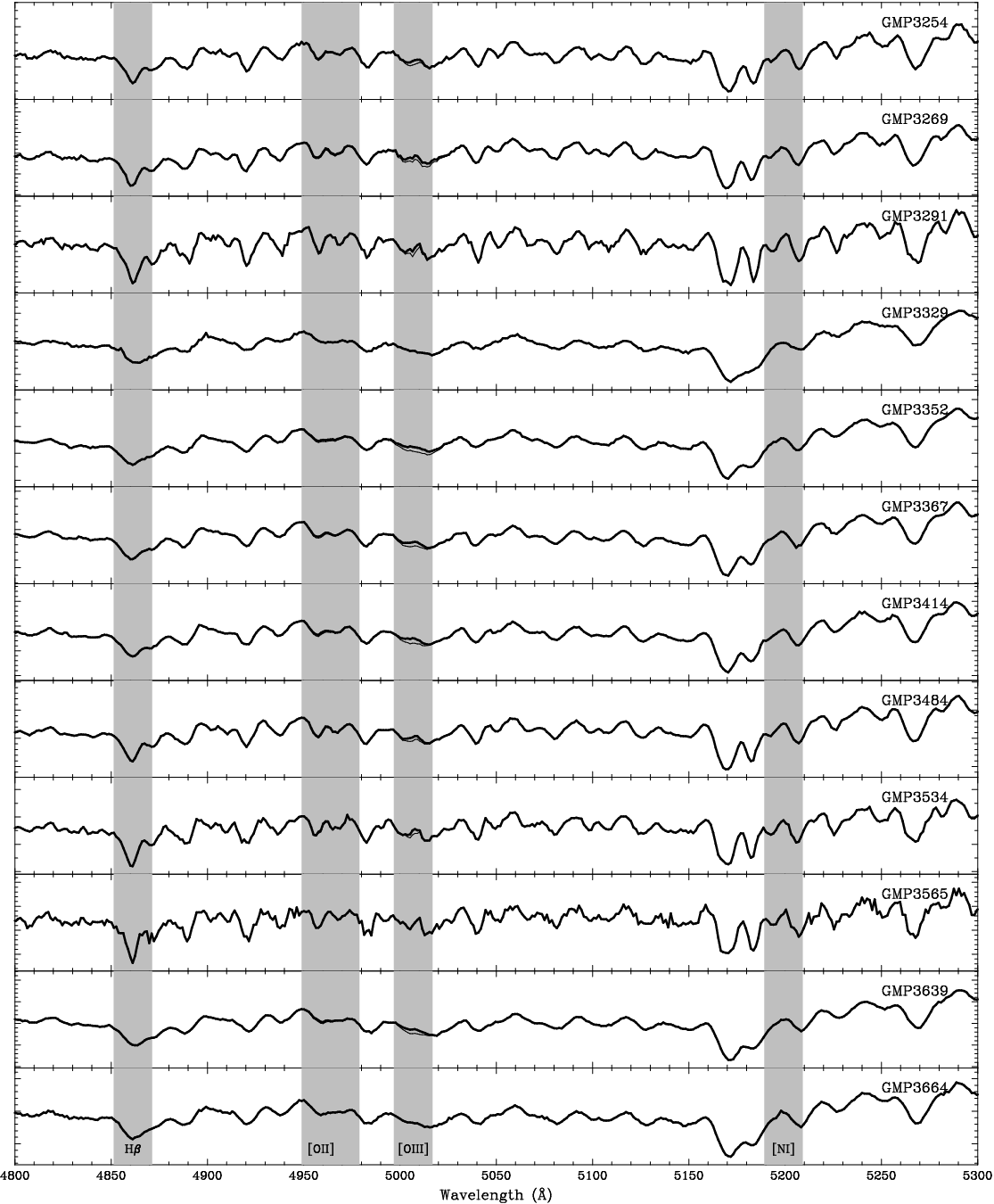}
\caption{Observed and emission-cleaned spectra of Coma ETGs.  Spectra
  from the 2\farcs7 apertures (thick lines) have been fit with
  Gaussian emission lines and model spectra from Vazdekis (in prep.)
  and then cleaned of emission (thin lines) using GANDALF
  \citep{Sarzi06}.  The emission-line index definitions of \hbeta,
  [\mbox{O\,\textsc{iii}}]$\lambda4959$ \AA, and
  [\mbox{O\,\textsc{iii}}]$\lambda5007$ \AA\ from \citet{G93} have
  been over-plotted as grey boxes.  A 20 \AA-wide `index' around the
  [\mbox{N\,\textsc{i}}]$\lambda\lambda5197.9,5200.4$ \AA\ doublet
  \citep{GE96} has also been over-plotted.\label{fig:residuals}}
\end{figure}

We have used GANDALF \citep{Sarzi06} to determine possible emission
contamination of our spectra by simultaneously fitting Gaussian
emission lines and stellar population model spectral templates
(Vazdekis, in prep., based on the spectra of
\citealt{SB06c})\footnote{Note that we use the kinematics measured
using the method described in \S\ref{sec:sigmaindex}, not those
determined with GANDALF, which are only used for fitting the model
templates to determine emission corrections.}.  We accept an emission
line to be significantly detected if the ratio of the amplitude of the
line to the expected noise $A/N>2$.  We find that while nearly of our
galaxies (except GMP 3565) have detectable
[\mbox{O\,\textsc{iii}}]$\lambda5007$ \AA\ emission, we do not detect
\emph{significant} \hbeta\ emission in \emph{any} of our galaxies in
the 2\farcs7 or `physical' \reo{2}\ apertures.  Nor do we detect any
significant [\mbox{N\,\textsc{i}}].  Because we have not detected
\hbeta\ emission in any galaxy, we have not bothered to correct for
emission in the higher-order Balmer lines.  We have measured line
strengths from the \emph{emission-cleaned} spectra rather than making
the corrections outlined in \citet{T00a} and \citet{K00}.

\subsection{Measuring line strengths}

For each object of interest (star or galaxy), we now measure the line
strengths on the (emission-cleaned) spectrum.  The spectrum is first
smoothed to the Lick/IDS resolution (Eq.~\ref{eq:idsres}).  Using the
systemic velocity given by LOSVD, the bandpasses are placed on the
spectrum.  For each index, LOSVD is then used to determine the offset
between the object and template spectra in a wavelength region that
extends 20 \AA\ from the extremes of the index definition.  This
places the index bandpasses \emph{precisely} on the Lick/IDS index
definition, as described above \citep[thereby following the AUTOINDEX
algorithm:][]{WFGB94,TWFBG98}.  Indexes and index errors are then
computed from the object spectrum and its variance spectrum using the
formalism described by \citet{G93}, namely that an index measured in
\AA\ is computed as
\begin{equation}
\mathrm{EW}=\int_{\lambda_{c_1}}^{\lambda_{c_2}}\!
\left(1-\frac{S(\lambda)}{C(\lambda)}\right)\;d\lambda
\label{eq:ew}
\end{equation}
and an index measured in magnitudes is computed as
\begin{equation}
\mathrm{Mag}=-2.5\log\left[{\left(\frac{1}{\lambda_{c_1}-\lambda_{c_2}}\right)
\int_{\lambda_{c_1}}^{\lambda_{c_2}}\!\frac{S(\lambda)}{C(\lambda)}\;
d\lambda}\right].
\end{equation}
Here $\lambda_{c_1}$ and $\lambda_{c_2}$ are the wavelength limits of
the central bandpass, $S(\lambda)$ is the observed flux per unit
wavelength in the object spectrum, and $C(\lambda)$ is the
linearly-interpolated pseudo-continuum:
\begin{eqnarray}
C(\lambda)&=&S_b\frac{\lambda_r-\lambda}{\lambda_r-\lambda_b}+
              S_r\frac{\lambda-\lambda_b}{\lambda_r-\lambda_b},\ \mathrm{where}\\
       S_b&=&\frac{\int_{\lambda_{b_1}}^{\lambda_{b_2}}S(\lambda)d\lambda}{\lambda_{b_2}-\lambda_{b_1}}\ \mathrm{and}\\
       S_r&=&\frac{\int_{\lambda_{r_1}}^{\lambda_{r_2}}S(\lambda)d\lambda}{\lambda_{r_2}-\lambda_{r_1}},
\end{eqnarray}
with $\lambda_b=(\lambda_{b_1}+\lambda_{b_2})/2$ and
$\lambda_r=(\lambda_{r_1}+\lambda_{r_2})/2$.  Errors are then computed
using the variance spectrum $V(\lambda)$:
\begin{eqnarray}
\sigma(\mathrm{EW})&=&\frac{S_c}{C_c}\left[\frac{V_c}{S_c^2}+
\frac{V_b}{C_c^2}\left(\frac{\lambda_r-\lambda_c}{\lambda_r-\lambda_b}\right)^2+
\frac{V_r}{C_c^2}\left(\frac{\lambda_c-\lambda_b}{\lambda_r-\lambda_b}\right)^2\right]^{1/2} \label{eq:ewerr}\\
\sigma(\mathrm{Mag})&=&\frac{2.5\times10^{0.4\mathrm{Mag}}}{\ln(10)(\lambda_{c_1}-\lambda_{c_2})}\sigma(\mathrm{EW}), \label{eq:magerr}
\end{eqnarray}
where $\lambda_c=(\lambda_{c_1}+\lambda_{c_2})/2$, $C_c=C(\lambda_c)$,
$S_c=\int_{\lambda_{c_1}}^{\lambda_{c_2}}\! S(\lambda)d\lambda$, and
\begin{eqnarray}
V_c&=&S_c^2 / \int_{\lambda_{c_1}}^{\lambda_{c_2}}\!
             \frac{S^2(\lambda)}{V(\lambda)},\\
V_b&=&S_b^2 / \int_{\lambda_{b_1}}^{\lambda_{b_2}}\!
             \frac{S^2(\lambda)}{V(\lambda)},\\
V_r&=&S_r^2 / \int_{\lambda_{r_1}}^{\lambda_{r_2}}\!
             \frac{S^2(\lambda)}{V(\lambda)}
\end{eqnarray}
\citep[cf.\ the discussion in][]{CGCG98}.  The implementation of this
algorithm in Python is called SPINDEX2\footnote{We note here that
  fractional pixels are handled in the same manner as Gonz\'alez, so
  indexes measured by SPINDEX2 are identical to those measured at the
  same bandpass wavelengths by SPINDEX in the VISTA image processing
  package.}; when coupled with LOSVD to measure systemic velocities
and velocity dispersions (\S\ref{sec:sigma}), the program is called
SPINDLOSVD.

\subsubsection{Reliability of estimated errors}

\begin{table}
  \caption{Error ratios from 2\farcs7-synthesised aperture line
    strengths}
  \label{tbl:errors}
  \begin{tabular}{lcc}
    \hline
    Index&\multicolumn{1}{c}{$\langle\sigma_c/\sigma_s\rangle$}&
    \multicolumn{1}{c}{std.\ dev.}\\
    \hline
    \cnone&1.13&0.68\\
    \cntwo&1.11&0.82\\
    Ca4227&2.21&1.27\\
    G4300&1.52&0.78\\
    Fe4383&1.27&0.87\\
    Ca4455&1.77&1.09\\
    Fe4531&1.54&0.66\\
    \ctwo&1.05&0.70\\
    \hbeta&1.24&0.74\\
    Fe5015&0.78&0.47\\
    \mgone&0.93&0.60\\
    \mgtwo&0.67&0.56\\
    \mgb&1.08&0.57\\
    Fe5270&1.55&0.93\\
    Fe5335&1.60&0.97\\
    \fe&1.12&0.95\\
    \hda&1.27&0.79\\
    \hga&1.50&0.81\\
    \hdf&1.40&0.40\\
    \hgf&1.64&1.17\\
    \hbetag&1.25&1.07\\
    \hline
  \end{tabular}

  Col.\ 1: Index name.  Col.\ 2: Median error ratio, in the sense
  error inferred from combined spectra divided by standard deviation
  of index strengths derived from individual exposures.  Col.\ 3:
  Standard deviation of error ratio.
\end{table}

To check the accuracy of the errors measured from the variance spectra
of the averaged spectra as given by Equations~\ref{eq:ewerr} and
\ref{eq:magerr} above, line strengths were measured from
one-dimensional spectra extracted from each individual exposure.  The
standard deviation of each index was then determined.
Table~\ref{tbl:errors} gives (1) the median and standard deviations of
the ratios of the errors in each index computed from the combined
spectra and (2) the standard deviations of the index strengths
computed from the individual exposures.  None of the differences is
significant, although some of the means differ from one (e.g., Ca4227
and \mgtwo).  Much of this scatter likely arises from interpolation
errors when extracting the one-dimensional spectra from the individual
exposures, a problem greatly ameliorated when three images are
combined during the extraction.  We therefore believe that the error
estimates for the LRIS absorption-line strengths are likely to be
reliable.

\subsection{Calibration onto the stellar Lick/IDS system: offsets and
  velocity-dispersion corrections}
\label{sec:indexcorrections}

\begin{table}
  \caption{Corrections required to bring LRIS stellar indexes onto
    Lick/IDS system}
 \label{tbl:calib}
 \begin{tabular}{lrl}
   \hline
   Index&$\langle\mathrm{(IDS-LRIS)}\rangle^\mathrm{a}$&
   \multicolumn{1}{c}{RMS}\\
   \hline
   \cnone&$0.0034\pm0.0008$&$0.0118$\\ 
   \cntwo&$0.0063\pm0.0005$&$0.0064$\\ 
   Ca4227&$0.034\phn\pm0.004\phn$&$0.057$\\ 
   G4300 &$-0.236\phn\pm0.006\phn$&$0.084$\\
   Fe4383&$-0.033\phn\pm0.006\phn$&$0.094$\\ 
   Ca4455&$-0.077\phn\pm0.004\phn$&$0.058$\\
   Fe4531&$-0.062\phn\pm0.006\phn$&$0.085$\\
   \ctwo &$-0.325\phn\pm0.012\phn$&$0.177$\\
   \hbeta&$0.043\phn\pm0.004\phn$&$0.056$\\ 
   Fe5015&$0.059\phn\pm0.010\phn$&$0.150$\\ 
   \mgone&$0.0139\pm0.0005$&$0.0082$\\ 
   \mgtwo&$0.0185\pm0.0004$&$0.0067$\\ 
   \mgb  &$0.018\phn\pm0.009\phn$&$0.139$\\ 
   Fe5270&$0.031\phn\pm0.006\phn$&$0.090$\\ 
   Fe5335&$0.176\phn\pm0.006\phn$&$0.091$\\ 
   Fe5406&$0.094\phn\pm0.003\phn$&$0.050$\\ 
   \hda  &$-0.143\phn\pm0.006\phn$&$0.083$\\
   \hga  &$0.612\phn\pm0.015\phn$&$0.219$\\ 
   \hdf  &$0.123\phn\pm0.004\phn$&$0.057$\\ 
   \hgf  &$0.174\phn\pm0.006\phn$&$0.085$\\ 
   \hline
 \end{tabular}

$^\mathrm{a}$Weighted mean
\end{table}

\begin{figure*}
\includegraphics[width=178mm]{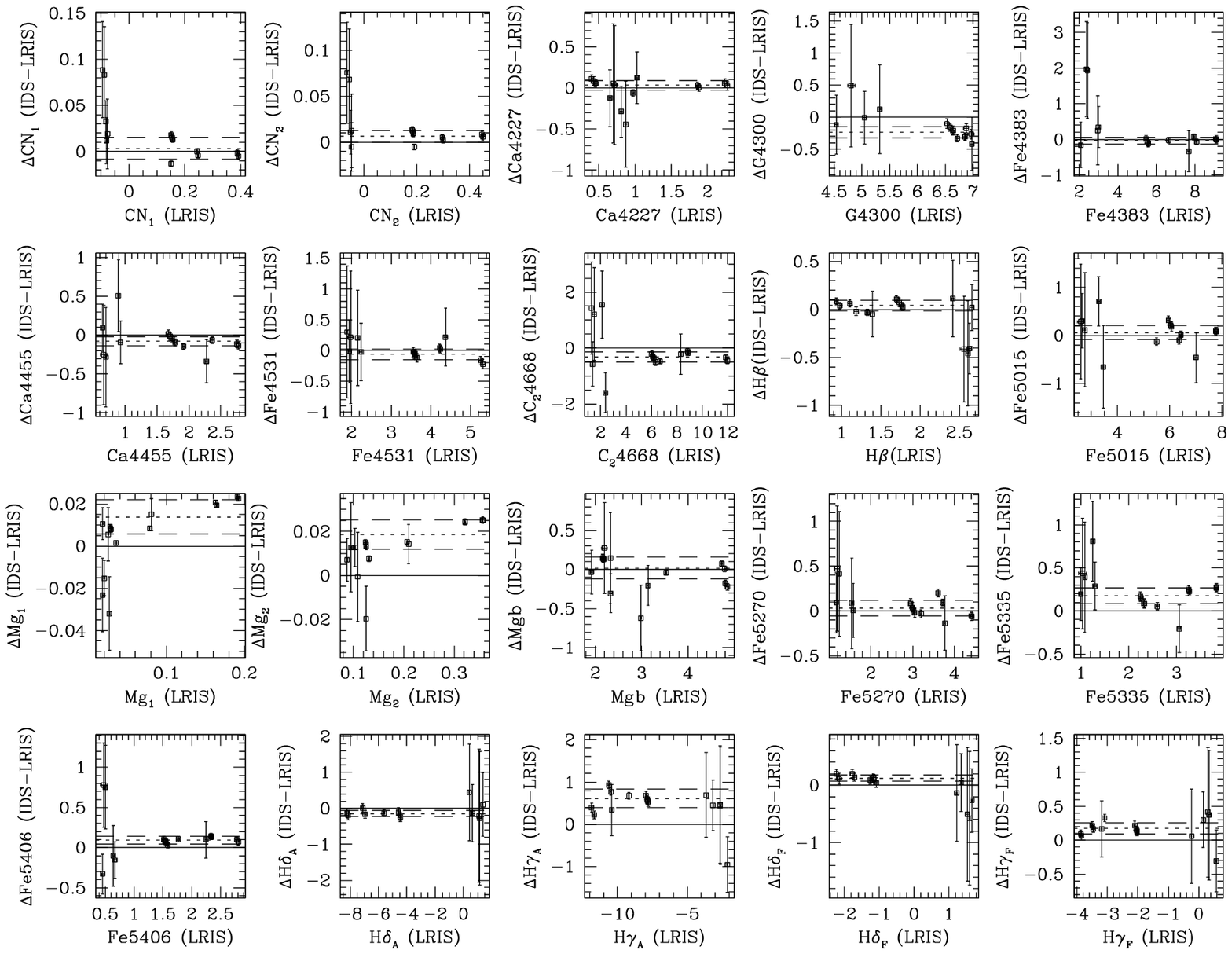}
\caption{Calibration onto the Lick/IDS system.  Line strengths of
  Lick/IDS stars taken with LRIS were measured with SPINDEX2 (after
  smoothing to the Lick/IDS resolution, \citealt{WO97}) and compared
  with the published Lick/IDS line strengths \citep{WFGB94,WO97}.
  Vertical error bars include uncertainties in the Lick/IDS standard
  star system \citep{WFGB94}, scaled using the `goodness'
  signal-to-noise parameter \citep{TWFBG98} and the number of
  observations of each star.  Offsets from the Lick/IDS system are
  seen to be small for most indexes, except for the well known offset
  of \mgtwo\ due to the spectral shape of the quartz lamp used to
  `flux' the IDS spectra \citep{G93}. Dotted and dashed lines
  indicate the weighted mean offsets and RMS deviations
  (Table~\ref{tbl:calib}); solid lines are lines of equality.  Note
  that some offsets might be better modelled as linear functions of
  line strength (e.g., \mgone\ and \mgtwo).\label{fig:compids}}
\end{figure*}

Because the flux calibration of the present spectra differs from that
of the Lick/IDS spectra \citep[which were not fluxed but divided by a
quartz lamp;][]{WFGB94}, small offsets may required to finally bring
absorption-line strengths of objects taken with LRIS onto the Lick/IDS
system.  These offsets are determined by comparing line strengths of
the LRIS stars measured with SPINDEX2 to the published values for
those stars, as shown in Figure~\ref{fig:compids}.  The weighted mean
offsets and root-mean-square deviations are given in
Table~\ref{tbl:calib}.  As seen by other authors
\citep[e.g.,][]{G93,K00}, most indexes have negligible offsets
\emph{within the typical errors of the Lick/IDS system}.  The
exceptions are \mgone\ and \mgtwo, for which the Lick/IDS zero-point
was set by the quartz lamp used to `flux' the spectra, and \hga, for
unknown reasons, although the inferred offset is consistent with that
found by both \citet{WO97} and \citet{K00}.  Fe5015 has a large
scatter for unknown reasons, but it is well within the typical
Lick/IDS error.  Note that \mgone\ and \mgtwo\ may be better fit by a
linear relation than a simple offset.

Before these offsets are applied to the line strengths of a galaxy, we
must first correct them for its velocity dispersion.  This is done for
the original 21 Lick/IDS line strengths \citep{WFGB94} using the
multiplicative corrections given by \citet{TWFBG98}.  For the four
higher-order Balmer line indexes defined by \citet{WO97}, we use the
multiplicative corrections given by \citet{LW05}.  For \hbetag\
\citep{J99}, the \hbeta\ correction given by \citet{TWFBG98} was used.
As a check on our method, we have compared our
velocity-dispersion-corrected indexes (before Lick/IDS system
correction) to indexes corrected using the formalism of
\citet{KIFvD06}, based on broadening templates but not the program
spectra, and have found offset and slope differences from our method
of less than $1\sigma$ for all indexes except Ca4227, Fe4531, and
\hdf\ (offset) and \mgb\ (offset and slope), where the deviation in
the latter index is small ($-0.13$ \AA) and only in the strongest
(highest-\mgb, highest-$\sigma$) objects.  Due to the
small-to-negligible differences, and more importantly, to be
consistent with previous studies, we use the \citet{TWFBG98}
corrections but note that Kelson et al.'s comments on the suitability
of \emph{additive} rather than \emph{multiplicative}
velocity-dispersion corrections should taken into consideration.

\section{Comparison of LRIS data with literature data sources}
\label{sec:others}

We now compare our systemic velocities (redshifts), velocity
dispersions, and line strengths with those found in the literature.
In order to determine the differences most accurately, the LRIS
indexes were measured on spectra with apertures matched as closely as
possible to the literature data (Sec.~\ref{sec:reduction}).  For the
study of \citet{Mehlert03}, however, the published \reo{10}\ apertures
were much too small for nearly all of the galaxies observed in the
present sample, so the gradient measures of \citet{Mehlert00} were
used to compute index strengths in 2\farcs7-diameter equivalent
circular apertures; Fe5270 and Fe5335 indexes and gradients were
kindly provided by Dr.~D.~Mehlert.  We also note that the line
strengths of \citet{TKBCS99} are not truly calibrated onto the
Lick/IDS system, but rather used the \citet{K00} offsets as a rough
correction.  This is particularly problematic for the \hda\ index, as
the \citet{K00} study did not include the bluest indexes.  Finally, we
used the 2\farcs7 velocity dispersion measurements to compare with the
results of \citet{MG05}, as velocity dispersion gradients in these
galaxies are likely to be nearly flat \citep[cf.][]{JFK96}.

\begin{table}
  \caption{Comparison with other absorption-line strength studies of
    the Coma Cluster, in the sense Literature$-$LRIS}
  \label{tbl:comparisons}
  \begin{tabular}{llrrrr}
    \hline
    Quantity&Source&\multicolumn{1}{c}{N}&\multicolumn{1}{c}{Offset}&
    \multicolumn{1}{c}{RMS}&\multicolumn{1}{c}{$\chi^2$}\\
    \hline
$cz_{hel}$&H01&3&$133.32\pm3.44$&5.95&171.20\\*
$cz_{hel}$&M02&8&$97.51\pm1.51$&4.28&70.65\\*
$cz_{hel}$&NFPS&7&$123.65\pm0.78$&2.07&557.06\\*
$cz_{hel}$&ALL&18&$120.00\pm0.42$&1.78&276.57\\[6pt]
$\log\sigma$&D84&4&$-0.022\pm0.010$&0.020&0.803\\*
$\log\sigma$&G92&1&$0.025\pm\phm{0.000}$&&\\*
$\log\sigma$&H01&3&$-0.021\pm0.006$&0.010&1.847\\*
$\log\sigma$&J99&8&$-0.004\pm0.003$&0.009&1.220\\*
$\log\sigma$&K01&2&$-0.001\pm\phm{0.000}$&&\\*
$\log\sigma$&MG05&6&$-0.014\pm0.003$&0.007&2.651\\*
$\log\sigma$&M00&3&$0.019\pm0.009$&0.015&1.594\\*
$\log\sigma$&M02&8&$-0.010\pm0.001$&0.004&11.490\\*
$\log\sigma$&NFPS&7&$-0.009\pm0.002$&0.005&4.338\\*
$\log\sigma$&SB06&5&$0.033\pm0.003$&0.006&10.820\\*
$\log\sigma$&ALL&47&$-0.004\pm0.000$&0.002&4.743\\[6pt]
\cnone&NFPS&7&$-0.011\pm0.002$&0.006&1.986\\*
\cnone&P01&5&$-0.040\pm0.003$&0.006&22.562\\*
\cnone&IDS&2&$0.010\pm\phm{0.000}$&&\\*
\cnone&ALL&14&$-0.024\pm0.001$&0.004&9.076\\[6pt]
\cntwo&NFPS&7&$-0.020\pm0.003$&0.008&2.079\\*
\cntwo&P01&5&$-0.026\pm0.003$&0.007&10.909\\*
\cntwo&SB06&5&$-0.008\pm0.004$&0.009&1.236\\*
\cntwo&IDS&2&$0.012\pm\phm{0.000}$&&\\*
\cntwo&ALL&19&$-0.018\pm0.001$&0.004&3.990\\[6pt]
Ca4227&NFPS&7&$0.189\pm0.027$&0.071&1.346\\*
Ca4227&P01&5&$-0.322\pm0.043$&0.097&2.947\\*
Ca4227&SB06&5&$-0.179\pm0.023$&0.052&2.527\\*
Ca4227&IDS&2&$-0.249\pm\phm{0.000}$&&\\*
Ca4227&ALL&19&$-0.095\pm0.009$&0.038&1.970\\[6pt]
G4300&NFPS&7&$0.067\pm0.040$&0.107&0.349\\*
G4300&P01&5&$0.108\pm0.088$&0.198&4.334\\*
G4300&SB06&5&$0.030\pm0.042$&0.095&1.447\\*
G4300&IDS&2&$0.505\pm\phm{0.000}$&&\\*
G4300&ALL&19&$0.057\pm0.015$&0.065&1.851\\[6pt]
Fe4383&NFPS&7&$-0.469\pm0.053$&0.140&2.252\\*
Fe4383&P01&5&$0.370\pm0.108$&0.242&2.694\\*
Fe4383&SB06&5&$0.057\pm0.057$&0.127&0.774\\*
Fe4383&T99&6&$0.547\pm0.027$&0.065&13.156\\*
Fe4383&IDS&2&$-0.214\pm\phm{0.000}$&&\\*
Fe4383&ALL&25&$0.309\pm0.010$&0.052&4.491\\[6pt]
Ca4455&NFPS&7&$-0.384\pm0.020$&0.053&8.421\\*
Ca4455&P01&5&$-0.091\pm0.058$&0.130&4.696\\*
Ca4455&SB06&5&$-0.092\pm0.032$&0.071&0.731\\*
Ca4455&IDS&2&$-0.257\pm\phm{0.000}$&&\\*
Ca4455&ALL&19&$-0.258\pm0.009$&0.040&4.641\\[6pt]
Fe4531&NFPS&7&$-0.275\pm0.032$&0.084&1.891\\*
Fe4531&P01&5&$-0.423\pm0.093$&0.207&1.968\\*
Fe4531&SB06&5&$-0.383\pm0.050$&0.113&2.648\\*
Fe4531&IDS&2&$0.153\pm\phm{0.000}$&&\\*
Fe4531&ALL&19&$-0.316\pm0.015$&0.063&1.959\\
    \hline
  \end{tabular}
\end{table}

\begin{table}
  \contcaption{}
  \begin{tabular}{llrrrr}
    \hline
    Quantity&Source&\multicolumn{1}{c}{N}&\multicolumn{1}{c}{Offset}&
    \multicolumn{1}{c}{RMS}&\multicolumn{1}{c}{$\chi^2$}\\
    \hline
\ctwo&FFI&1&$-0.286\pm\phm{0.000}$&&\\*
\ctwo&M02&8&$0.012\pm0.062$&0.175&2.036\\*
\ctwo&NFPS&7&$0.079\pm0.061$&0.162&0.253\\*
\ctwo&P01&5&$0.565\pm0.106$&0.237&12.830\\*
\ctwo&SB06&5&$-0.962\pm0.106$&0.237&4.382\\*
\ctwo&T99&6&$-0.277\pm0.030$&0.074&8.030\\*
\ctwo&IDS&2&$-1.053\pm\phm{0.000}$&&\\*
\ctwo&ALL&34&$-0.198\pm0.010$&0.058&4.580\\[6pt]
\hbeta&FFI&1&$-0.104\pm\phm{0.000}$&&\\*
\hbeta&J99&4&$0.178\pm0.048$&0.096&1.206\\*
\hbeta&K01&2&$0.217\pm\phm{0.000}$&&\\*
\hbeta&M00&3&$0.434\pm0.055$&0.095&10.795\\*
\hbeta&M02&8&$0.008\pm0.015$&0.042&6.926\\*
\hbeta&NFPS&7&$-0.141\pm0.015$&0.038&2.804\\*
\hbeta&P01&5&$0.026\pm0.048$&0.107&1.821\\*
\hbeta&SB06&5&$-0.042\pm0.026$&0.058&0.481\\*
\hbeta&IDS&2&$0.230\pm\phm{0.000}$&&\\*
\hbeta&ALL&37&$-0.014\pm0.004$&0.022&3.537\\[6pt]
Fe5015&FFI&1&$-0.582\pm\phm{0.000}$&&\\*
Fe5015&M02&8&$-0.565\pm0.047$&0.133&2.919\\*
Fe5015&NFPS&7&$-0.057\pm0.043$&0.115&0.802\\*
Fe5015&P01&5&$-0.471\pm0.105$&0.236&2.015\\*
Fe5015&SB06&5&$-0.499\pm0.069$&0.153&3.814\\*
Fe5015&IDS&2&$-0.072\pm\phm{0.000}$&&\\*
Fe5015&ALL&28&$-0.347\pm0.013$&0.070&2.181\\[6pt]
\mgone&J99&4&$0.006\pm0.001$&0.002&4.607\\*
\mgone&M02&8&$0.002\pm0.001$&0.003&0.574\\*
\mgone&NFPS&7&$0.015\pm0.001$&0.003&5.605\\*
\mgone&P01&5&$-0.000\pm0.001$&0.002&10.325\\*
\mgone&IDS&2&$0.002\pm\phm{0.000}$&&\\*
\mgone&ALL&26&$0.006\pm0.000$&0.001&4.972\\[6pt]
\mgtwo&D84&4&$0.012\pm0.002$&0.004&3.719\\*
\mgtwo&G92&1&$0.003\pm\phm{0.000}$&&\\*
\mgtwo&J99&8&$0.006\pm0.001$&0.003&2.262\\*
\mgtwo&K01&2&$-0.005\pm\phm{0.000}$&&\\*
\mgtwo&M02&8&$0.008\pm0.001$&0.003&3.495\\*
\mgtwo&NFPS&7&$-0.013\pm0.001$&0.003&2.605\\*
\mgtwo&P01&5&$-0.036\pm0.001$&0.003&36.954\\*
\mgtwo&IDS&2&$0.007\pm\phm{0.000}$&&\\*
\mgtwo&ALL&37&$-0.004\pm0.000$&0.001&7.404\\[6pt]
\mgb&FFI&1&$-0.028\pm\phm{0.000}$&&\\*
\mgb&J99&4&$0.267\pm0.050$&0.101&1.773\\*
\mgb&K01&2&$-0.247\pm\phm{0.000}$&&\\*
\mgb&M00&3&$-0.152\pm0.059$&0.102&3.260\\*
\mgb&M02&8&$-0.016\pm0.015$&0.041&2.834\\*
\mgb&NFPS&7&$-0.190\pm0.016$&0.043&4.636\\*
\mgb&P01&5&$-0.259\pm0.046$&0.103&5.239\\*
\mgb&SB06&5&$-0.085\pm0.052$&0.116&1.620\\*
\mgb&IDS&2&$0.367\pm\phm{0.000}$&&\\*
\mgb&ALL&37&$-0.086\pm0.004$&0.025&3.036\\[6pt]
Fe5270&M00&3&$-0.070\pm0.063$&0.109&0.164\\*
Fe5270&M02&8&$-0.079\pm0.016$&0.045&2.010\\*
Fe5270&NFPS&7&$-0.115\pm0.019$&0.049&1.163\\*
Fe5270&P01&5&$-0.051\pm0.052$&0.117&1.119\\*
Fe5270&SB06&5&$0.050\pm0.036$&0.082&0.780\\*
Fe5270&IDS&2&$0.248\pm\phm{0.000}$&&\\*
Fe5270&ALL&30&$-0.072\pm0.005$&0.028&1.295\\
    \hline
  \end{tabular}
\end{table}

\begin{table}
  \contcaption{}
  \begin{tabular}{llrrrr}
    \hline
    Quantity&Source&\multicolumn{1}{c}{N}&\multicolumn{1}{c}{Offset}&
    \multicolumn{1}{c}{RMS}&\multicolumn{1}{c}{$\chi^2$}\\
    \hline
Fe5335&M00&3&$0.030\pm0.070$&0.121&1.395\\*
Fe5335&M02&8&$-0.091\pm0.023$&0.064&1.775\\*
Fe5335&NFPS&7&$-0.210\pm0.021$&0.054&4.050\\*
Fe5335&P01&5&$-0.381\pm0.060$&0.135&1.942\\*
Fe5335&SB06&5&$0.135\pm0.041$&0.092&1.805\\*
Fe5335&IDS&1&$0.264\pm\phm{0.000}$&&\\*
Fe5335&ALL&29&$-0.114\pm0.006$&0.035&2.273\\[6pt]
\fe&J99&4&$0.048\pm0.046$&0.092&0.542\\*
\fe&K01&2&$-0.213\pm\phm{0.000}$&&\\*
\fe&M00&3&$-0.019\pm0.047$&0.081&0.978\\*
\fe&M02&8&$-0.084\pm0.014$&0.039&2.793\\*
\fe&NFPS&7&$-0.160\pm0.014$&0.037&3.689\\*
\fe&P01&5&$-0.214\pm0.040$&0.089&2.014\\*
\fe&SB06&5&$0.091\pm0.028$&0.062&2.321\\*
\fe&IDS&1&$-0.004\pm\phm{0.000}$&&\\*
\fe&ALL&35&$-0.087\pm0.004$&0.022&2.240\\[6pt]
\hda&NFPS&7&$0.327\pm0.075$&0.200&1.183\\*
\hda&P01&5&$1.140\pm0.101$&0.225&15.226\\*
\hda&SB06&5&$0.239\pm0.065$&0.144&4.029\\*
\hda&T99&6&$0.204\pm0.024$&0.058&7.200\\*
\hda&IDS&2&$0.266\pm\phm{0.000}$&&\\*
\hda&ALL&25&$0.263\pm0.010$&0.051&5.914\\[6pt]
\hga&NFPS&7&$-0.478\pm0.059$&0.156&1.754\\*
\hga&P01&5&$-0.028\pm0.095$&0.213&9.560\\*
\hga&SB06&5&$-0.367\pm0.062$&0.138&2.049\\*
\hga&T99&6&$-0.092\pm0.024$&0.058&2.771\\*
\hga&IDS&2&$-0.706\pm\phm{0.000}$&&\\*
\hga&ALL&25&$-0.161\pm0.010$&0.049&3.559\\[6pt]
\hdf&NFPS&7&$-0.139\pm0.046$&0.121&0.477\\*
\hdf&P01&5&$0.097\pm0.068$&0.151&2.277\\*
\hdf&SB06&5&$0.117\pm0.037$&0.082&2.381\\*
\hdf&IDS&2&$0.346\pm\phm{0.000}$&&\\*
\hdf&ALL&19&$0.045\pm0.014$&0.061&1.491\\[6pt]
\hgf&NFPS&7&$-0.308\pm0.026$&0.070&3.158\\*
\hgf&P01&5&$0.218\pm0.059$&0.131&4.310\\*
\hgf&SB06&5&$-0.040\pm0.036$&0.081&0.706\\*
\hgf&IDS&2&$-0.498\pm\phm{0.000}$&&\\*
\hgf&ALL&19&$-0.139\pm0.011$&0.049&2.559\\[6pt]
\hbetag&J99&6&$0.144\pm0.024$&0.059&1.912\\*
\hbetag&M02&8&$-0.046\pm0.011$&0.032&5.253\\*
\hbetag&ALL&14&$-0.002\pm0.007$&0.028&3.821\\
    \hline
  \end{tabular}

Col.\ 1: Quantity of interest.  Col.\ 2: Source of literature data.
These abbreviations are shown in Table~\ref{tbl:literature}, except
ALL is the combination of all studies with data for that quantity.
Col.\ 3: Number of galaxies in common.  Col.\ 4: Weighted mean
difference and (random) error of mean, in sense other$-$LRIS.
Col.\ 5: Root-mean-square of difference.  Col.\ 6: $\chi^2$ of
differences.
\end{table}

\begin{figure*}
\includegraphics[width=178mm]{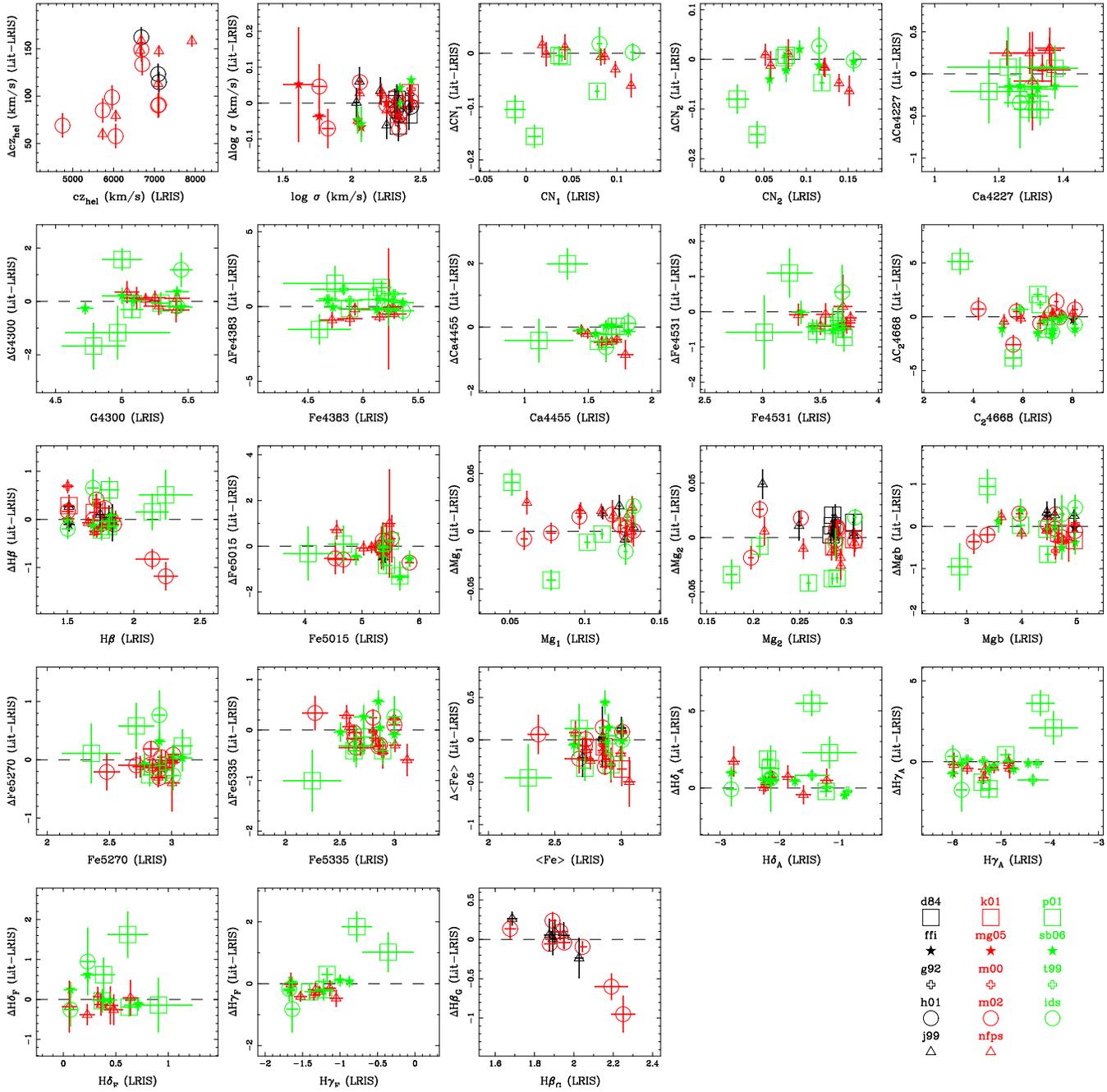}
\caption{Comparison with literature values.  Differences are defined
  as literature$-$LRIS. Key in lower-right; abbreviations are shown in
  Table~\ref{tbl:literature}.\label{fig:comparisons}}
\end{figure*}

Figure~\ref{fig:comparisons} shows the differences in the sense
literature$-$LRIS as a function of LRIS index strength for all of the
indexes measured on the LRIS spectra as well as heliocentric velocity
$cz_{hel}$ and the logarithm of the velocity dispersion $\log\sigma$.
Table~\ref{tbl:comparisons} gives the mean offsets and the errors in
the means ($\sigma_M$), in both cases weighted by the errors of the
differences (the quadratic sum of the LRIS and literature index
errors); the root-mean-squared (RMS) deviation; and the $\chi^2$ of
the sample difference, computed as
\begin{equation}
  \chi^2 = \frac{1}{N}\sum_{i=1}^{N}\frac{d_i^2}{\sigma_i^2}
\end{equation}
where $d_i$ is the difference of index strengths of galaxy $i$ between
a given literature study and LRIS, and $\sigma_i$ is the error of the
difference.  Note that $N<12$ in all cases, as no single literature
study contains all of the galaxies in the LRIS sample.  The `ALL'
entries are error-weighted mean differences for all differences --
i.e., the average difference of all literature index values with
respect to the LRIS index values.

As we claim that we are precisely on the Lick/IDS system, we can view
Table~\ref{tbl:comparisons} and Figure~\ref{fig:comparisons} as guides
to the success of the authors of each study in achieving the same
calibration.  In general, we find that \citet{TKBCS99} and \citet{P01}
are not well-calibrated onto the Lick/IDS system, while \citet{M02}
and \citet{SB06a} do a much better job, with some exceptions discussed
below.  We make no attempt to adjust other studies to our calibration.
We note here that our \hbeta\ strengths are in line with most other
studies, excluding the \citet{Mehlert00} sample, which are clearly
much too strong (see below).

\begin{table}
  \caption{\hbeta\ absorption-line strengths of GMP 3329 (NGC 4874) from all
    sources}
  \label{tbl:ngc4874}
  \begin{tabular}{lrr}
    \hline
    Source&\multicolumn{1}{c}{\hbeta\ (Lit)}&\multicolumn{1}{c}{\hbeta\ (LRIS)}\\
      \hline
      FFI&$1.410\pm0.130$&$1.514\pm0.037$\\
      IDS&$1.307\pm0.196$&$1.503\pm0.040$\\
      J99&$1.780\pm0.120$&$1.512\pm0.036$\\
      K01&$1.800\pm0.210$&$1.513\pm0.033$\\
      M00&$2.140\pm0.123$&$1.503\pm0.040$\\
      M02&$1.674\pm0.132$&$1.503\pm0.040$\\
      NFPS&$1.519\pm0.151$&$1.578\pm0.035$\\
      SB06&$1.496\pm0.100$&$1.503\pm0.040$\\
      \hline
  \end{tabular}

Col.\ 1: Source of literature data, as in Table~\ref{tbl:comparisons}.
Col.\ 2: \hbeta\ strength of GMP 3329 from literature.  Col.\ 3:
\hbeta\ strength of GMP 3329 from LRIS spectra, through the equivalent
aperture as described in text.
\end{table}

We are given further confidence in our calibration by examining the
literature values of the \hbeta\ index strengths of the cD galaxy GMP
3329 (=NGC 4874), shown in Table~\ref{tbl:ngc4874}.  The
error-weighted \hbeta\ index strength in the literature, excluding the
significant outlier from \citet{Mehlert00}, is $1.57\pm0.05$ \AA.
This is within $1\sigma$ of our \hbeta\ index strength for this
galaxy, suggesting that (at least for this bright galaxy) we are on
the Lick/IDS system as well as is possible.  However, the \hbeta\
strength of GMP 3329 (=NGC 4874) in the \citet{Mehlert00,Mehlert03}
studies appears to be much stronger (by $0.64\pm0.13$ \AA) than the
mean of all other literature data, including the LRIS measurement; in
fact, the \citet{Mehlert00,Mehlert03} \hbeta\ strength of this galaxy
is higher by more than $0.3$ \AA\ than \citet{J99} and \citet{K01},
the strongest other available measurements.  It is likely that the
explanation for this excess \hbeta\ strength is over-subtraction of
light of this galaxy due to a slit that was too short (Mehlert,
priv.~comm.).  In other words, the signal taken from the end of the
slits used to correct for the sky brightness was contaminated by light
from the galaxy itself.  Given the large size of this galaxy and its
surface brightness profile, it is likely that this subtraction results
in a too-strong \hbeta\ absorption-line strength (see
Sec.~\ref{sec:reduction} for details on how we have dealt with this
issue).  Finally, we note that the SSP-equivalent age of GMP 3329
inferred from the data of \citet{SB06a} is
$\logt=1.02_{-0.14}^{+0.20}$, higher than the age inferred from the
LRIS data, $\logt=0.90_{-0.04}^{+0.05}$ (Table~\ref{tbl:allcoma}),
even though the \hbeta\ strengths from the two studies are identical
within the errors.  This is due to the difference in the measured
\mgb\ strengths: \citet{SB06a} find $\mgb=4.58\pm0.26$, while we find
$\mgb=4.95\pm0.04$ (Table~\ref{tbl:allcoma}).  We do not know the
cause of the difference in this index for this galaxy; all other
galaxies in common have very similar \mgb\ strengths in the two
samples, and even in GMP 3329 the strengths of the \hbeta, Fe5270, and
Fe5335 indexes all match well between the samples.

Detailed examination of Figure~\ref{fig:comparisons} shows that the
\citet{M02} measurements of GMP 3291 and GMP 3534 are significantly
discrepant in \hbeta\ (and \hbetag).  The cause for this discrepancy
is not understood, but could be due to slit (in the case of the LRIS
spectra) or fibre (in the case of the \citealt{M02} spectra)
misplacement.  Examination of the LRIS slit-alignment images taken
immediately before the spectroscopic exposures (Fig.~\ref{fig:coma})
suggests LRIS slit misplacement is unlikely.  It is also not due to
errors in the velocity dispersion corrections, as both of the galaxies
have $\sigma<75\;\kms$.  Given that the \hbeta\ strengths of both of
these galaxies in the \citet{M02} study are extremely low (GMP 3534
has $\hbeta=1.02$ \AA\ in that study, well below the oldest stellar
population models; e.g., Fig.~\ref{fig:hbmgbfe_lris_w94}), it is
possible that emission fill-in could be the culprit, but no
significant emission is detectable in either galaxy in either study.
(Note that a handful of galaxies, typically of low mass, from
\citealt{M02} have \hbeta\ strengths that cause them to fall below the
oldest stellar population models.)  We also note that GMP 3291 is also
discrepant with respect to the \citet{P01} measurements, but usually
in the opposite sense from the comparison with \citet{M02}.  The LRIS
measurements of GMP 3565 are strongly discrepant with respect to the
\citet{P01} measurements for nearly all indexes.  We have confirmed
through our slit-alignment images that we have definitely targeted GMP
3565, so LRIS slit misplacement is unlikely to be the culprit for the
discrepancies.

\setcounter{table}{2}
\begin{table*}
  \vbox to220mm{\vfil Landscape table to go here
  \vfil}
  \caption{}
  \label{tbl:allcoma}
\end{table*}

Finally, in Table~\ref{tbl:allcoma} we present absorption-line index
strengths and stellar population parameters for all Coma galaxies for
six samples: J99, M00, M02, LRIS, NFPS, and SB06 (see
Table~\ref{tbl:literature} for definitions).  For the M00, SB06, and
LRIS samples, we synthesised $2\farcs7$-diameter equivalent circular
apertures (\S\ref{sec:reduction}); for the other samples, the indexes
are taken as published.  For simplicity, we present only \hbeta, \mgb,
Fe5270, and Fe5335 index strengths and errors, as these were the
indexes used to compute the stellar population parameters.  For
completeness, we repeat the LRIS indexes, stellar population
parameters, and errors here from previous tables.  We have not
attempted to provide `best' indexes or parameters for galaxies with
multiple measurements given the systematic differences between the
samples and differences in aperture size (Tables~\ref{tbl:comparisons}
and \ref{tbl:literature}).

\end{appendix}

\clearpage

\end{document}